\documentclass[pdflatex,sn-apa]{sn-jnl}

\usepackage{tabls}
\usepackage{subcaption}
\usepackage{float}
\usepackage{placeins}
\usepackage{mathtools} 
\usepackage{hyperref} 
\usepackage{url}
\usepackage{tabularx} 
\usepackage{tabu}
\usepackage{rotating}
\usepackage{caption}
\usepackage{float} 
\usepackage{multirow,booktabs,setspace}
\usepackage{tikz}

\usepackage{graphicx}%
\usepackage{amsmath,amssymb,amsfonts}%
\usepackage{amsthm}%
\usepackage{mathrsfs}%
\usepackage[title]{appendix}%
\usepackage{xcolor}%
\usepackage{textcomp}%
\usepackage{manyfoot}%
\usepackage{booktabs}%
\usepackage{algorithm}%
\usepackage{algorithmicx}%
\usepackage{algpseudocode}%
\usepackage{listings}%
\usepackage{pdfpages}

\graphicspath{ {./images/} }
\DeclareCaptionLabelSeparator*{spaced}{\\[2ex]}
\theoremstyle{thmstyleone}%
\theoremstyle{thmstyletwo}%

\theoremstyle{thmstylethree}%

\begin{document}

\title[Making the quantum world accessible to young learners through Quantum Picturalism: An experimental study]{Making the quantum world accessible to young learners through Quantum Picturalism: An experimental study}

\author*[1,2]{\fnm{Selma} \sur{Dündar-Coecke}}\email{selma.coecke@gmail.com}
\author[1]{\fnm{Caterina} \sur{Puca}}\email{caterina.puca@quantinuum.com}
\author[1,3]{\fnm{Lia} \sur{Yeh}}\email{lia.yeh@cs.ox.ac.uk}
\author[1,4]{\fnm{Muhammad Hamza} \sur{Waseem}}\email{hamza.waseem@quantinuum.com}
\author[5]{\fnm{Emmanuel} \sur{M. Pothos}}\email{emmanuel.pothos.1@city.ac.uk}
\author[1]{\fnm{Thomas} \sur{Cervoni}}\email{thomas.cervoni@quantinuum.com}
\author[6]{\fnm{Sieglinde M.-L.} \sur{Pfaendler}}\email{sieglinde.pfaendler1@ibm.com}
\author[1]{\fnm{Vincent} \sur{Wang-Ma\'{s}cianica}}\email{vincent.wang@quantinuum.com}
\author[1]{\fnm{Peter} \sur{Sigrist}}\email{peter.sigrist@quantinuum.com}
\author[8]{\fnm{Ferdi} \sur{Tomassini}}\email{ferdi@mothquantum.com}
\author[1]{\fnm{Vincent} \sur{Anandraj}}\email{vincent.anandraj@quantinuum.com}
\author[1]{\fnm{Ilyas} \sur{Khan}}\email{ilyas@quantinuum.com}
\author[3,7]{\fnm{Stefano} \sur{Gogioso}}\email{stefano.gogioso@cs.ox.ac.uk}
\author[3]{\fnm{Aleks} \sur{Kissinger}}\email{aleks.kissinger@cs.ox.ac.uk}
\author[1,9]{\fnm{Bob} \sur{Coecke}}\email{bob.coecke@quantinuum.com}

\affil[1]{\orgname{Quantinuum}, \orgaddress{\street{17 Beaumont Street}, \city{Oxford} \postcode{OX1 2NA}, \country{United Kingdom}}}

\affil[2]{\orgdiv{Centre for Educational Neuroscience}, \orgname{University College London}, \orgaddress{\street{20 Bedford Way}, \city{London} \postcode{WC1H 0AL}, \country{United Kingdom}}}

\affil[3]{\orgdiv{Department of Computer Science}, \orgname{University of Oxford}, \orgaddress{\street{Wolfson Building, Parks Road}, \city{Oxford} \postcode{OX1 3QD}, \country{United Kingdom}}}

\affil[4]{\orgdiv{Department of Physics}, \orgname{University of Oxford}, \orgaddress{Clarendon Laboratory, \street{Parks Road}, \city{Oxford} \postcode{OX1 3PU}, \country{United Kingdom}}}

\affil[5]{\orgdiv{Department of Psychology}, \orgname{City, University of London}, \orgaddress{Social Sciences Building, \street{32-38 Whiskin Street}, \city{London} \postcode{EC1R 0JD3}, \country{United Kingdom}}}

\affil[6]{\orgname{IBM Research GmbH}, \orgaddress{\street{Säumerstrasse 4}, \postcode{CH–8803} \city{Rüschlikon}, \country{Switzerland}}}

\affil[7]{\orgname{Hashberg Ltd}, \orgaddress{\street{71-75 Shelton Street}, \postcode{WC2H 9JQ} \city{London}, \country{United Kingdom}}}

\affil[8]{\orgname{Moth Quantum}, \orgaddress{\street{Somerset House, West Wing}, \postcode{WC2R 1LA} \city{London}, \country{United Kingdom}}}

\affil[9]{\orgname{Perimeter Institute}, \orgaddress{\street{31 Caroline Street North}, \city{Waterloo}, \state{Ontario} \postcode{N2L 2Y5}, \country{Canada}}}


\abstract{The abstract serves both as a general introduction to the topic and as a brief, non-technical summary of the main results and their implications. Authors are advised to check the author instructions for the journal they are submitting to for word limits and if structural elements like subheadings, citations, or equations are permitted.}


\abstract{\textbf{Background:} The educational value of a fully diagrammatic approach in a scientific field has never been explored. We present Quantum Picturalism (QPic), an entirely diagrammatic formalism for all of qubit quantum mechanics. This framework is particularly advantageous for young learners as a novel way to teach key concepts such as entanglement, measurement, and mixed-state quantum mechanics in a math-intensive subject. This eliminates traditional obstacles without compromising mathematical correctness --- removing the need for matrices, vectors, tensors, complex numbers, and trigonometry as prerequisites to learning. Its significance lies in that a field as complex as Quantum Information Science and Technology (QIST), for which educational opportunities are typically exclusive to the university level and higher, can be introduced at high school level. In this study, we tested this hypothesis, examining whether QPic reduces cognitive load by lowering complex mathematical barriers while enhancing mental computation and conceptual understanding. The data was collected from an experiment conducted in 2023, whereby 54 high school students (aged 16-18) underwent 16 hours of training spread over eight weeks. 

\textbf{Results:} The post-assessments illustrated promising outcomes in all three specific areas of focus: (1) whether QPic can alleviate technical barriers in learning QIST, (2) ensures that the content and teaching method are age appropriate, (3) increases confidence and motivation in science and STEM fields. There was a notable success rate in terms of teaching outcomes, with $82\%$ of participants successfully passing an end-of-training exam and $48\%$ achieving a distinction, indicating a high level of performance. Findings highlight participants’ deeper grasp of the abstract core principles of QIST, especially in areas where traditional approaches reliant on the Hilbert Space formalism would have been inadequate. The unique testing and training regime effectively reduced the technical barriers typically associated with traditional approaches, as hypothesized. Furthermore, participating in the QPic training program helped motivate students to engage with STEM subjects.

\textbf{Conclusion:} By lowering the entry barrier, QPic has the potential to facilitate new cross-skilling opportunities for both developers and learners. This study demonstrates that a complex theory, often perceived as intimidating or highly mysterious, can be taught rigorously in a fun and engaging way. It could also play a pivotal role in mitigating workforce shortages by accelerating the integration of learners and developers into emerging computational domains. 
}

\keywords{Quantum theory, quantum information science and technology, quantum physics, education, high school, cognition, teaching, learning, STEM, STEM workforce}

\maketitle

\section{Introduction}\label{sec:Introduction}
We are now witnessing a second revolution in quantum technologies, where the theoretical foundations established during the first quantum revolution are being applied to harness and control quantum systems for practical uses in computation, sensing, and communication~\citep{dowling2003quantum}. Unlike the initial wave, which focused primarily on experimental evidence supporting the then-proposed theory of quantum mechanics, this new era involves the development of quantum information theory and technologies (QIST) that leverage quantum phenomena (such as quantum measurement, entanglement, and coherence~\cite{QISConcepts2023v2}) to achieve computational power and precision. Innovations such as quantum computers, quantum sensors, and networks are poised to set transform numerous industries, spanning fields from chemistry and medicine to human cognition \citep{pothos2022quantum}.

To grow talent and leadership in this rapidly evolving field, it is of utmost importance to prepare an inclusive pipeline to tackle the skills shortage in quantum science \citep{dsit2023quantum}. Quantum mechanics stands as a captivating focal point for young learners, drawing them in with its enigmatic allure and often propelling them towards pursuing physics at the university level \citep{johansson2017undergraduate}. However, the accelerating expansion of the quantum ecosystem calls for broader participation and accessibility, as well as increased opportunities for early exposure to QIST, even in secondary education \citep{nationalworkforce}.  

Nevertheless, most current educational opportunities are typically exclusive to the university level. Of the 58 undergraduate and graduate QIST courses in the US surveyed in \cite{meyer2024QIScontent}, over $90\%$ cover or review each of the following math topics: Dirac notation (bra-ket), Complex numbers, Unitary matrices, Inner product, Dimension of Hilbert space, finite-dimensional vector spaces, Hermitian matrices, and Eigenvalues/eigenvectors. At the university level, this often results in disappointed students having to face the consequences of the “shut up and calculate” approach \citep{johansson2018shut}. At the high school level, despite recent advancements in high school courses teaching quantum concepts, a macro-level analysis of curriculum documents across 15 countries \citep{stadermann2019analysis} reveals that existing teaching efforts primarily focus on the fundamental principles of quantum, ranging from wave-particle duality to discrete energy levels. However, the learning outcomes of these topics largely remain theoretical; in many curricula, content tends to be confined to historical and philosophical discussions regarding the nature of science. This bypasses the multiple mathematical prerequisites posing barriers to accessibility, such as vector spaces, complex numbers, linear algebra, and probability theory, but inevitably leads to a restricted and unsubstantial comprehension of the subject. 

Our goal is to assess methods for eliminating these barriers and to develop a more suitable approach for learners. One of the key methods we are developing is Quantum Picturalism (QPic), a novel visual mathematical framework that transforms complex quantum concepts and algorithms into intuitive diagrammatic representations. QPic offers an accessible alternative to traditional algebraic methods, enabling learners to conceptualize core quantum principles without the heavy reliance on the prerequisite-intensive Hilbert space formalism (HilbS) \citep{von2018mathematical}. Its foundation relies on graphical calculus for category theory, a mathematical toolbox aimed at representing categorical operations, along with their real-world applications, via diagrams. 

QPic has been formulated over two decades of research, pioneered by \cite{abramsky2004categorical} and developed together with many others \citep{coecke2007quantum, CPaqPav, coecke2012environment, coecke2008interacting, CDKZ, chiri1, cnonsig}. 
It has initially gained significance in academia, becoming an integral part of graduate and undergraduate courses at institutions worldwide. For over a decade, it has been taught across multiple courses in the departments of computer science, software engineering, and mathematics at the University of Oxford. It then has been incorporated into programs at other universities\footnote{Including the University of Amsterdam, University of Edinburgh, University of Cambridge, and Indiana University Bloomington. Additionally, it has been a central topic at international summer schools, conferences, and tutorials, demonstrating its growing influence across diverse academic and research settings globally. Graduate courses rooted in QPic have been taught at the University of Oxford for the past decade. Presently, five courses at the University of Oxford teach and examine QPic: the Quantum Processes and Computation graduate course in Computer Science, the Quantum Software graduate course (the successor to the former Categorical Quantum Mechanics course) in Computer Science, the Quantum Computing graduate course in Software Engineering, the Quantum Computing course in the Department for Continuing Education, and the Introduction to Quantum Information graduate course in Mathematics. Courses over the past year at other universities with multiple lectures of QPic in the curriculum include the Full-stack Quantum Computing Master's in Quantum Computer Science course at the University of Amsterdam, the Introduction to Quantum Programming and Semantics undergraduate course at the University of Edinburgh, the Advanced Topics in Category Theory graduate course at the University of Cambridge, and the Programming Quantum Computers graduate course in summer 2023 at Indiana University Bloomington. Additionally, it has been the central topic of the Picturing Quantum Weirdness summer school at the University of Gdańsk in 2023 in Poland; a school at the Abdus Salam International School of Theoretical Physics (ICTP) in Trieste, Italy in 2023; two tutorials at the IEEE International Quantum Computing and Engineering Conference in 2023 in the US and in 2024 in Canada; a tutorial at the Quantum Initiative Rhineland-Palatinate (QUIP) International Winter School on Quantum Machine Learning 2024 at the Fraunhofer Institute for Industrial Mathematics ITWM in Germany; and a dozen lectures for the upcoming E-learning African International School on Quantum Science and Technology (ELAIS-QST) 2024/25 organized by the Kwame Nkrumah University of Science and Technology in Ghana. Note that this list does not include the growing implementation of QPic in the US or the increasing interest it is generating in other countries worldwide.}, as well as a wide range of
outreach initiatives involving high school students and teachers\footnote{Including the KYMA program, aimed at introducing Greek high school students to the world of Quantum Information, through a collaboration between secondary and university instructors, supported by the Hellenic Governmental Authorities and Hellenic Mathematical Society; the Quantum Physics in Pictures course taught by Muhammad Hamza Waseem and hosted by Khwarizmi Science Society, Pakistan; a two-day summer camp for high schoolers hosted at the University of Dalhousie; the 2025 Intro to Quantum Teacher Workshop for middle and high school teachers ahead of QSEEC; the QIST in the CSU Faculty Workshop.}.

Central to QPic's framework is the ZX calculus, introduced in \cite{coecke2008interacting,coecke2011interacting}, which focuses on the diagrammatic reasoning of quantum computations.
In the ZX calculus, every diagram represents a specific matrix.
Each rule of the calculus is an equality of diagrams which represent the same matrix.\footnote{Whenever any part of a diagram matches one side of a rule equation, you can replace it with the other side, to get a new overall diagram equal to the original.
\cite{hadzihasanovic2018two} showed the ZX calculus is \emph{complete}: For all of qubit quantum mechanics, any two diagrams representing the same matrix can always be proven equal using only the rules of the calculus.
Just 8 rules, each simple and conceptually meaningful, are enough to derive all possible rules~\citep{jeandel2018beyondCliffT, vilmart2019nearmin} --- even for mixed-state quantum mechanics~\cite{selinger2007cpm,coecke2008cpm}, enabling concepts such as decoherence and the Stern-Gerlach experiment to be taught in this course.
Since, the ZX calculus has been proven to be complete for all of finite-dimensional Hilbert space~\cite{poor2023completeness, wang2024completenessqufinitezxwcalculus, poor2024zxcalculus}.} This presents a new formalism to understand quantum computation without handling of complex and very large matrices, which grow exponentially in size with each additional qubit.
To provide readers interested in the underlying principles with a more comprehensive understanding, Appendix A presents a concise introduction to QPic, highlighting its key features through practical examples and outlining the methodological differences between QPic and the traditional HilbS formalism.

\begin{figure}[h!]
    \centering\includegraphics[scale=0.83]{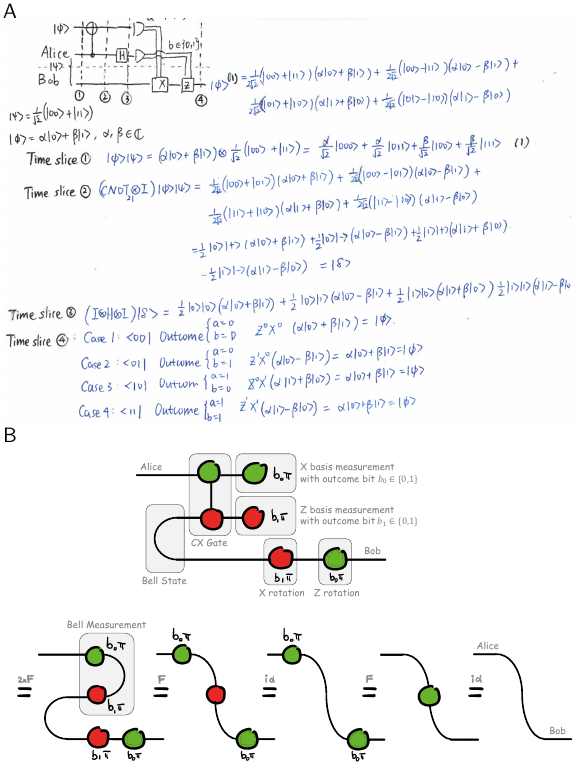}
    \caption{Figure 1A shows calculations in bra-ket notation by (then) Master's student Sarah Meng Li at the University of Waterloo, demonstrating the correctness of the quantum teleportation protocol as part of their graduate course, QIC 710: Introduction to Quantum Information Processing. This exemplifies how the HilbS formalism produces a more arcane description of the quantum teleportation protocol. In contrast, Figure 1B presents the diagrammatic version of the same protocol using QPic. Each step of the proof is labelled with the corresponding diagrammatic rule of the ZX calculus. This diagram elucidates a clearer visualisation of the flow of quantum information, offering more room for an intuitive representation of the operations involved in the quantum teleportation protocol, specifically the transfer of quantum information from Alice to Bob.}
    \label{fig:teleportation-comparison}
\end{figure}

As an example, we illustrate here quantum teleportation in the ZX calculus. This phenomenon, typically regarded as complex, becomes more accessible when explained using QPic, as shown in Figure \ref{fig:teleportation-comparison}.

The ZX calculus has been widely applied and extended, both in academic research and industry settings, actively used in quantum computing research at major companies, such as Google, PsiQuantum, and Quantinuum, and serving as a cutting-edge tool across many areas, such as quantum circuit optimisation \citep{de2020fast, EPTCS318.9, kissinger2019reducing}, quantum error correction \citep{gidney2019efficient, de_Beaudrap_2020, huang2023graphical, bombin2024unifying}, measurement-based quantum computing \citep{coecke2008interacting, kissinger2019universal,  
duncan2009graph}, quantum simulation \citep{https://doi.org/10.4230/lipics.tqc.2022.5}, quantum foundations \citep{coecke2011phase, backens2016complete, https://doi.org/10.4230/lipics.calco.2019.19}, and quantum natural language processing \citep{coecke2020foundationsneartermquantumnatural, Meichanetzidis_2021}. 

Very recent developments in the use of QPic, which are of high relevance to society in their own right, include the development of quantum-enabled explainable AI  \citep{tull2024towards, qdisccirctheory, qdisccircexperiment}, aiming to effectively bridge the gap between theoretical understanding and practical application. Furthermore, due to its category-theoretic foundations, QPic shares formal similarities with graphical calculi employed in various other disciplines, including linguistics \citep{CSC, coecke2020mathematicstextstructure, sadrzadeh2017quantization}, computer science \citep{pavlovic2013monoidal}, machine learning \citep{koziellpipe2024hybridquantumclassicalmachinelearning, gavranovic2024fundamental}, electronics \citep{boisseau2021string}, game theory \citep{ghani2018compositional}, control theory \citep{bonchi2021survey} and many more.

Despite its promise, the educational efficacy and instructional potential of QPic have yet to be rigorously tested. This gap in research was more evident considering that previous evaluations of QPic predominantly focused on university level and specific academic disciplines, raising uncertainties about its adaptability and effectiveness across broader demographics. To address this, we conducted an experimental study during the Spring and Summer terms of 2023, in which high school students underwent training using QPic [the preliminary details can be found in \citep{D_ndar_Coecke_2023}]. Our focus was particularly on high school students who typically lacked a foundational understanding of prerequisite mathematics. The study concentrated on three key aspects:
\begin{itemize}
   
 \item Whether QPic can alleviate the technical barriers inherent in traditional approaches, making QIST more accessible to younger learners. 
 \item Ensuring the content remains within the Zone of Proximal Development (ZPD) \citep{lloyd1999lev} of this age range.
  \item Increasing the students’ confidence, interest, and exposure to science and STEM fields.

\end{itemize}

This study aims to present the evidence gathered from this investigation. In the following section, we discuss the potential educational benefits of this formalism from an interdisciplinary perspective, highlighting its applicability across various fields. 
Methodology section elaborates on the experimental approach, outlining the procedures implemented during experiment. This was followed by the results section, which presents exam scores, analyses of students demographics, students' feedback on the course, and other findings that inform the discussion section, reflecting on the insights gained throughout the experimental journey. Finally, the conclusion section will outline potential future research directions and emphasize the importance of this research program for diverse learners.

\subsection{Impact of QPic formalism on Education}\label{sec:Impact}
For over a century, algebraic formalism has been the medium for understanding and teaching quantum theory, serving as its foundational language. With the introduction of QPic, we now offer a visual mathematical formalism that simplifies quantum operations and makes the content more practical for both learners and practitioners. Given its novel approach, it is expected to face criticism, as any revolutionary method challenges long-standing conventions, inviting debate from academic community. Although this study serves as proof of concept, we recognize the necessity of substantial data required from diverse age and ability groups across various cultures to thoroughly discuss its educational efficacy. We should highlight that several experiments replicating the present study are currently underway across different countries.\footnote{Our findings in the UK, as reported in this paper, have already received media attention \citep{theguardianPhysicistCoecke} for their promising demonstration of how pictorial methods can enhance educational outcomes in QIST.}

In general, the use of visual support in the field of science education is not unusual, with a growing interest in complementing learning material with tools like multimedia and schematic diagrams \citep{ainsworth2006deft, kozma2003material, mayer2003promise}. Numerous well-established studies highlight the critical role and importance of using diagrams for educational purposes. Despite the extensive body of research demonstrating that diagrams convey information beyond descriptive texts, instructional visuals have historically been used primarily as supplementary aids to support text or symbolic representations \citep{bobek2016creating, herrlinger2017when}, with studies showing that visual aids not only enhance comprehension when paired with text but can also replace text \citep{alesandrini1984pictures, tversky2013visualizing}. Extensive evidence suggests that diagrams enhance learning by providing intuitive explanations \citep{bobek2016creating, mielicki2015affordances}, improving memory retention, and catering to a diverse learning style \citep{richards2002fundamental, verdi1997organized}. They are widely used to simplify complex concepts and promote creativity and problem-solving skills. 

Research in science education has particularly evolved to examine the types of visual representations, ranging from multiple representations \citep{eilam2008learning} to multimodal representations  \citep{marquez2006multimodal}. However, the primary focus has been on how students make sense of abstract ideas when knowledge is transformed between different modes -from descriptive texts to depictive pictographics and vice versa. Much of this research has centered around interpreting visual representations (learning from representations) rather than constructing representations (learning with representations) \citep{perez2010graphicacy, tippett2016recent}. 

QPic leverages this constructive approach and enables learners to create and manipulate representations to build quantum concepts. Despite modern pedagogical approaches increasingly endorsing the integration of visual aids into educational programs, QPic stands alone as the first computational formalism capable of transforming technical content into a visual form while retaining all essential mathematical information. Therefore it offers newfound potential as an educational tool that can be used effectively to perform computations as an alternative to textual and symbolic representations.

Unlike commonly used visual aids, QPic qualifies as a mathematically sound diagrammatic method, enabling quantum reasoning directly in the diagrammatic setting. Therefore, the language allows overcoming traditional algebraic prerequisites for learning QIST concepts. Such an application in the field of education sciences is unprecedented, primarily because no comparably expressive languages have existed until recently.

One notable example of a topic where diagrams are commonly used to make learning beginner-friendly is in computer science, where a common approach to teaching binary logic (e.g. AND, OR, and NOT gates) at the high school level and above is through drawing circuit diagrams. In recent years, a growing number of university courses have employed the QPic approach to teach quantum information science. However, despite such courses having existed as far back as the year 2012, the educational value of QPic has not been evaluated before. Our study presents a pioneering effort in demonstrating its effectiveness, particularly among novice learners - a more challenging demographic to teach complex concepts. 

In the field of physics, this becomes more crucial, as QIST, traditionally regarded as one of the most challenging scientific fields, is typically reserved for advanced university courses due to its complex mathematical framework relying on the HilbS formalism. Computation with matrices is typically taught in a Linear Algebra course, which often neglects the tensor products essential for presenting matrices when dealing with two or more qubits. This introduction usually occurs in the late first year or second year of university in STEM fields. Meanwhile, computing with bra-kets is typically taught in the third or fourth year of university for physics undergraduates.

Linear algebra and proficiency with matrix operations have traditionally been considered essential prerequisites for learning quantum theory. However, these mathematical demands create significant barriers to entry, particularly for young learners, discouraging many from pursuing such educational and career opportunities at the critical age at which they decide their future plans after secondary school. Consider matrix operations as an example. In England, Wales, and Northern Ireland, matrices are only in the curriculum of the highest level of mathematics course, Further Mathematics~\citep{mathsgovuk}. Only $4.7\%$ of the A-level students in state schools in England took Further Mathematics in 2021, underscoring how few secondary school students have the opportunity to consider learning QIST through traditional pathways ~\citep{imperialfurthermathsreport}.

In Scotland, matrices are included only in the curriculum of the most advanced mathematics course, Advanced Higher Mathematics~\citep{sqaadvhighmathsspec}. According to their 2023 exam report, many students struggled with the matrix multiplication question, highlighting the challenges posed by this topic~\citep{sqaadvhighmathscoursereport}. Furthermore, Advanced Higher Mathematics is taken by significantly fewer students over 10 times fewer than those who take National 5 Mathematics, reflecting the limited exposure of this mathematical concept among the wider student population~\citep{sqaattainment2024}.

Although multiple proposals aimed at introducing quantum science to younger students, including those in secondary education \citep{stadermann2019analysis}, many current approaches are still not able to effectively deliver mathematical prerequisites, such as linear algebra and complex numbers, without sacrificing rigorous conceptual understanding \citep{krijtenburg2017insights}. Formal approaches to teaching quantum physics at the secondary school level have a high risk of rendering the subject incomprehensible and inaccessible. On the other hand, overly informal methods can result in significant misconceptions among students \citep{muller2002teaching}.

It is important to note that any mathematical formalism of quantum theory consists of a set of axioms that interact in specific ways. In principle, dedicated students can learn or be taught the rules of this formalism -- without ever needing to reference quantum physics itself.
However, this does not mean the formalism has no connection to quantum physics. Rather, the formalism provides a language to express physical reality. This, of course, applies to both QPic and Hilbert space formalisms. The two formalisms are equally expressive for finite-dimensional quantum systems \citep{coecke2017picturing, hadzihasanovic2018two, poor2024zxcalculus}; in other words, any application of the Hilbert space approach can be replaced by its QPic counterpart and vice versa. Of the two formalisms, what makes QPic more approachable for young students is not its divorce from physical reality -- which is not the case anyway -- but its fewer educational prerequisites. 

A few approaches currently exist to teach mathematically accurate QIST at the high school level, highlighting the need for further development and rigorous evaluation of these methods. Below, we list a few current formalisms for representing the mathematical information of quantum circuits and protocols, as well as courses implementing them.
\begin{itemize}
    \item The QPic approach evaluated in this work was first introduced in~\cite{coecke2006kindergarten, coecke2010quantum} and proposed more comprehensively in~\cite{D_ndar_Coecke_2023}.
    \item The first iteration of the year long Qubit by Qubit course by The Coding School nonprofit in 2020, taken by students around the world, focused the first semester on teaching matrices, vectors, complex numbers, and introductory Python programming; in subsequent years, the proponents of this approach have emphasized shifting to much fewer mathematical prerequisites based on student feedback~\citep{QxQ}.
    \item The five-week Quantum Quest course for students age 16-20, jointly organised by the QuSoft research centre in Amsterdam and the Cluster of Excellence CASA of the Ruhr-Universität Bochum, utilizes vectors and the bra-ket notation. Operations are described based on how they act upon statevectors, instead of utilizing matrices~\citep{QuantumQuest}.
    \item The misty states approach, which represents qubit states as shapes passing through circuits where superposition states are mists of shapes, was introduced in the ``Q is for Quantum'' book by~\cite{Rudolph2017QisforQuantum}, and has been recently expanded upon in an approach called the Quantum Abacus~\citep{german2024abacus}. This was utilized as part of a two-day outreach program called CTech for female students in Virginia in their final two years of high school~\citep{economou2020teachqis}, which they further developed as a semester course for first-year undergraduate students as a mandatory course of the multidisciplinary QISE minor degree at Virginia Tech~\citep{economou2022hqw}.
\end{itemize}
It is important to recognize that these approaches to teaching QIST are not mutually exclusive, and each approach has its own set of strengths and weaknesses, depending on the specific quantum concepts being taught. It is possible that learning through different ways can yield new understanding, so long as this does not overwhelm the students with too much notation. For instance,~\cite{german2024abacus} compares two approaches (QPic and Quantum Abacus) by demonstrating proofs of the GHZ state and quantum teleportation in both frameworks. They discuss how QPic uses graph rewriting techniques to provide insights into quantum computations while maintaining circuit equivalence. In contrast, the Quantum Abacus approach resembles the ``slow-motion'' replays in televised sports, offering a step-by-step view of how a quantum state passes through a circuit.

While these comparisons illustrate how different methods can complement each other in learning complex concepts, we argue that the potential benefits of the QPic formalism for a diverse range of learners are deeply rooted in its cognitive advantages. One explanation worth discussing here relates to the Gestalt school of thought, which posits that images are not perceived as collections of separate fragments but as integrated wholes with internal structures. Principles of perception, including ‘grouping’, ‘continuation’, ‘similarity’, ‘prägnanz (simplicity)’, ‘proximity’, and ‘common fate’ are particularly relevant to how QPic facilitates understanding: learners do not have to process components in isolation, but perceive them through the overall structure and the relationships between parts, providing a more intuitive grasp of complex concepts. For example, elements that move or change together are perceived as being related (common fate). After a short training period, viewers can fill in the gaps when perceiving incomplete components by arranging elements in a way that follows natural flows through logical and coherent relationships (continuity and grouping). The method enables the interpretation of ambiguous components in the simplest form possible (prägnanz), based on the principle that the mind naturally seeks out patterns and coherent shapes, making it easier to grasp the structure and message of a diagram in QPic -- as an illustration, the principles necessary for solving exam questions are shown in Appendix D. 

The way this method helps learners organize and integrate new information is by reducing cognitive load and making abstract concepts more accessible. We propose such grasp is achieved in two stages: first, learners need to detect and identify the local properties and the direction of movement or change within a display, and second, they re-assemble the topographical information into a coherent whole. By breaking down complex information into more manageable visual components and then integrating them holistically, learners can more efficiently manage their mental processes -a dual-stage cognitive process that emphasizes the importance of both local feature detection and global integration in learning. 

This dual-stage cognitive processing could initially be independent of language faculties in the brain -there is evidence supporting the view that the recognition of visual phenomena involves several brain areas not necessarily associated with linguistic tasks, particularly the visual cortex involved in visual processing \citep{tolmie2020lifespan}. Although both orthographic and diagrammatic representations initially require perceptual skills, the ways textual and pictorial messages are formed into coherent units in semantic space differ due to multiple conceptual stores where verbal and nonverbal concepts are neuroanatomically segregated \citep{mayer2024past, thierry2006dissociating}. This suggests the existence of a dual coding pathway, aligning with theory proposing that verbal and nonverbal information are processed by functionally distinct yet interconnected long-term memory systems \citep{clark1991dual}. These double dissociations indeed appear detectable at the behavioural level, with students showing significant performance advantages in accessing the meaning of pictorial representations compared to textual ones \citep{clark1991dual, sadoski2004dual}.

This line of research is particularly relevant when considering that QPic may largely tap into the nonlinguistic part of cognition, which is evolutionarily an earlier developed human capability. Evidence indeed from early cave paintings, for instance, shows that humans developed nonlinguistic communication long before the advent of linguistic systems, particularly writing, indicating that visual representations have been a fundamental mode of human cognition and communication \citep{donald1991origins, lewis2002mind, murray1988history}. Visual expression played a crucial role in conveying meaning, recording events, and transmitting knowledge, serving as a fundamental mode of human cognition and communication in early human societies. This perspective aligns with the notion that the human brain has evolved to process visual information efficiently and that this capacity predates the development of written languages and symbolic notation. Leveraging QPic in educational contexts could tap into these cognitive pathways, potentially opening a new mathematical mechanism for making complex quantum processes more accessible by relying on the brain’s natural strengths. We recognize that while past research offers valuable insights for this point, the unique features of QPic necessitate further investigation. A follow-up neuroimaging study is underway to explore this argument.

\section{Methods}\label{sec:Methods}
This intervention trial was conducted online through weekly one-hour live tutorial sessions over a 12-week period, from June 5th to August 11th, 2023. The study comprised eight weeks of training, with a one-week break following the fourth week. After the training phase, the final three weeks were devoted to administering the post-training exam. 

We adopted an exploratory approach, collecting data using a post-test design to evaluate the effectiveness of the QPic method. We sought to explore evidence for any potential benefits of QPic, focusing on the three key aspects outlined above. A purposive sampling approach was used to recruit 75 volunteer high-school students aged 16, 17, and 18. The sample size was limited by the number of tutors available to provide live online tutorials during this period. 

A detailed description of the methodology, including participant profiles, materials, experimental procedure, and the assessment process, can be found in Appendix B.

\section{Results}\label{sec:Results}
We report some exploratory analyses aimed at understanding the basic patterns and trends in the data. Anonymised data, whereby each student was identified by a unique code, were analysed using SPSS and Matplotlib in Python, within a Jupyter notebook. 

The first section presents general demographics, providing an overview of the background of the participants, while subsequent sections address the three  primary aims of the experiment. 
\subsection{Descriptive statistics / Demographics}
\begin{figure}[H]
\centering
\includegraphics[scale=0.7]{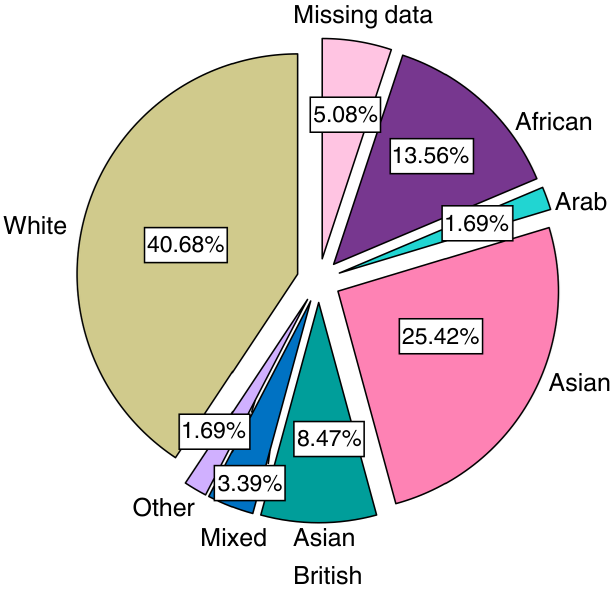}
\caption{The ethnic background composition of the participants.}
\label{fig:ethnic}
\end{figure}

The sample comprised 54 participants, with the majority aged 17 (31; 59.6\%), followed by 25\% (13) aged 16, and 5 of them (9.6\%) aged 18 years old. Gender distribution indicated 23 (44.2\%) were female, one identified as 'they', and 26 (50\%) were male, with three cases of missing data. The composition of ethnic background was diverse, as illustrated in Figure \ref{fig:ethnic}. 


Over half of the participants (55.9\%) reported English as their sole mother tongue, while the remainder were bilingual, speaking at least two languages spanning European, Asian, and Arabic origins. Geographically, 24.07\% of participants were from London, with the remaining individuals joining from various cities such as Cardiff, Cambridge, Manchester, totalling 24 cities across the UK and representing 37 different schools. A small portion of 5.08 \% is not characterised in this way due to missing information.
 
School type, as outlined in Table \ref{Table 1}, provides potentially relevant information about participants’ demographic trends and social mobility, especially important in the UK, where there are significant differences in funding, admission criteria, and educational outcomes associated with each type of school. Among our 51 respondents, more than half (55.6\%) attended state schools, which are funded by the government and usually do not require entrance exams; such schools reflect the broad educational experience of a majority of students. 

\begin{table}[h]
\captionsetup{justification=centering}
\caption{Frequency and percentage distribution of school types.}
\centering
\label{Table 1}
\begin{tabular}{lll}
\toprule
\emph{School type}
& \emph{Frequencies}
& \emph{Percent} \\
\midrule
State  & 30 &  55.6  \\
Grammar & 14 & 25.9 \\
Private Boarding & 2 & 3.7 \\
State Boarding & 3 & 5.6 \\
Private independent & 2 & 3.7 \\
Missing data & 3 & 5.5 \\
\emph{Total} & \emph{54} & \emph{100} \\
\bottomrule
\end{tabular}
\end{table}

Grammar schools, attended by 25.9\% of the respondents, are state-funded secondary schools that select students based on academic ability, requiring them to pass an entrance exam known as the 11-plus. These schools are often seen as more academically rigorous and competitive. 

Private boarding schools (3.7\% of the respondents attended) are independent schools where students live on campus during the term. They are funded through tuition fees paid by the families of the students, 

State boarding schools (5.6\% of participants) offer boarding facilities while providing state-funded education. However, parents are usually charged for boarding expenses. 

Private independent schools (3.7\% of participants) are funded by tuition fees and donations. These schools often have selective admission processes, including entrance exams or interviews, and typically offer smaller class sizes, more resources, and a wider range of extracurricular activities. 
 
\begin{figure}[H]
\centering
\includegraphics[width=130 mm, height=70 mm, trim=0 20mm 0 0]{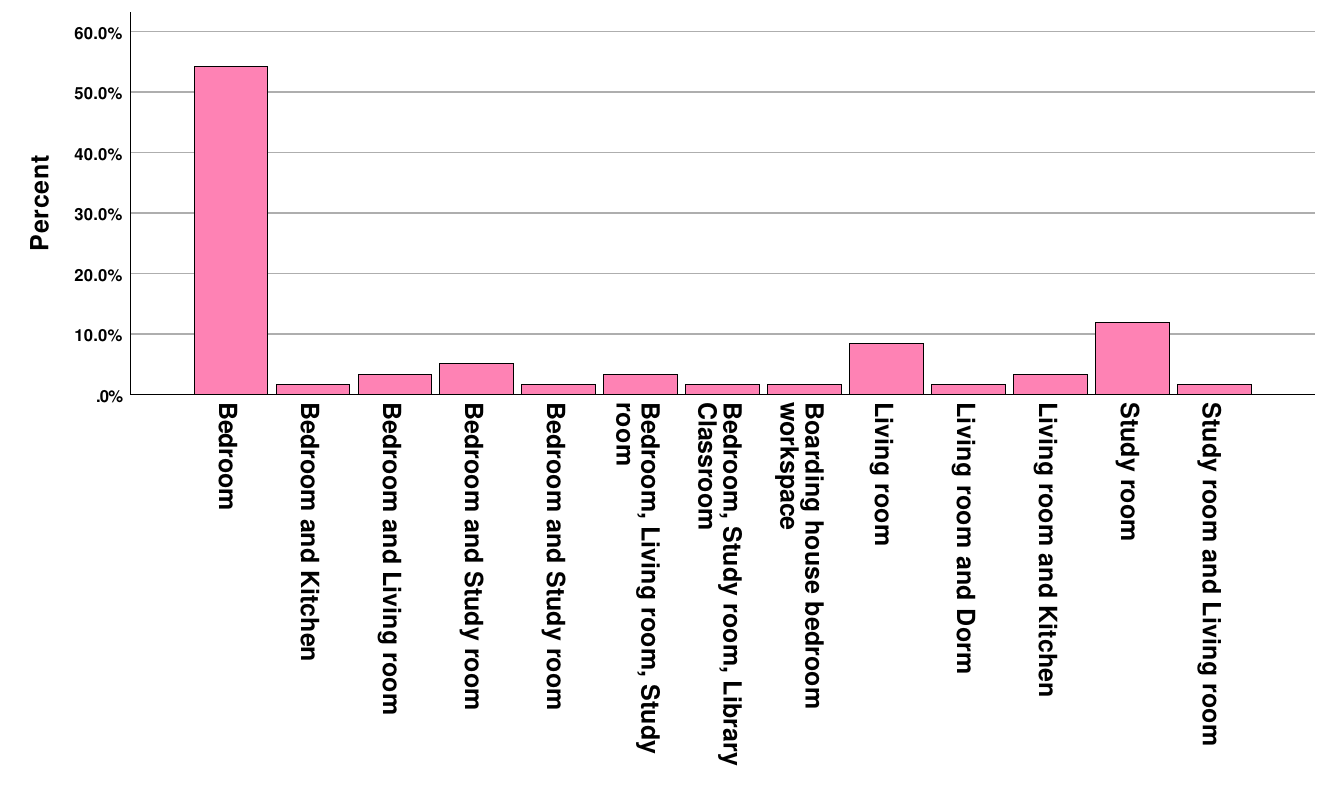}
\vspace{15pt}
\caption{Venue used for tutorials.}
\label{fig:attended venue}
\end{figure}

Further, to understand more about other socioeconomic indicators we note that over half of the participants (57.5\%) reported having more than one sibling, and 57.6\% came from homeowner families, while the remaining were renters. Most participants regularly preferred to attend live tutorials from their bedrooms, as shown in Figure \ref{fig:attended venue}. Approximately 10\% of participants had access to a dedicated study room. There was 1 case of missing data. However, a significant majority (72.9\%) reported being able to access private study space whenever necessary.

 \subsection{Descriptive statistics / Knowledge and Background}

Most participants (66.1\%) were either in the process of completing their GCSEs or had just completed them as their highest level of education. Additionally, 18.6\% had completed an A-level in Mathematics, and 11.9\% had an A-level in Further Mathematics. 

\begin{figure}[H]
\centering
\includegraphics[scale=0.6, trim = 0 20mm 0 0]{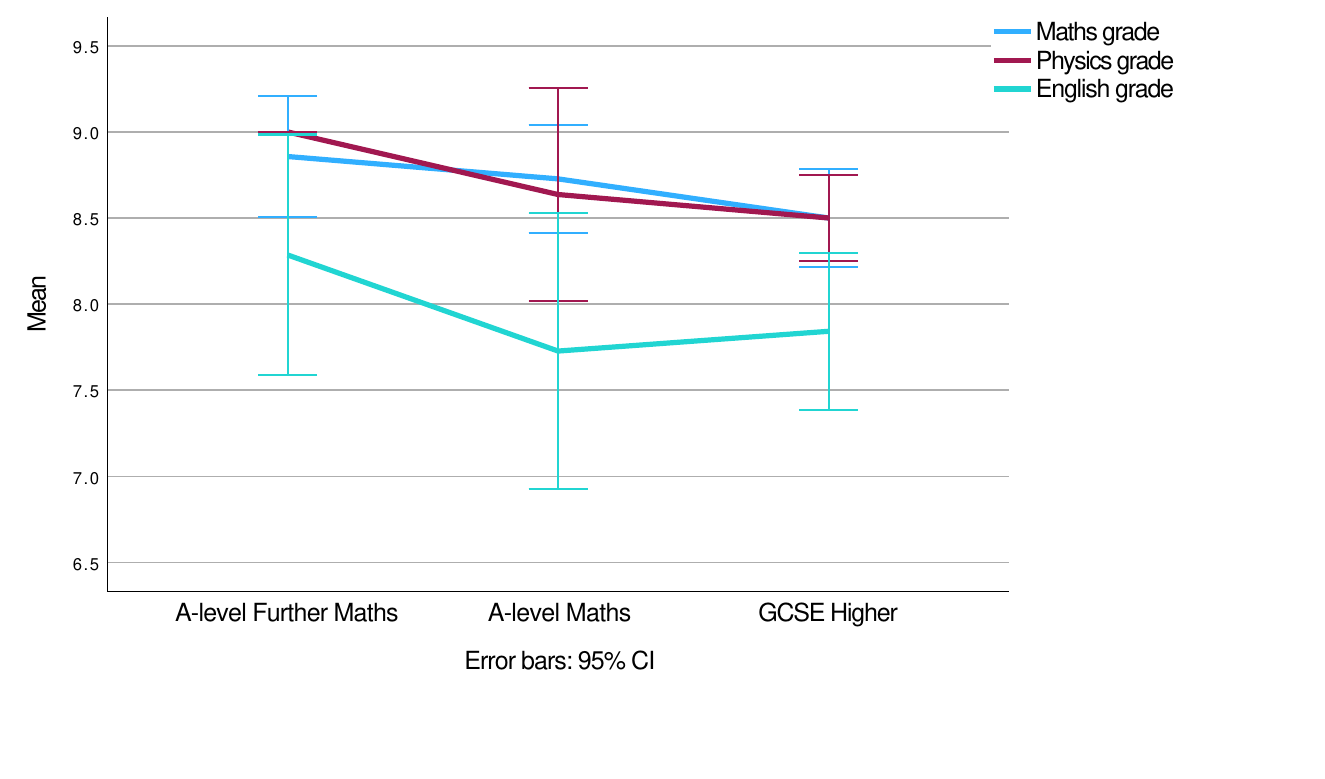}
\caption{Variations in mean scores for Mathematics, Physics and English grades.}
\label{fig:variations in mean scores}
\end{figure}

The chart in Figure \ref{fig:variations in mean scores} illustrates the variations in mean scores for Mathematics, Physics and English grades among participants based on their highest level of education achieved (CI: 95\%). 

Looking at Figure \ref{fig:variations in mean scores}, overall:
\begin{itemize}
    \item The mean scores for Physics and Mathematics exhibit more similarity compared to the means for English and Mathematics/Physics, suggesting less variation between the Mathematics and Physics grades. Larger confidence Intervals (CIs) for English grades also confirm this. 
\item  Students who have completed A-level Further Mathematics demonstrate the highest mean scores in English, Mathematics, and Physics indicating a potential correlation between advanced mathematical abilities and in other subjects. 
\item Students with A-level Mathematics also show higher mean scores in Mathematics and Physics compared to those with GCSEs qualifications, although this trend is not observed in English.
\item Conversely, students who have completed A-level Mathematics have slightly lower mean scores in English compared to those with A-level Further Mathematics but still outperform students who have completed GCSEs.
\end{itemize}

Hours spent studying per week varied between 1 to 4 hours, as shown in Table \ref{table: hours}, indicating that on average, one or two hours of study sufficed for most participants. We should highlight that despite the complexity of the topics, even for participants requiring four hours of study, this observation is promising in relation to the complexity of the topic. 

\begin{table}[h]
\centering
\captionsetup{justification=centering}
\caption{Hours spent to study per week.}
\label{table: hours}
\begin{tabular}{ll}
\toprule
\emph{Hours/week} & \emph{Percent} \\
\midrule
1-2 & 64.4 \\
3-4 & 30.5 \\
4-5 & 1.7 \\
\emph{Total} & \emph{96.6} \\
\bottomrule
\end{tabular}
\end{table}


 
Clearly, all students juggled various commitments during their training, including exam preparations, part time jobs, internships, dedicated sport activities, and camp attendances. See Table \ref{table: commits}.

\begin{table}[h]
\centering
\captionsetup{justification=centering}
\label{table: commits}
\caption{Other commitments over the course period.}
\label{table: commits}
\begin{tabular}{cccccc}
\toprule
 & \emph{Exam preparation}
 & \emph{Part-time job}
 & \emph{Internship}
 & \emph{Sport activity}
 & \emph{Camps/leisure} \\
\midrule
\% & 100 & 20.3 & 11.9 & 10.2 & 57.6 \\
\bottomrule
\end{tabular}
\end{table}

\begin{figure}
    \centering
\includegraphics[scale=0.5, trim = 0 10mm 0 0]{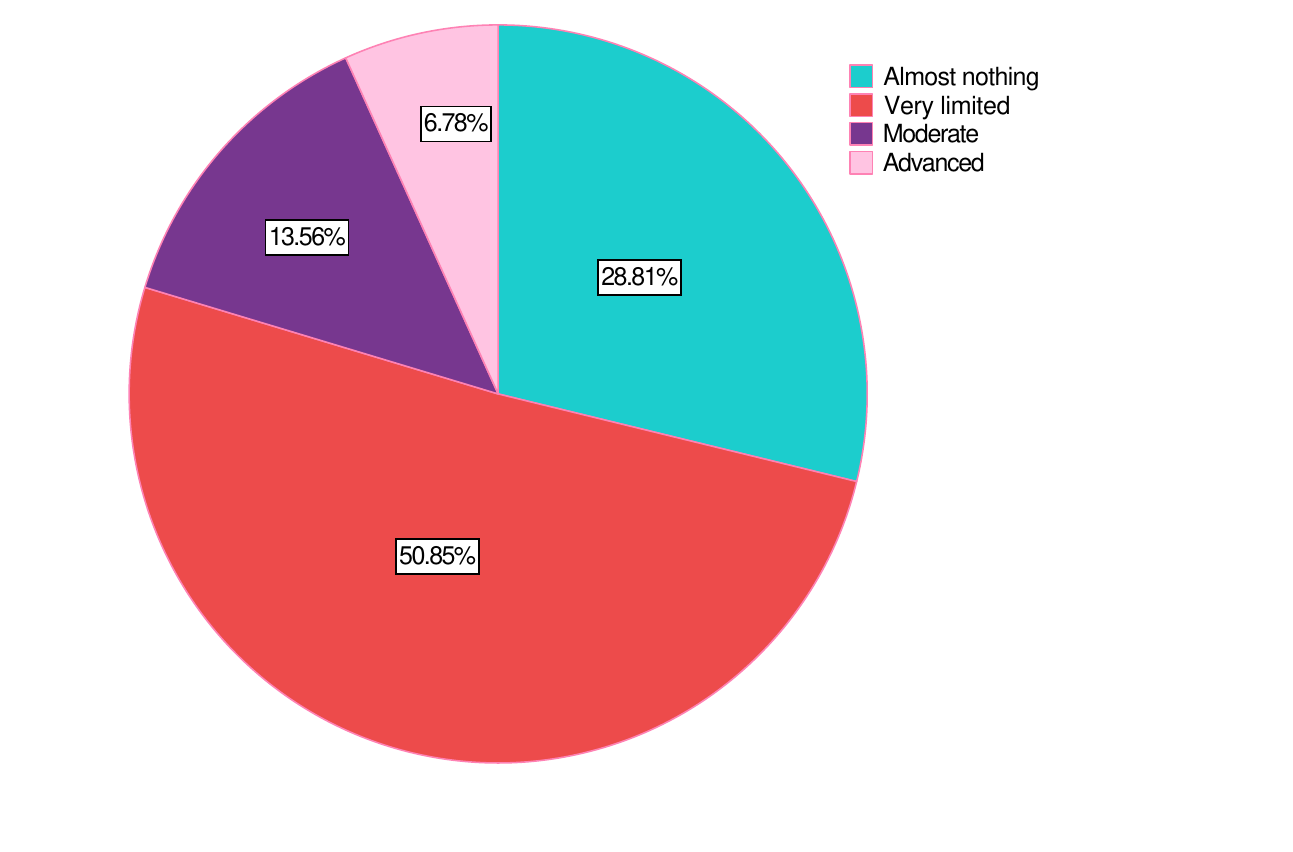}
\caption{Prior knowledge before attending training.}
    \label{fig:prior knowledge}
\end{figure}

Their prior knowledge of QIST before attending the training is depicted in Figure \ref{fig:prior knowledge}, revealing that 79\% of participants had little to no knowledge about QIST, while 6.8\% considered themselves to possess advanced knowledge. It is worth noting that later correlations show a high correlation between QIST knowledge and Physics grade only.  

\subsection{Whether QPic provides an effective method for presenting QIST to young learners}
We aimed to assess this by focusing primarily on tutor communication and the overall learning experience of the participants. 

The majority rated tutor communication and their learning experience as good (37.3\%) or very good (27.1\%), with 33.9\% of them finding it okay/acceptable. However, one participant found the communication not very engaging. All participants rated the tutorial booking system as good or excellent. 
 
Regarding negative experiences during tutorials, the following feedback was received:
\begin{itemize}
\item Eight participants encountered technical problems, such as internet connection issues.
\item Four participants felt that tutorial content was redundant.
\item Three participants believed that the number of tutorial episodes was insufficient. They requested more.
\item One participant found that topics were vaguely explained. 
\item One participant suggested more practice questions similar to those in the exam should have been included.
\item One participant suggested more complex tutorials or additional extension materials were needed. 
\item One participant expressed discomfort with interacting with tutors: While dealing with technical issues, overly complex topics were introduced too quickly. 
\item Another participant stated while most tutorials were well-paced some lacked extension questions, and in particular quantum teleportation diagrams were rushed. 
\end{itemize}

When asked to rate the tutorials, the majority found them to be good (52.5\%) or excellent (30.5\%), with 16.9\% finding them acceptable - there were no participants who provided negative feedback.
 
Although extra Q\&A sessions were offered by the team to follow up tutorials, 61\% of participants did not take part. Reasons for non-participation in any Q\&A sessions included: other commitments (47.5\%), tutorials were sufficient (10.2\%), feeling intimidated by discussions (3.4\%), no particular reason (39\%)
 
Participants were asked to identify which topics they found the clearest and best understood. They had the option to select more than one topic, allowing them to indicate multiple areas where they felt particularly confident. Table \ref{tab:tutorials frequency} shows the topics/tutorials that participants most frequently rated as the easiest to comprehend. Participants were allowed to rate multiple tutorials based on their level of comprehension.  

\begin{table}[h]
\centering
\captionsetup{justification=centering}
\caption{Best understood/rated tutorials.}
\label{tab:tutorials frequency}
\begin{tabular}{lll}
\toprule
 \emph{Tutorial/week}
 & \emph{Frequency}
 & \emph{Percent} \\
\midrule
1 - Quantum in pictures: Wires and Boxes & 43 & 79.6 \\
2 - Introduction to quantum teleportation & 24 & 44.4 \\
3 - The Quantum world is a world of spiders & 50 & 92.6 \\ 
4 - Quantum Computing & 29 & 53.7 \\
5 - Advanced Quantum Teleportation & 13 & 24.1 \\ 
6 - Keeping Einstein Happy & 9 & 16.6 \\ 
7 - Quantum versus Ordinary & 15 & 27.7 \\
8 - Everything just with Pictures & 13 & 24.1 \\
\bottomrule
\end{tabular}
\end{table}

In terms of perceived exam difficulty, one participant found it too easy, while 27.1\% found it too difficult, with the majority considering the final exam to be just right (71.2\%). Furthermore, 89.8\% of them stated that topics covered in the final exam were representative of the course content. One participant mentioned that “Topics were good but there was not much practice of those exam-type questions”, indicative of the structured and unanticipated nature of our assessment approach. Another participant suggested “Could have covered more understanding of how the system connects to physics.” Lastly, one participant noted, “Largely, but the third question was a conceptual question rather than the manipulation of the diagrams.” 
 
When asked if they felt prepared to answer the questions of the final exam, 62.7\% of participants responded affirmatively. Regarding the most difficult question, 42.5\% identified the third question as the most difficult, 3.4\% chose the second question as the most challenging, and one participant found all questions difficult, with 52.5\% not making any comments. 
 
Participants were asked to describe the format of the Quiz solutions, and their responses were highly positive. The percentages presented reflect the proportion of participants who described the format, as detailed in Table \ref{tab:quiz format}: 
\begin{table}[h]
\captionsetup{justification=centering}
    \caption{Description of quiz format by participants  (\%) }    
    \begin{tabular}{lc}  
  \toprule
  Description & \% \\
  \midrule
   Fun & 6.8 \\
   Engaging & 20.3  \\
   Necessary & 25.4  \\
   Prefer live solutions & 11.9 \\ 
   Fun, Engaging & 10.2 \\ 
   Engaging, Necessary & 11.9  \\
   Fun, Engaging, Necessary & 5.1  \\
  \bottomrule
\end{tabular}
\label{tab:quiz format}
\end{table}

The majority of participants described the quiz format as `necessary' with 25.4\% rating it as such. Additionally, 20.3\% found the format `engaging', while a smaller portion, 11.9\%, preferred `live solutions'. The remaining participants described the quiz format as `fun' and `engaging,' reflecting a positive overall reception.

\subsection{Whether QPic ensures the content is within the ability of students in this age range
}

We began with some correlations, between a range of variables, with only significant correlations presented in Tables \ref{tab:my_label} and \ref{tab: other corr} below.

\begin{table}[h]
\centering
\captionsetup{justification=centering}
    \caption{Significant correlations between school grades.}
    \label{tab:my_label}
\begin{tabular}{lllll}  
  \toprule
  & \emph{Age}
  & \emph{Physics}
  & \emph{English}
  & \emph{Mathematics} \\
    &
  & \emph{Grade}
  & \emph{Grade}
  & \emph{Grade} \\
  \hline  
   QIST Knowledge & & \begin{tabular}{@{}l@{}} .322* \\ .015 \end{tabular} & & \\ 
   Engaging tutor communication and teaching & \begin{tabular}{@{}l@{}}-.263* \\ .044 \end{tabular} & & & \\
   Feeling prepared to answer the exam questions & & \begin{tabular}{@{}l@{}}-.337* \\ .011 \end{tabular} & &  \\ 
   Family housing status: Homeowners vs. renters & \begin{tabular}{@{}l@{}}.302* \\ .028 \end{tabular} & \begin{tabular}{@{}l@{}}-.342* \\ .013 \end{tabular} & \begin{tabular}{@{}l@{}}-.342* \\ .013 \end{tabular} & \begin{tabular}{@{}l@{}}-.550** \\ .001\end{tabular} \\
   Gender &  & \begin{tabular}{@{}l@{}}.346* \\ .009 \end{tabular} &  & \begin{tabular}{@{}l@{}}.346* \\ .009\end{tabular} \\
  Physics Grade &  &  &  \begin{tabular}{@{}l@{}}.584** \\ .001 \end{tabular} & \begin{tabular}{@{}l@{}}.715** \\ .001 \end{tabular} \\
  English Grade &  & \begin{tabular}{@{}l@{}} .584** \\ .001 \end{tabular} &  & \begin{tabular}{@{}l@{}} .621** \\ .001 \end{tabular}\\
  \bottomrule
\end{tabular}
{* $P < 0.5$, ** $p=.001$}
\end{table}

QIST knowledge correlated with Physics but not with Mathematics grade, which aligns with an intuition that a background in physics would bolster understanding in QIST. This may also support the suggestion that QIST comprehension can be achieved independently of advanced mathematical knowledge. However, it is important to note this is just a correlation, and further investigation is warranted. 

We also considered the feeling of preparedness for exam questions. This feeling negatively correlated only with physics grade, suggesting that participants who felt less prepared tended to have higher physics grades, while those who felt more prepared tended to have lower physics grades. The question regarding feeling of preparedness was correlated negatively with the final exam difficulty, as might be expected,  [$ r(59)=-0.77,(p<.001)$]. This disparity between perceived preparedness and actual exam performance in A levels needs further investigation to understand the underlying factors at play, but it is possibly just the result of students deciding to devote more or less effort to the quantum course vs. their A levels.

Another notable finding was the negative correlation between age and engagement with tutor communication and teaching. Younger participants tended to find tutor communication more engaging and rated it more positively 

There were positive correlations between gender, Physics, and Mathematics grades. The correlations indicate that as we go from females (labeled with a 1) to males (labeled with 2), performance in Physics and Mathematics improved. Independent samples t-test confirmed significant results for both Mathematics grades [$t(36.033) = -2.577$, $p<.001$, CI(-.911, -.128)] and Physics grades [$t(31.966) = -2.664, df (p<.001)$, CI(-.911, -.128)], with male participants scoring higher than female counterparts in both subjects. There is clearly work that needs to be done, to understand the role of gender in determining the performance levels in these subjects. Note, this trend is consistent with the literature on gender differences in STEM fields, with the general finding being that males tend to outperform females in physics and mathematics \citep{hyde2009gender, elsequest2010cross, wang2017gender}. Such an outcome might be influenced by various socio-cultural factors and educational practices, as discussed by \cite{hyde2009gender}. Interestingly, this trend does not extend to English grades, indicating that gender may have a differential impact on performance across different subjects.

Another noteworthy finding was the strong difference in all of Mathematics, Physics, and English A level grades, between participants with families in owned (higher scores) vs. rented accommodation. As we go from renters to homeowners, participants from homeowner families tended to achieve higher grades in Mathematics, Physics, and English as highlighted by the negative correlations. Possible explanations include stronger academic motivation among students from more advantaged family backgrounds, along with potential disparities in access to resources or support systems. For example, students from homeowner families may receive additional assistance or interventions to improve their skills in these subjects. Using independent samples t-tests to compare scores from the three subjects, we observed for Mathematics $t(20.303) = -4.652, p<.001$, CI(.499 - 1.258), Physics $t(24.979) = -2.575, p<.001$, CI(.120 - .974)], and English $t(50) = 2.577, p-.01$, CI(.203, 1.635). Note, the family status variable indicated differences in additional commitments too: we created a variable by adding 0/1 indicator variables, across four categories: additional commitments regarding an exam, a part time job, an internship, some support activity, or some leisure activity (e.g., a club commitment). However, there was no difference in overall additional commitments between students from home owning families vs. families renting their accommodation ($p=.87$).  

Further, we continued examining correlations between grades and other socioeconomic parameters: 
\begin{itemize}
    \item All students had commitments to exam preparation during the course period.
 \item 20.4\% had additional part-time job commitments.
\item 11.9\% dedicated time to an internship.
\item 10.2\% were involved in supportive activities. 
\item 57.6\% had commitments to camps and other leisure activities.
\item 25.4\% juggled at least three types of commitments simultaneously.
  \end{itemize}
Mathematics grades negatively correlated with those involved in additional camp or leisure activities ($r = .269, p<.045$), but not with any other variable, though it is unclear why leisure and camp commitments may impact Mathematics grades specifically, but not Physics and English grades. \\

\begin{table}[h]
\centering
\captionsetup{justification=centering}
    \caption{Significant correlations between other variables}
    \label{tab: other corr}
\begin{tabular}{lllll}  
  \toprule
   & \begin{tabular}{@{}l@{}}
   \emph{The course} \\ \emph{strengthened} \\ \emph{my motivation}
   \end{tabular}
   & \begin{tabular}{@{}l@{}}
   \emph{The course} \\ \emph{motivated me} \\ \emph{pursuing STEM career}
   \end{tabular}
   & \begin{tabular}{@{}l@{}}
   \emph{How much} \\ \emph{did you enjoy} \\ \emph{the course?}
   \end{tabular}
   & \begin{tabular}{@{}l@{}}
   \emph{Completed} \\ \emph{school year}
   \end{tabular}
   \\
  \hline
QIST Knowledge & \begin{tabular}{@{}l@{}}.261* \\ .046 \end{tabular} &  &  & \begin{tabular}{@{}l@{}} .277* \\ .047 \end{tabular} \\ 
    \hspace{-7pt}
   \begin{tabular}{@{}l@{}}
   How engaging \\ were tutors' \\ communication and \\ teaching styles
   \end{tabular}
    & \begin{tabular}{@{}l@{}} .285* \\ .029 \end{tabular} & \begin{tabular}{@{}l@{}} .419** \\ .001 \end{tabular} & \begin{tabular}{@{}l@{}} .605** \\ .001 \end{tabular} &  \\ 
Hours spent a week & & & \begin{tabular}{@{}l@{}} .307* \\ .020 \end{tabular} &  \\ 
   \bottomrule
\end{tabular}
{* $P < 0.5$, ** $p=.001$}
\end{table}

Positive correlations between QIST knowledge and motivation suggested that, as QIST knowledge increased, participants' motivation also strengthened, confirming a positive impact of the course on motivation. 

We asked participants to evaluate how engaging they found the tutors’ communication and teaching style. The majority rated it as either good (40.7\%) or very good (29.6\%). Additionally, there was a highly significant correlation between this evaluation and the item “the course motivated me to pursue a STEM career.” This indicates that the experiment and teaching style significantly increased participants’ motivation to pursue a STEM career, demonstrating a sizeable positive impact ($r(54)=.419, p=.001$) on participants -- this result attests to the achievement of one of the primary goals of the experiment in supporting young people entering the STEM fields. Significant correlations between this item and ‘the course strengthened my motivation’ ($r(54)=.261, p=.02$) further validate this outcome. 

Interestingly, the number of hours spent studying per week showed no correlation with most variables. One exception was the question about how much one enjoyed the course ($r(57)=.307, p=.02$). 

\begin{figure}[H]
    \centering
\includegraphics[scale=0.5, trim = 0 10mm 0 0]{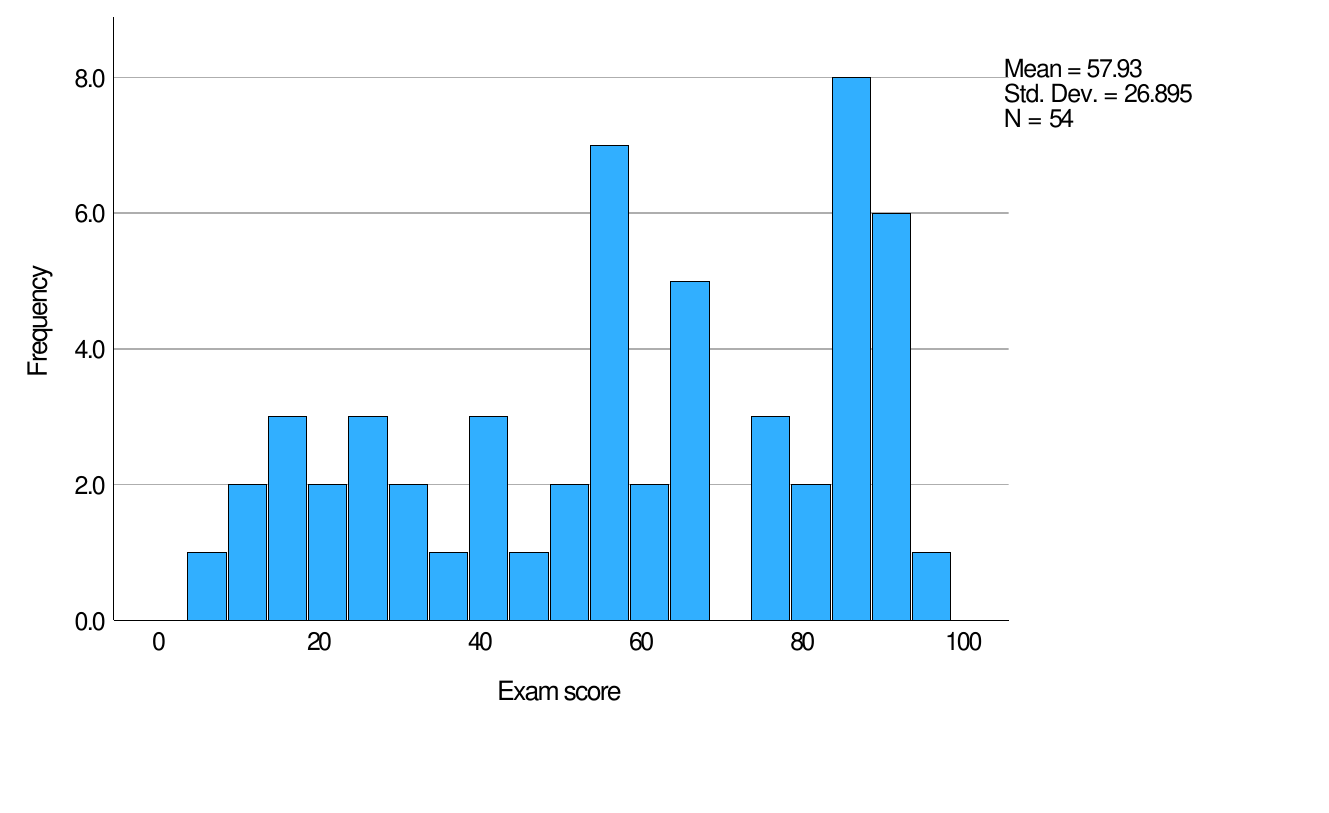}
 \caption{Distribution of raw exam marks.}
    \label{fig:raw marks}
\end{figure}

 \paragraph{Exam scores}
The key dependent variable of the present study concerns the overall, final score in the quantum exam, following the training. With a pass mark of 50\%, minimally we are interested in the extent to which there is evidence of above chance performance. This proved to be the case, as shown by a one-sample t-test against chance, $t(53)= 2.166, p=.035$, with a mean exam score of $57.93$ and standard deviation of $26.895$. The distributional characteristics of the exam scores in Figure \ref{fig:raw marks} shows a broad range of individuals, in terms of performance, from a sizable group of pupils with solid, over 70\%, to many who performed well below chance. 

Upon reviewing the raw data, 20 participants achieved a distinction ($>70$), while 7 participants attained a merit (60-69), and 9 participants scored within the pass range ($50-59$), with 3 participants falling into the near fail category ($40-49$). Furthermore, 15 participants received a fail grade ($0-39$). However, two canary questions were included in the exam, a practice that has been implemented at the Oxford Computer Science exam tradition for over a decade. These questions were designed serve as indicators of participants’ overall understanding, with successful passing dependent on their successful completion. As detailed in Appendix D, canary questions acted as a critical benchmark in our assessment, typically setting a standard that ensures comprehension of the fundamental concepts or skills necessary for successful completion of the training. Therefore, those who were unable to solve these specific questions would not have been able to achieve passing grades, regardless of their performance on other sections of the exam. Figure 6 shows the final mark distribution, for all scripts and for scripts passing the canary test respectively. They are plotted in 10-mark buckets. Conditional on the canary test, the distribution shifts to higher marks,  with no scripts marked below 10 and very few scripts marked below 20.

\begin{figure}[H]
    \centering
\includegraphics[scale=0.3]{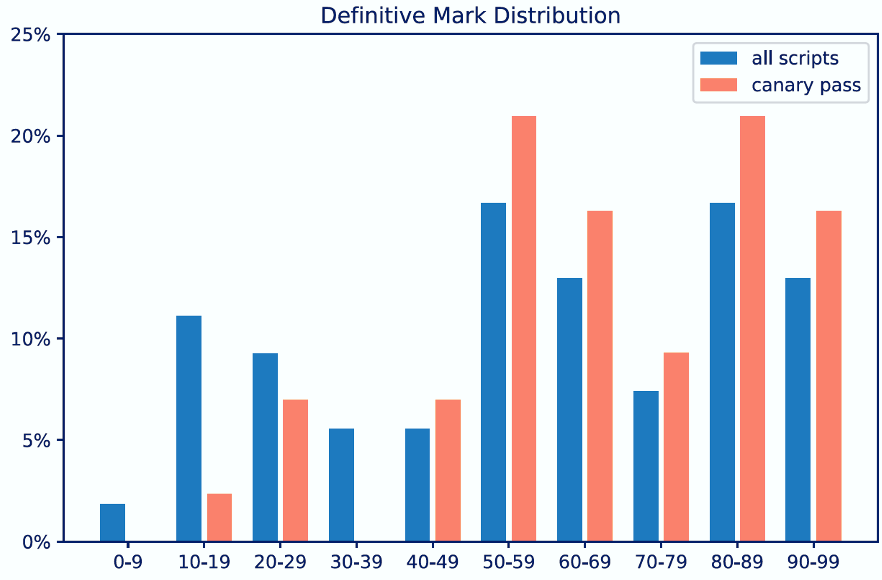}
  \caption{Distribution of marks based on canary and sentinel questions.}
   \label{fig:marks canary}
\end{figure}

Based on the inclusion of canary and sentinel questions in the assessment, the pass/fail rates among participants were: 82\% of participants successfully passed the exam, reflecting a substantial majority who met the minimum criteria set by the canary questions. Additionally, 48\% of participants achieved a distinction, indicating a high level of performance among those who passed. 

\paragraph{Markers' consistency}

\begin{figure}[H]
    \centering    
\includegraphics[scale=0.2]{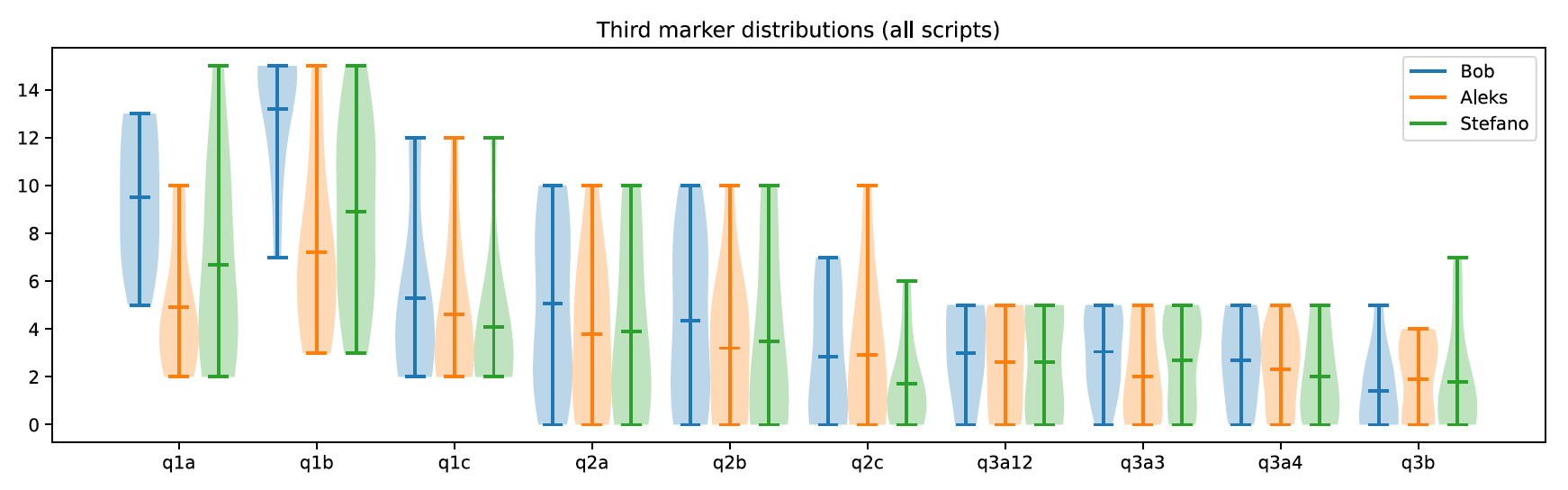}
 \caption{Distribution of raters' marks.}
  \label{fig:raters mark}
\end{figure}

We examined the consistency between the raters’ markings to evaluate how closely they agreed. Both markers, Bob Coecke and Aleks Kissinger, have extensive teaching and marking experiences with undergraduate and postgraduate students. However, the marking process was double-blinded, meaning that the raters were unaware of each other’s assessments. In cases where there was a discrepancy of more than 9\% between Coecke and Kissinger's marks for 10 students, Stefano Gogioso got involved as the third marker. Gogioso also has substantial experience in teaching and marking both QPic and HilbS exams, and his marks were considered alongside those of both raters. The violin plot in Figure \ref{fig:raters mark}  displays the distribution of marks for each question as assigned by three raters across all methods. 

The violin plot in Figure \ref{fig:raters mark}    shows the distribution of marks for each question, as assigned by Coecke (blue, left), Kissinger (orange, middle) and Gogioso (green, right) to triple-marked scripts only. Despite Gogioso's marks distributions being based on only a subset of scripts (10 participants), score distributions across the three raters look very similar, with the exception of questions 1a, 1b, 2c and 3b. 

It appears there are some systematic differences between Coecke and Kissinger. In questions 1a and 1b, Coecke’s scores are higher than Kissinger's scores, and for question 1b they are concentrated towards high marks (as shown by the violin plot being thicker at the top for Coecke’s distribution and at the bottom for Kissinger’s distribution). For these two questions, Gogioso’s marks span the full range of Coecke’s and Kissinger’s marks, and are more evenly distributed. These are the questions that contributed the majority of marks difference between Coecke and Kissinger, prompting the need for triple marking. That is, in the cases where the difference between Coecke and Kissinger was more than 9\% points, Gogioso was called to third-mark (this is why Gogioso marked far fewer scripts than either of the two original markers). Such a marking approach is consistent with standard practice in e.g. higher education. For questions 2c and 3b, the distributions of Coecke’s marks and Kissinger’s marks are not actually too different in their bulk, but a different maximum mark makes one of the two violins taller. In question 2c, Gogioso’s mark distribution is slightly more concentrated towards the bottom (i.e. Gogioso’s marks were slightly lower), while in question 3b they have a distribution similar to that of Coecke’s marks, but with a higher maximum mark assigned to one script.

We statistically confirmed the impression that Coecke assigned higher marks than Kissinger (overall one-way repeated measures ANOVA, $F(2,18)=19.3, p<.001$ and corresponding pairwise comparison between Coecke and Kissinger $t(53)=4.57, p<.001$; note the degrees of freedom are different between the omnibus ANOVA and the paired-samples comparison, because of missing values concerning Gogioso’s marking). 

\begin{figure}[H]
    \centering
\includegraphics[scale=0.2]{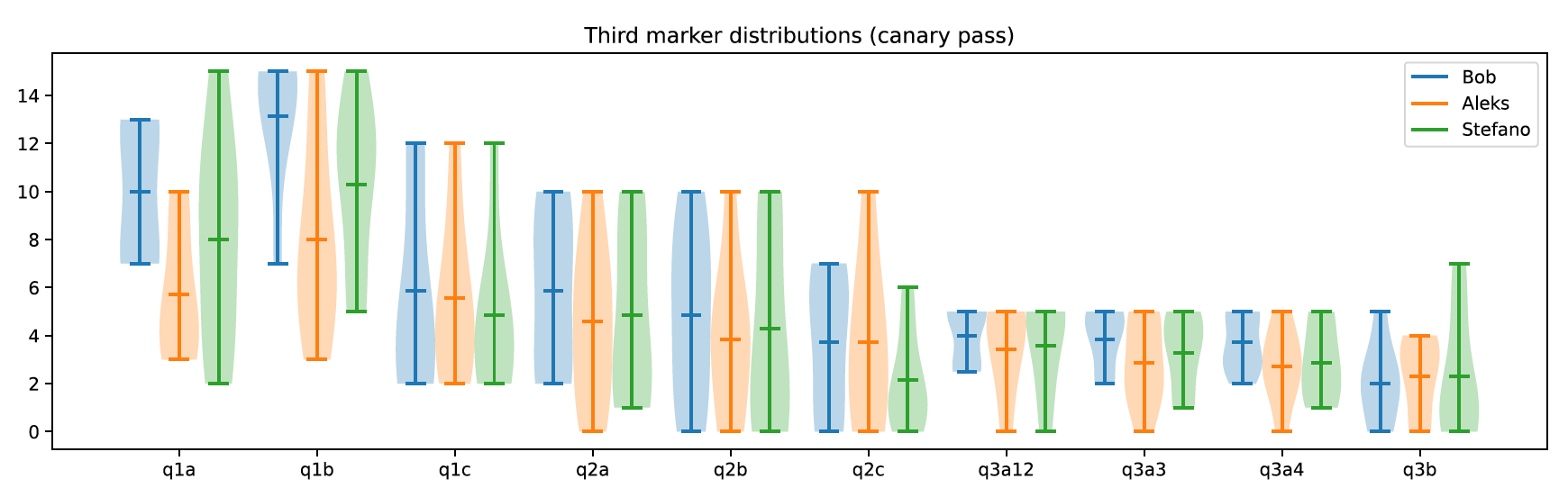}
\caption{Distribution of raters' marks based on canary questions.}
    \label{fig: raters marks canary}
\end{figure}

Figure \ref{fig: raters marks canary} contains the same information as the plot in Figure \ref{fig:raters mark}, with an important differentiation. Figure \ref{fig: raters marks canary} illustrates those scripts passing the canary test. 

Considerations are similar to Figure \ref{fig:raters mark}, with the exception that Coecke and Gogioso’s marks are more significantly concentrated towards the top for question 3a in triple-marked scripts. Some variations on these questions is to be expected, since high marks in parts 1 and 2 of question 3a were used as the canary test. However, it is notable that these variations come mostly from Coecke’s and Gogioso’s distributions, while Kissinger’s distribution remains fairly uniform.

\begin{figure}[H]
    \centering
\includegraphics[scale=0.2]{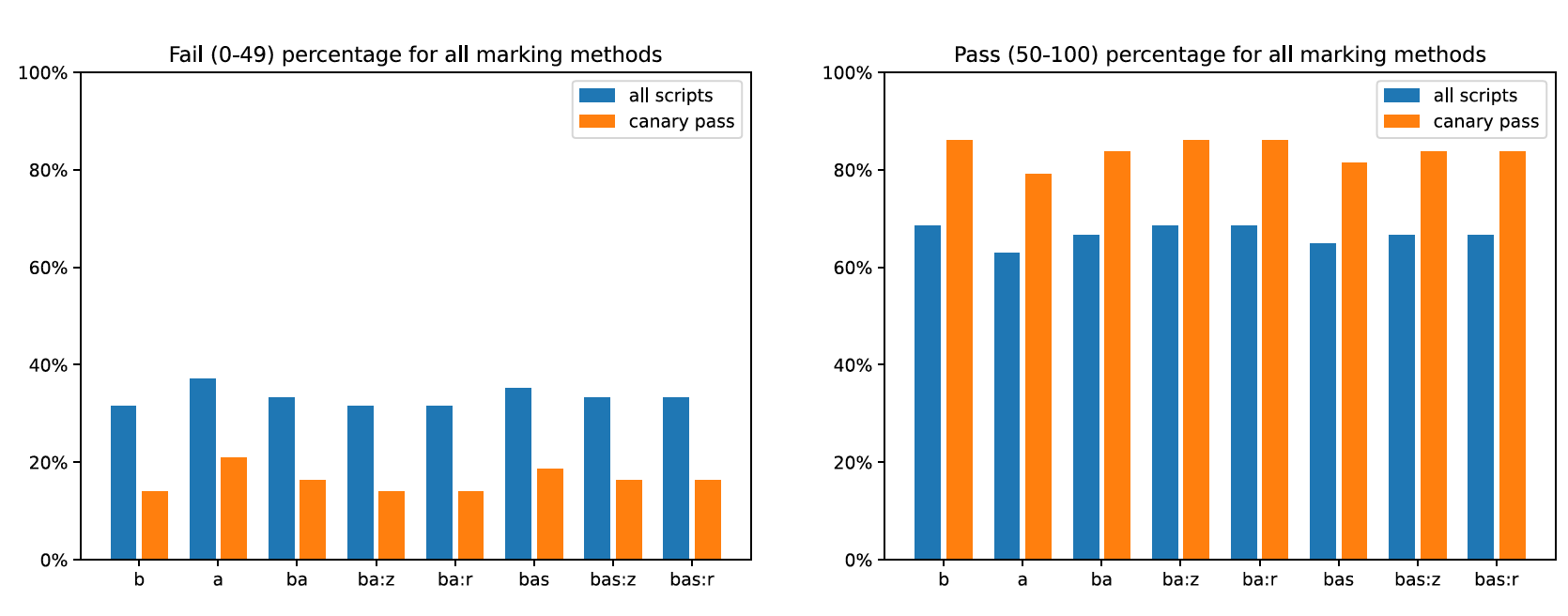}
\caption{Pass and fail percentages for all marking methods. X axis labels are: \texttt{b} for Bob alone; \texttt{a} for Aleks alone; \texttt{ba} for Bob and Aleks; \texttt{ba:z} for Bob and Aleks (z-score normalization); \texttt{ba:r} for Bob and Aleks (range normalization); \texttt{bas} for Bob, Aleks and Stefano; \texttt{bas:z} for Bob, Aleks and Stefano (z-score normalization); \texttt{bas:r} for Bob, Aleks and Stefano (range normalization).}
    \label{fig:pass and fail}
\end{figure}

Figure \ref{fig:pass and fail} plots demonstrate that the pass/fail percentages are very close with all scoring methods -both for canary and raw data. Around 65\% if computed over all scripts, and around 80\% if computed over the scripts passing the canary test.

\paragraph{Potential factors affecting participants marks}
Regarding the scores from the final exam, after training, there are three supplementary questions we can ask: whether these scores are dependent on age, school type, and location of residence. 
\begin{itemize}
    \item \emph{The developmental dynamics within the studied age range} (16 to 18 years), it is important to note that all participants were recruited within this limited range, with an average mean of 16.84 years and a standard deviation of $0.58$. A correlation between age and exam scores did not reveal evidence for a reliable trend ($r(54)=-0.19$). An alternative approach to exploring the dependence of exam scores on age would be to treat the four different ages amongst our participants as a nominal variable and consider possible differences via a one-way, between participants ANOVA. As can be seen in Figure \ref{fig: exam scores} below, there is a trend for lower exam scores, with participants who were 18 years old versus the younger ones (16, 17), but without any evidence for statistical reliability, $F(1,48)=1.45, p=.24$. Note, the differences in error bars might indicate a violation of the homogeneity of variance assumption, but this was not the case (using Levene’s test, $p=.78$). While we are reluctant to interpret a non-significant trend, it can be perhaps speculated that older participants had more varied and pressing responsibilities, such as A Level exams.  

\begin{figure}[H]
    \centering
\includegraphics[scale=0.6, trim = 0 30mm 0 0]{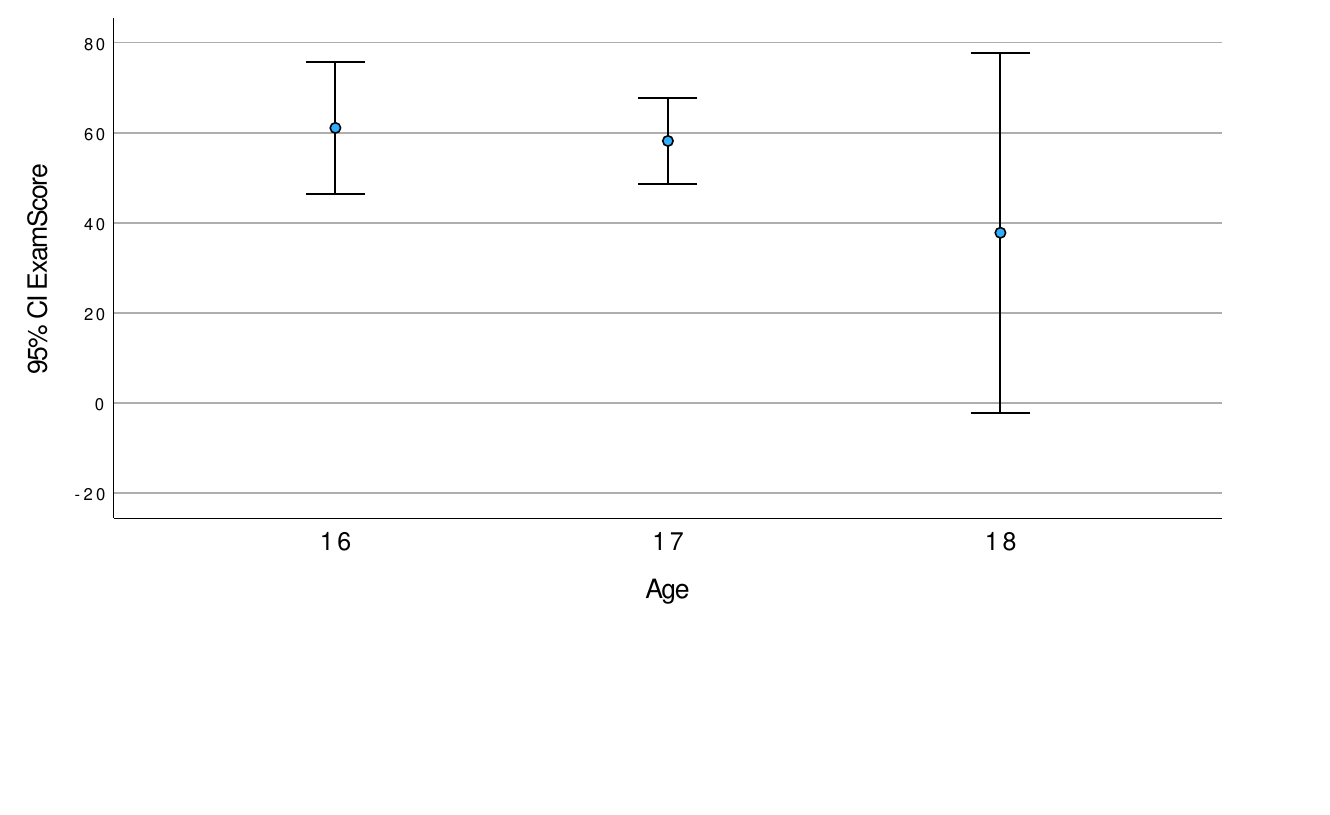}
\caption{Exam scores by age}
\label{fig: exam scores}
\end{figure}


\item Regarding location of residence, most participants (13/54) were from London, but there was a sizable number of participants with missing information (16/54). As 24 different locations were represented, we decided to aggregate into the following categories, for analyses: London, larger cities and towns with a university tradition (Cambridge, Bristol, Manchester etc.), and other cities and towns. Figure 11 below provides a visual impression of how scores differ, across the three types of location of residence. The trend for lower scores for participants from ‘other cities and towns’ was not significant, $F(2, 35)=0.96, p=.39$ (note, the residual degrees of freedom are lower in this comparison, because of missing values). It is perhaps unsurprising, as for many locations we had only one participant, which precludes systematic analyses. 

\begin{figure}[H]
\centering
\begin{minipage}{.5\textwidth}
  \centering
  \includegraphics[width=1.2\linewidth]{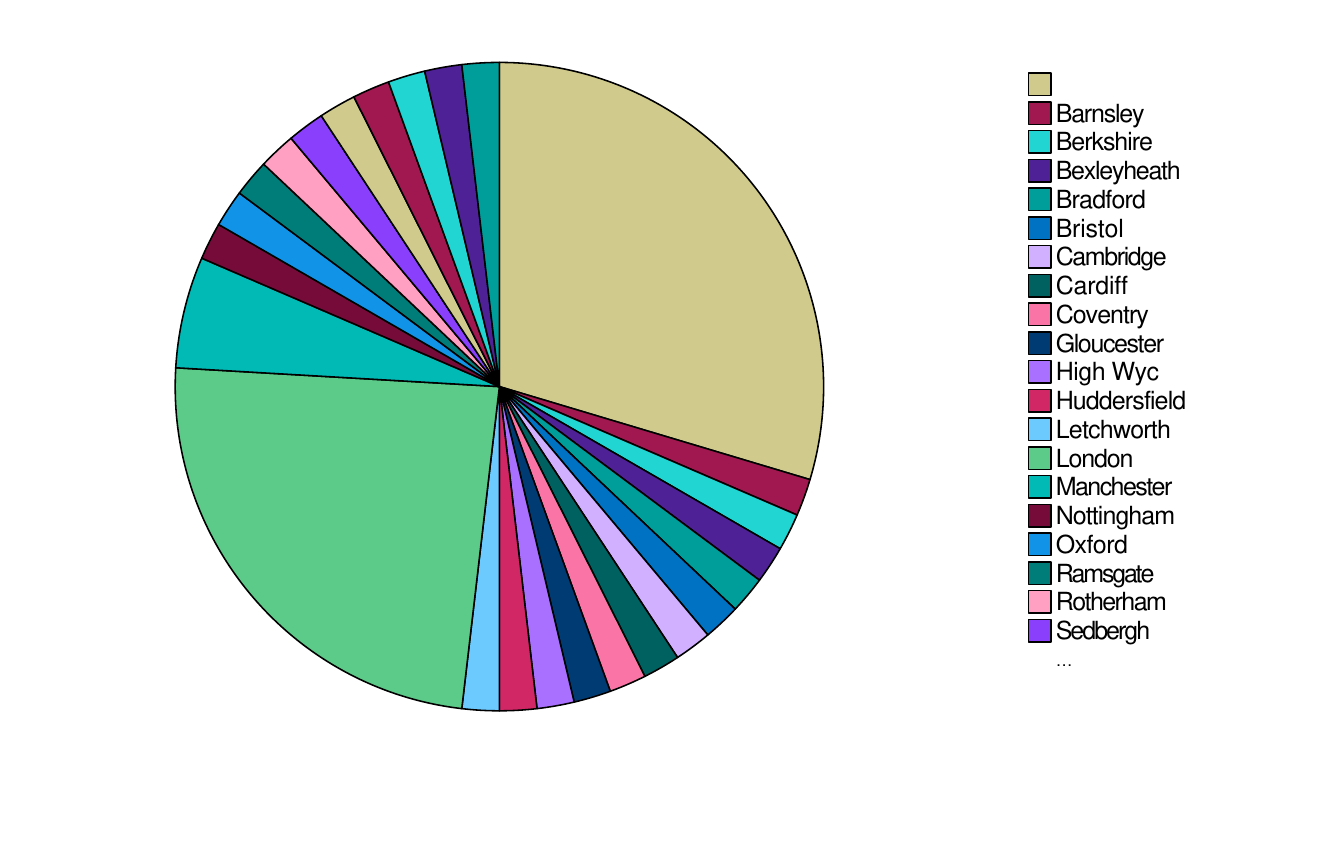}
\end{minipage}%
\begin{minipage}{.5\textwidth}
  \centering
  \includegraphics[width=0.8\linewidth]{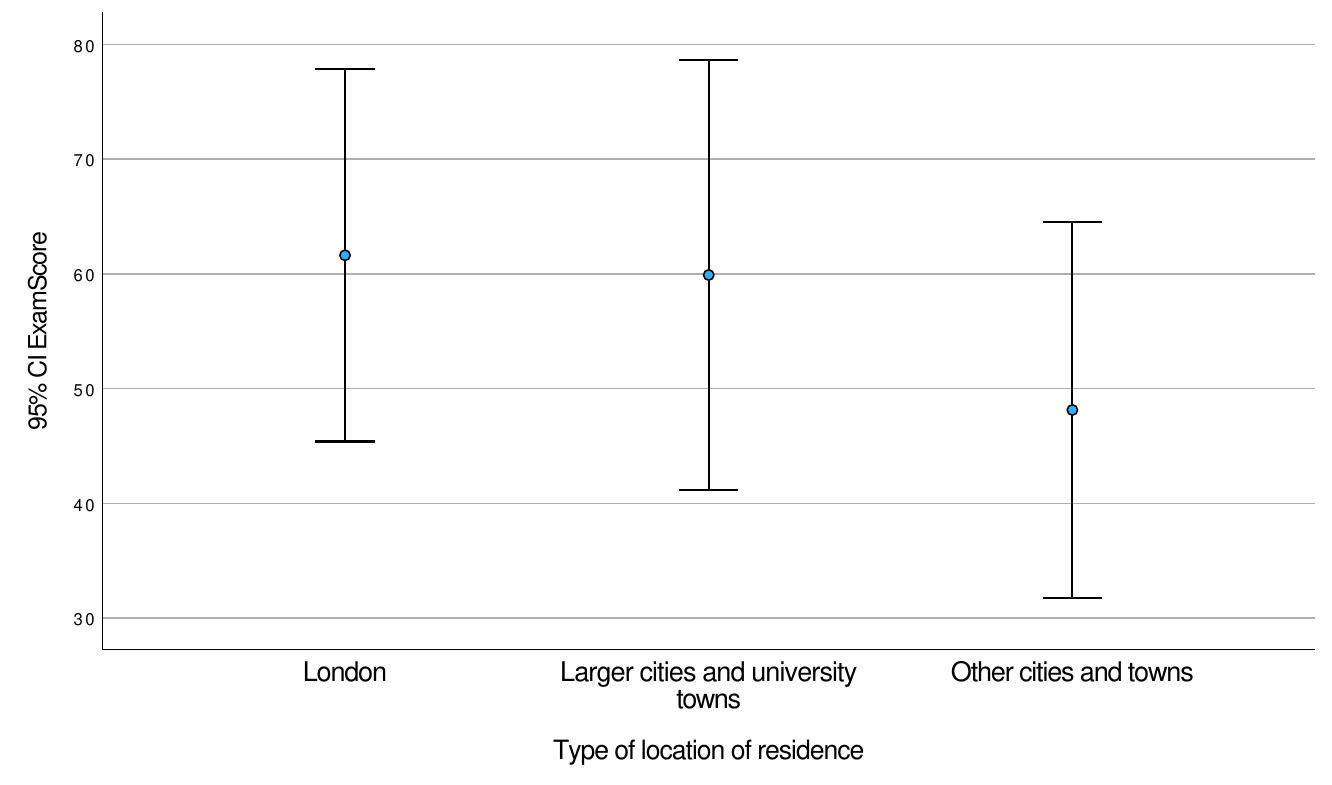}
\end{minipage}

\caption{Exam score distribution across the three types of residential locations.}
\end{figure}

\item Regarding school type, we had participants across five types of school: state, grammar, private boarding, state boarding, and private independent schools. However, participant numbers across the five school types varied considerably, as shown in Figure \ref{fig:school type}.  

\end{itemize}

\begin{figure}[H]
    \centering
    \includegraphics[scale=0.5, trim = 0 10mm 0 0]{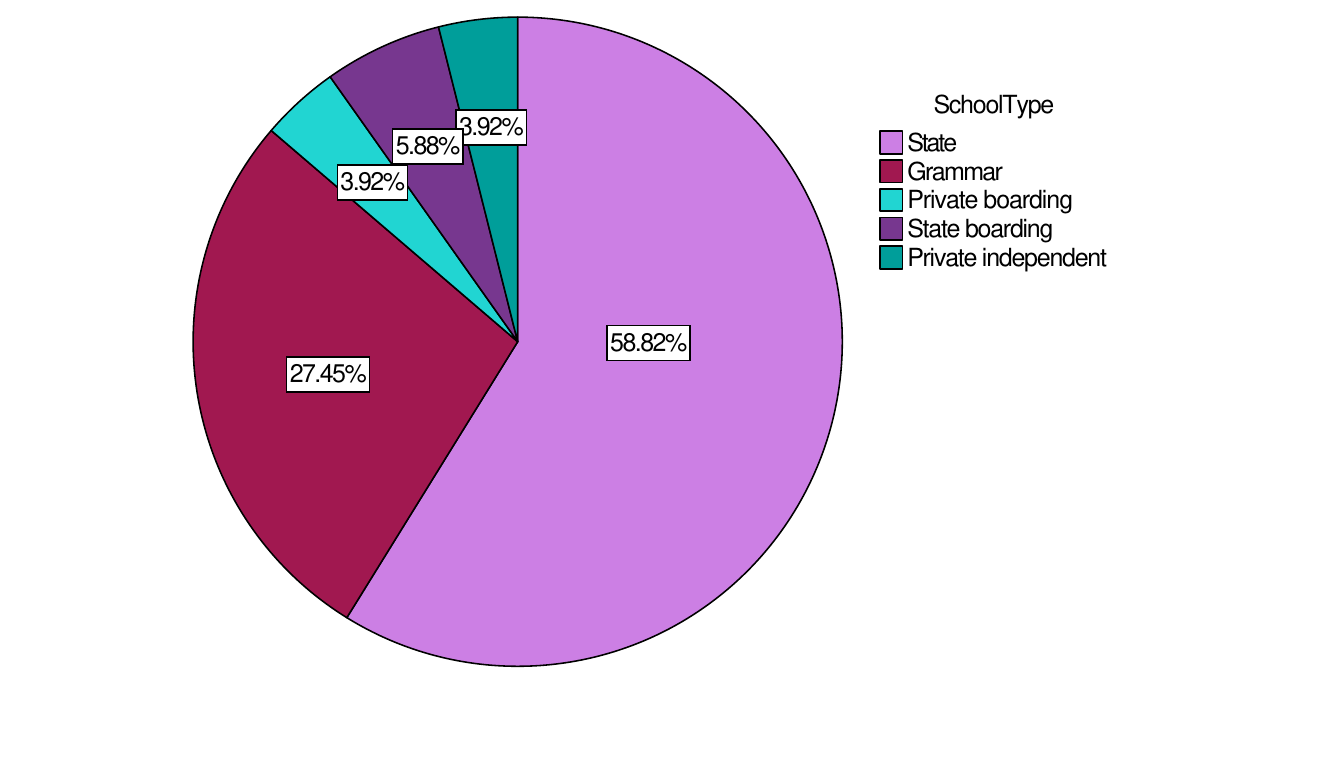}
    \caption{Participants by school type.}
    \label{fig:school type}

\end{figure}

Accordingly, so as to be able to run some meaningful analyses, we collapsed the categories of private boarding, state boarding, and private independent as in Table \ref{tab: school type}. A pedagogical justification for doing so would be that both private boarding and private independent schools would share some qualities in terms of e.g. socioeconomic characteristics of their students, while private and state boarding schools would be similar by virtue of the schooling format.

\begin{table}[h]
\centering
\captionsetup{justification=centering}
    \caption{School types}
\label{tab: school type}
\begin{tabular}{llllll}  
  \toprule
   \emph{School type} & \emph{N} & \emph{Mean}  & \emph{Min} & \emph{Max} & \emph{SD} \\
  \hline
  State &  30 & 54.4  & 6 & 95 & 28.4 \\ 
  Grammar & 14 & 59.7 & 29 & 93 & 22.4 \\ 
  Private boarding & 2 & 54.4 & 15 & 92 & 54.4 \\
  State boarding  & 3 & 58.7 & 19 & 92 & 36.9 \\
  Private independent  & 2 & 76 & 66 & 86 & 14.1 \\
   \bottomrule
\end{tabular}

\end{table}

\begin{figure}[H]
    \centering    
\includegraphics[trim = 0 40mm 0 0, scale=0.6]{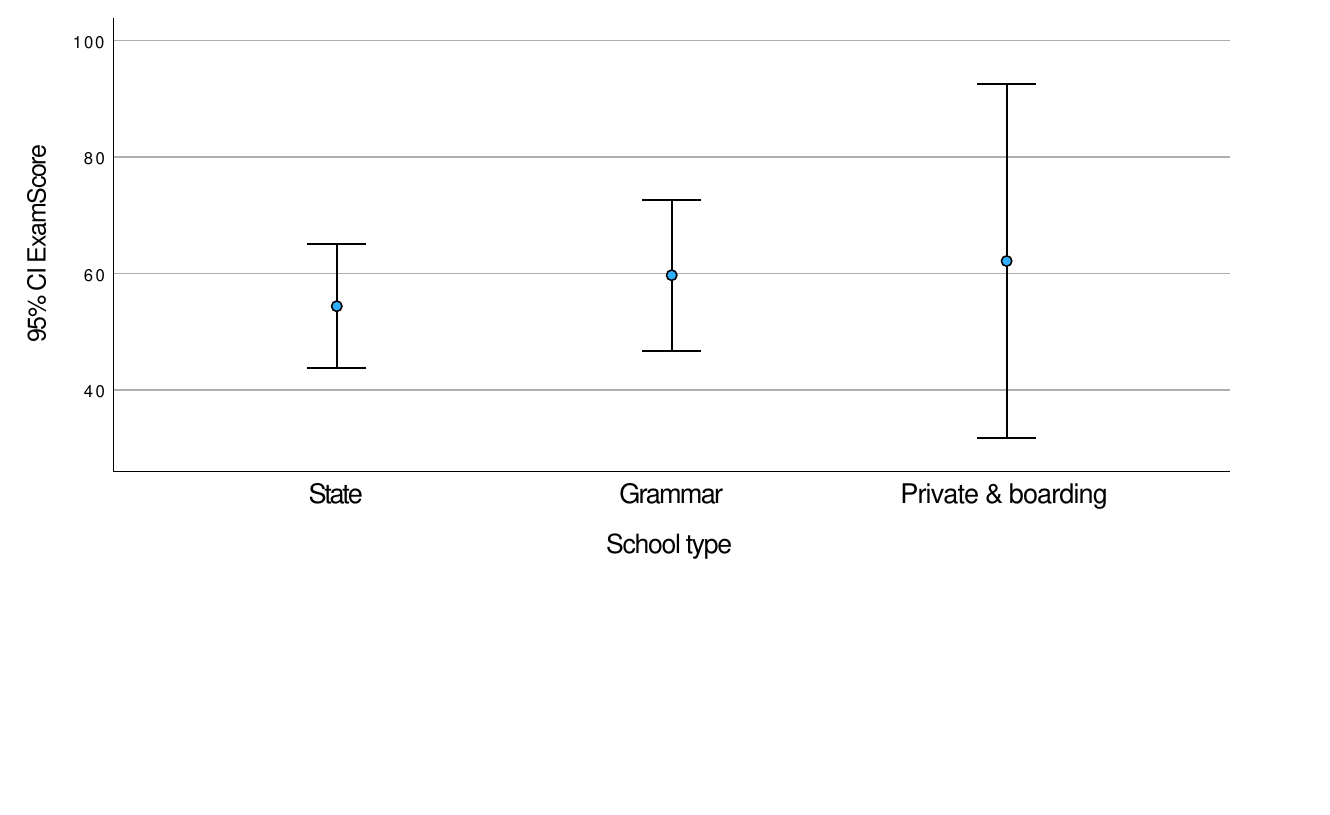}
\caption{Mean performance on the final QPic exam, by school type}
\label{fig: mean performance}
\end{figure}

 Figure \ref{fig: mean performance} shows means and error bars for average performance across the three categories. As can be seen, the means are nearly identical and a one-way between participants ANOVA failed to reach significance, $F(2, 48)=0.32, p=.72$ (note, Levene’s test offered no evidence for violations of the assumption of homogeneity of variance, $p=.56$). Further examination shows that, while mean performance for state boarding and private independent participants was nearly identical to that from state and grammar school participants, we have some extremely high variability regarding private boarding participants (this is partly due to the low N in this category).

\subsection{Whether QPic Increased participants’ confidence and motivation in science and STEM fields.}

We aimed to assess this by focusing primarily on participants’ satisfaction and suggestions for improvement. In response to the question regarding enjoying the course, opinions varied. While two participants expressed minimal enjoyment, 12 were uncertain, whereas the majority either enjoyed (47.5\%) or extremely enjoyed (28.8\%) the course. 
 
Regarding future iterations, participants provided good feedback for enhancing the course experience. Suggestions included increasing the number of tutorial sessions throughout the week, incorporating more examples, and expanding the number of quizzes. 
 
Additionally, participants recommended utilizing social networking platforms such as Discord (39\%) and Canvas (10.2\%) for enhanced interaction. However, a notable portion (47.5 \%) preferred communications via email. 

Two key questions aimed to evaluate the course’s impact on participants’ perspectives: 
\begin{itemize}
    \item ``Did the course strengthen your motivation and confidence to study QIST?”
\item  ``Did this course motivate you to pursue a career in STEM and increase motivation?”
\end{itemize}
Responses to these questions were documented in the following table.

\begin{table}[h]
\centering
\captionsetup{justification=centering}
\caption{Impact of the course on participants (\%)}
\label{tab:agree/disagree}
\begin{tabular}{lll}
 \toprule
  & \emph{
  \begin{tabular}{@{}l@{}}
  The course strengthened \\ my motivation and confidence
  \end{tabular}
  }
  & \emph{\begin{tabular}{@{}l@{}}
  The course motivated me \\ to pursue  a STEM career
  \end{tabular}
  } \\
 \hline
 Disagree  & 3.4 &  1.7  \\
Not sure & 28.8 & 11.9 \\
Agree & 37.3 & 32.2 \\
Strongly agree & 30.5 & 54.2 \\
\bottomrule
\end{tabular}
\end{table}

The majority of the participants either agreed or strongly agreed that the course strengthened their motivation and confidence to study QIST. Particularly, a substantial proportion of participants strongly agreed that the course motivated them to pursue a career in STEM fields. These highlight the significant impact of training on participants’ perspectives and indicate its effectiveness in cultivating interest not only in QIST but also in STEM fields broadly, potentially leading to enhanced engagement and career aspirations in these domains. 
\subsection{Exam scores and demographic variables}
It is interesting to consider whether better performance in the final exam increased participants’ interest in pursuing a STEM career. As well, we can ask whether there are gender, school type, or location differences concerning the final exam scores. 
We adopted an exploratory approach, starting in Table \ref{tab: big table} with correlations between exam scores and the motivation question, the `enjoy the course' question, school grades (for Mathematics, Physics, and English), and finally the maximum number of study hours per week. 

\begin{table}[h]
\centering
\captionsetup{justification=centering}
    \caption{Significant correlations}
    \label{tab: big table}
\begin{tabular}{ m{2.0cm} m{1.0cm} m{1.35cm} m{1.25cm} m{1.0cm}  m{1.0cm} m{1.0cm} m{1.5cm} m{1.5cm}}  
  \hline
  & \emph{Exam score}
  & \emph{The course motivated me to pursue a STEM career}
  & \emph{How much did you enjoy the course}
  & \emph{Maths grades at school}
  & \emph{Physics grades at school}
  & \emph{English grades at school}
  & \emph{Maximum hours spent studying per week} \\
  \hline
   Exam score
   & 1
   & \begin{tabular}{@{}l@{}}.002 \\ .991 \end{tabular}
   & \begin{tabular}{@{}l@{}}-.091 \\ .581 \end{tabular}
   & \begin{tabular}{@{}l@{}} .267 \\ .111 \end{tabular}
   & \begin{tabular}{@{}l@{}} .195 \\ .247\end{tabular}
   & \begin{tabular}{@{}l@{}}.168 \\ .322\end{tabular}
   & \begin{tabular}{@{}l@{}}.142 \\ .403 \end{tabular} \\

    \vspace{2mm}N
    & \vspace{2mm} 54
    & \vspace{2mm} 39
    & \vspace{2mm} 39
    & \vspace{2mm} 37
    & \vspace{2mm} 37
    & \vspace{2mm} 37
    & \vspace{2mm} 37 \\ 

    \vspace{2mm} The course motivated me to pursue a STEM career 
    & \begin{tabular}{@{}l@{}}  .002 \\ .991 \\ \end{tabular} \vspace{1mm}
    & \begin{tabular}{@{}l@{}} 1 \\ \\ \end{tabular} \vspace{1mm}
    & \begin{tabular}{@{}l@{}} .784** \\ $<$.001 \\ \end{tabular} \vspace{1mm}
    & \begin{tabular}{@{}l@{}} .182 \\ .280 \\ \end{tabular} \vspace{1mm}
    & \begin{tabular}{@{}l@{}} .193 \\ .251 \\ \end{tabular} \vspace{1mm}
    & \begin{tabular}{@{}l@{}} .160 \\ .346 \\ \end{tabular} \vspace{1mm}
    & \begin{tabular}{@{}l@{}} .143 \\ .397 \\ \end{tabular} \vspace{1mm}\\
     
    \vspace{2mm} N
    & \vspace{2mm} 39 
    & \vspace{2mm} 39 
    & \vspace{2mm} 39 
    & \vspace{2mm} 37 
    & \vspace{2mm} 37 
    & \vspace{2mm} 37 
    & \vspace{2mm} 37  \\ 
      
    \vspace{2mm} How much did you enjoy this course? 
    & \vspace{0mm} \begin{tabular}{@{}l@{}} -.091 \\ .581 \end{tabular} 
    & \vspace{0mm} \begin{tabular}{@{}l@{}} .784** \\ $<$.001 \end{tabular} 
    & \vspace{0mm} \begin{tabular}{@{}l@{}} 1 \\\end{tabular} 
    & \vspace{0mm} \begin{tabular}{@{}l@{}}  .014 \\ .932 \end{tabular} 
    & \vspace{0mm} \begin{tabular}{@{}l@{}} .067 \\ .694 \end{tabular} 
    & \vspace{0mm} \begin{tabular}{@{}l@{}} -.062 \\ .714 \end{tabular} 
    & \vspace{0mm} \begin{tabular}{@{}l@{}} .293 \\ .078 \end{tabular}\\

    \vspace{2mm} N 
    & \vspace{2mm} 39 
    & \vspace{2mm} 39 
    & \vspace{2mm} 39 
    & \vspace{2mm} 37 
    & \vspace{2mm} 37 
    & \vspace{2mm} 37 
    & \vspace{2mm} 37  \\ 
 
    \vspace{2mm} Maths grades at school
    
    & \begin{tabular}{@{}l@{}} .267 \\ .111 \end{tabular} \vspace{1mm}
    & \begin{tabular}{@{}l@{}} .182 \\ .280 \end{tabular} \vspace{1mm} 
    & \begin{tabular}{@{}l@{}} .014 \\ .932 \end{tabular} \vspace{1mm}
    & \begin{tabular}{@{}l@{}} 1 \\ \end{tabular} \vspace{3mm}
    & \begin{tabular}{@{}l@{}} .756** \\ $<$.001 \end{tabular} \vspace{1mm} 
    & \begin{tabular}{@{}l@{}} .637** \\ $<$.001 \end{tabular} \vspace{1mm} 
    & \begin{tabular}{@{}l@{}} -.282 \\ .096 \end{tabular} \vspace{1mm}\\
 
    N \vspace{2mm} 
    & 37 \vspace{2mm}
    & 37 \vspace{2mm}
    & 37 \vspace{2mm}
    & 37 \vspace{2mm}
    & 37 \vspace{2mm}
    & 37 \vspace{2mm}
    & 36 \vspace{2mm}\\ 
 
    \vspace{2mm} Physics grades at school 
    & \vspace{0mm} \begin{tabular}{@{}l@{}} .195 \\ .247 \end{tabular} 
    & \vspace{0mm} \begin{tabular}{@{}l@{}} .193 \\ .251 \end{tabular} 
    & \vspace{0mm} \begin{tabular}{@{}l@{}}  .067 \\ .694 \end{tabular} 
    & \vspace{0mm} \begin{tabular}{@{}l@{}} .765** \\ $<$.001 \end{tabular} 
    & \vspace{0mm} \begin{tabular}{@{}l@{}} 1 \end{tabular} 
    & \vspace{0mm} \begin{tabular}{@{}l@{}} .486** \\ .001  \end{tabular} 
    & \vspace{0mm} \begin{tabular}{@{}l@{}} -.094 \\ .584 \end{tabular} \\
 
    \vspace{2mm} N 
    & \vspace{2mm} 37 
    & \vspace{2mm} 37 
    & \vspace{2mm} 37 
    & \vspace{2mm} 37 
    & \vspace{2mm} 37 
    & \vspace{2mm} 37 
    & \vspace{2mm} 36  \\ 
  
    \vspace{2mm} English grades at school 
    & \vspace{0mm} \begin{tabular}{@{}l@{}} .168 \\ .322 \end{tabular} 
    & \vspace{0mm} \begin{tabular}{@{}l@{}} .160 \\ .346 \end{tabular} 
    & \vspace{0mm} \begin{tabular}{@{}l@{}} -.063 \\ .714 \end{tabular} 
    & \vspace{0mm} \begin{tabular}{@{}l@{}}  .637** \\ $<$.001 \end{tabular} 
    & \vspace{0mm} \begin{tabular}{@{}l@{}}.486** \\ .002  \end{tabular} 
    & \vspace{0mm} \begin{tabular}{@{}l@{}} 1  \end{tabular} 
    & \vspace{0mm} \begin{tabular}{@{}l@{}} -.184 \\.282 \end{tabular} \\

    \vspace{2mm} N
    & \vspace{2mm} 37 
    & \vspace{2mm} 37 
    & \vspace{2mm} 37 
    & \vspace{2mm} 37 
    & \vspace{2mm} 37 
    & \vspace{2mm} 37 
    & \vspace{2mm} 36  \\ 

    \vspace{2mm} Maximum hours spent studying per week
    & \begin{tabular}{@{}l@{}} .142 \\ .403 \end{tabular} \vspace{1mm}
    & \begin{tabular}{@{}l@{}} .143 \\ .397 \end{tabular} \vspace{1mm}
    & \begin{tabular}{@{}l@{}} .293 \\ .078 \end{tabular} \vspace{1mm}
    & \begin{tabular}{@{}l@{}} -.282 \\ .096 \end{tabular} \vspace{1mm}
    & \begin{tabular}{@{}l@{}} -.094 \\ .584 \end{tabular} \vspace{1mm}
    & \begin{tabular}{@{}l@{}} -.184 \\ .282 \end{tabular} \vspace{1mm}
    & \begin{tabular}{@{}l@{}} 1 \end{tabular} \vspace{2mm}\\ 

    \vspace{2mm} N 
    & \vspace{2mm} 37 
    & \vspace{2mm} 37 
    & \vspace{2mm} 37 
    & \vspace{2mm} 36 
    & \vspace{2mm} 36 
    & \vspace{2mm} 36 
    & \vspace{2mm} 37  \\ 
\hline
\end{tabular}
{** Correlation is significant at the 0.01 level (2-tailed)}
\end{table}

As can be seen, the only statistically reliable result concerned an increased motivation to pursue a STEM career, based on the enjoyment of the course ($r(39)=.78, p<.001$). Interestingly, there was no corresponding association between STEM career motivation and the actual exam result, with a correlation of nearly zero ($r(39)=-.002, p=.99$). Regarding gender, school type, and location (all coded as above), we considered the same set of correlations separately for the different levels of each variable, with one variable level examined at a time (this approach is an informal way of examining interactions between the original variables in the correlations analysis and gender, school type, and location; it can help obtain a first impression of the direction of any possible interactions, if they exist). Regarding gender, the correlation between enjoyment and STEM career motivation was slightly stronger for male participants ($r(19)=.85, p<.001$) than female participants ($r(19)=.70, p<.001$); no other associations emerged as notable in these subgroups (here and elsewhere, we are interested in associations concerning exam scores, motivation for STEM careers and the rest of the variables). Regarding school type, for state, grammar, and private and boarding schools, we observed correlations of .76 (N=21), .78 (N=10), and .98 (N=8), respectively (all significant). Additionally, a very high correlation emerged between motivation for STEM career and max study hours, for private and boarding schools ($r(5)=.92, p=.028$). Finally, an equivalent analysis for location offered broadly similar conclusions. For participants for whom we had no location information, were based in London, were based in larger cities and university towns, and were based in other cities and towns, the correlations between STEM career motivations and course enjoyment were, respectively, $r(12)=.81 (p=.002), r(10)=.65, p=.044, r(7)=26, p=.58$, and $r(10)=.92, p<.001$. No other correlations were significant. 

Overall, we think it is an interesting finding that motivation for a STEM course (a key dependent variable in the present research) and enjoyment of the course were associated both across the sample as a whole and within many specific groups, based on the relevant demographic variables. By contrast, exam scores as such correlated poorly with both the STEM motivation variable and most other variables. In fact, it is almost curious that exam scores did not correlate with max study hours per week, as well as school grades. This does suggest that the innovative teaching method of QPic can proceed in the absence of traditional competencies in science subjects; indeed, it is possible that QPic might unlock creativity and skill in students, who might otherwise be challenged by the traditional teaching approaches in science. The pattern of correlations is such that we do not think more detailed analyses are warranted (e.g., explore the significance of interaction terms with gender, school type, or location in corresponding regression analyses).

\section{Discussion}\label{sec:Discussion}
Until recently, the only way to conceptualize quantum processes was through philosophical discussion or algebraic representations within HilbS formalism - while powerful and mathematically rigorous, this approach lacks intuitive benchmarks for individuals not deeply versed in abstract mathematics. We propose a new formalism for conceptualizing and communicating quantum processes: a fully diagrammatic language that represents these relationships entirely through visual means, as outlined above and in Appendix A. 

The benefits of such a formalism are multifaceted, offering significant potential to impact various fields, especially education. Drawing on a decade of research, we hypothesized that QPic could make learning more engaging and accessible, even for students lacking the necessary mathematical background. This study provides the first evidence on the educational benefits of QPic, with the data revealing insights into three primary aspects of its effectiveness as an educational tool, which are discussed below. 

\paragraph{Whether QPic is an effective formalism for presenting QIST to young learners}

While we did not measure whether QPic could aid long-term acquisition of the relevant knowledge or whether participants could continue solving similar problems after a long break from the course, the data showed an immediate training effect. We evaluated the effectiveness based on exam outcomes, participants’ feedback, and tutor observations. 

The data illustrated positive outcomes for this question. For example, the majority of participants rated tutor communication and their learning experience as good or very good, underscoring the effectiveness of the training. All participants rated the tutorials well-managed, indicating that logistical aspects contributed to a smooth learning process. Overall, the majority of participants (83\%) rated the tutorials as good or excellent, with no negative feedback, showing that from the participants' point of view, the diagrammatic approach of QPic was well-received and effective in conveying the complex content. In terms of perceived difficulty, the majority of participants (71.2\%) considered the exam's difficulty level just right, and the vast majority (89.8\%) felt the topics covered in the final exam were representative, indicating adequate preparations. When asked if they felt prepared to answer the exam questions, more than half of the sample (62.7\%) responded affirmatively, with some (42.5\%) indicating the third question as challenging, highlighting the discriminative nature of the exam questions. 

The positive reception of tutor communication, learning experience, and tutorial quality, coupled with the good levels of exam outcomes across a representative course content, underscores that QPic provides an accessible yet precise framework for understanding and working with QIST, as well as bridges the gap between abstract mathematical formalism and intuitive visual comprehension. As hypothesized, QPic has the potential to act as a collaborative formalism, allowing people to solve problems or work together more effectively by providing a common visual ground. Quantum circuits are one example: while an algebraic approach requires interpreting long sequences of operations and transformations, QPic visually lays out gates and qubits, allowing for more efficient mental processing. Integrating information into a holistic visual form makes content easier to process and significantly reduces the  cognitive load. Another example is visualizing the superposition principle through overlapping waves, which can be more straightforward than algebraic descriptions. This enhances communication and QPic potentially serves as a universal language that transcends linguistic and cultural barriers, facilitating clearer communication between individuals from diverse backgrounds.

\paragraph{Whether the content is within the Zone of Proximal Development of this age range
}

The experiment aimed to investigate whether the content falls within the ZPD for high school students when delivered using QPic. The ZPD refers to the range of skills and knowledge that learners can acquire with appropriate guidance and support and we wanted to determine if QPic provides the necessary tools to help young learners to acquire these skills despite elementary mathematical background. A positive outcome would confirm that adequate guidance can indeed enhance their understanding of the content. Conversely, an unsuccessful outcome would suggest that the experimental setup may not have provided sufficient support to bridge the gap between the standard high school curriculum and the more specialized demands of QIST

The data illustrated that age differences among participants (16 to 18) had no significant effect on their scores, suggesting minimal developmental variation in their understanding of the course material in this age range. This may also support our hypothesis that QPic potentially enables high-schoolers to learn QIST, as indicated by high pass rates and absence of age-related differences in comprehension. Moreover, there was a trend indicating higher exam scores for participants who were 16 and 17 years old compared to the older ones (18), although this difference was not statistically significant. While we are cautious about interpreting a non-significant trend, the error bars in  Figure 11 indicated that older participants had more varied scores, subject to further investigation in follow up studies. 

Post-training exam scores is the key dependent variable of this study and we are interested in the extent to which there is evidence of above-chance performance. This proved to be the case. The distributional characteristics of the exam scores showed a broad range of individual performances, but with a sizable group of participants with solid scores over 70\%, to many who performed well above chance. 

Interestingly, the study uncovered several noteworthy outcomes. Contrary to expectations, participants’ prior knowledge or school grades in subjects like Mathematics, Physics, and English did not correlate with their performance in learning. Instead, socioeconomic status (SES) factors emerged as significant predictors of post-test performance. For instance, students from homeowner families generally outperformed those from renter families, and there were notable differences in performance between students attending private schools versus state schools. Despite the small number of participants from private schools, their performances were significantly higher than those from other school types. 

While these findings are exploratory and require further investigation, our data suggest that SES factors may play a more influential role in shaping performance outcomes in QIST education than prior academic achievement in this specific subject. This highlights the complexity of educational equality and the potential impact of SES disparities on students’ academic success in STEM fields.

Another notable finding was the negative correlation between age and engagement with tutor communication and teaching. Younger participants tended to find tutor communication more engaging and rated more positively. Additionally, we observed positive correlations between gender and grades in Physics and Mathematics, with male participants scoring significantly higher than females in both subjects. This trend aligns with existing literature \citep{wang2017gender, wang2023gender} on gender differences in STEM fields, with males typically outperforming females. 

We also considered feelings of preparedness for exam questions. This feeling negatively correlated with Physics grades, potentially indicating that participants who felt less prepared tended to have higher Physics grades, while those who felt more prepared tended to have lower physics grades. The same question also negatively correlated with the final exam difficulty. This disparity between perceived preparedness and actual exam performance in A levels needs further investigations to understand the underlying factors at play, but we think this was possibly due to students deciding to devote more or less effort to the course versus their A levels. 

Overall, looking at these findings, while QPic can potentially help high school students learn QIST, various factors such as SES, school type (private vs non-private), and gender differences merit further scrutiny. Further investigation with comparative groups is underway to explore these factors in more depth and to refine the teaching approach to greater effectiveness across diverse learner groups. 

\paragraph{Whether QPic increases participants’ confidence and motivation towards science and STEM fields.}

Our research concluded that motivation for STEM courses, a pivotal dependent variable, correlated significantly with enjoyment of the course across the entire sample and within various demographic groups. In contrast, traditional metrics such as exam scores showed poor correlations with STEM motivation and other variables of interest. Surprisingly, there was little correlation observed between exam scores and factors like maximum study hours per week or overall school grades. 

These results suggested that QPic’s innovative teaching method can potentially foster motivation and enjoyment in STEM subjects and science independently of traditional academic competencies and previous grades. It may also imply that QPic can potentially stimulate creativity and skills among students who might struggle with conventional science teaching approaches. The pattern of correlations observed is robust on the interactions with gender, school type, and school location. 

To assess this, we primarily focused on participants’ satisfaction and suggestions for improvement. In response to the questions regarding enjoying the course, opinions varied, but the majority enjoyed (47.5\%) or extremely enjoyed (28.8\%) the course.

Some of the suggestions provided by participants included increasing the number of tutorial sessions throughout the week, incorporating more examples, expanding the number of quizzes, and utilizing social networking platforms such as Discord for enhanced interactions.  

The majority of the participants either agreed or strongly agreed that the course strengthened their motivations and confidence to continue studying QIST. Particularly, a substantial proportion of participants strongly agreed that the course motivated them to pursue a career in STEM fields, potentially leading to enhanced engagement and career aspirations in these domains. Interestingly, there was no corresponding correlation between STEM career motivation and actual exam results (with a correlation of nearly zero). 

Regarding gender, the correlations between enjoyment and STEM career motivation was slightly stronger for male participants. Additionally, a very high correlation emerged between motivation for a STEM career and max study hours for private and boarding schools participants. 

Overall, we think it is an interesting finding that motivation for a STEM course and enjoyment of the course were associated both across the sample as a whole and within many specific groups, based on relevant demographic variables. By contrast, exam scores correlated poorly with both the STEM motivation and most other variables, underscoring that QPic’s innovative teaching method  not only enhances understanding of QIST but also positively impacts students’ motivation and engagement in STEM education even in the absence of traditional competencies in science as evidenced by higher grades in corresponding subjects. Our teaching method was restricted to only online interactions (live tutorials), we could not incorporate many aspects of face-to-face teaching and other online interactions, such as through Discord or Canvas, were precluded due to ethics concern. Despite these limitations, the positive outcomes are very promising and indicate that it is possible that QPic might unlock creativity and skill in participants who might otherwise be challenged by traditional teaching approaches in science.

\section{Conclusions}\label{sec:Conclusions}
This experiment provided us with the first evidence of how lowering  entry barriers can enhance understanding. Our ongoing interest lies in understanding developmental trends in problem-solving skills. Notably, we found no significant developmental differences between 16- and 18-year-olds in our study. However, this trend may shift when we lower the age threshold to 11, prompting further investigation into how problem solving abilities evolve with age and how educational content in QIST can be tailored to be age-appropriate.

Even though it remains unclear which specific systems quantum technologies will ultimately rely on, the fact is that classical systems are insufficient for handling tasks. As a rapidly growing experimental field, QPic has the potential to contribute to workforce shortages in the quantum technology sector by offering a balanced method that simplifies concepts  without compromising the depth of training.

Further, we aim to explore how cultural factors influence comprehension and problem-solving, examining cross-cultural differences within the same age groups. Given QPic largely requires nonverbal processing, variations across countries should be minimal. We also assess its its impact in socioeconomically disadvantaged vs. developed nations. Currently, six projects are replicating or adapting the experiment across six countries, offering insights into the universality and adaptability of innovative STEM educational tools.

\section{Declarations}

\subsection{Availability of data and materials}
The datasets used and analysed during the current study are available from the corresponding author on reasonable request.
\subsection{Competing interests}
The authors declare that they have no competing interests. 
\subsection{Funding}

The experiment was made possible through funding provided by Quantinuum, with significant collaboration from colleagues at the University of Oxford, City University of London, and IBM. 
\subsection{Authors' contributions}
All authors contributed to the research, with varying involvement in manuscript preparation. Each author played a distinct role in the implementation of the experiment, production of materials, data collection, data analysis, public relations, and dissemination. 
\subsection{Acknowledgements}
We would like to extend our gratitude to the participating students, their supportive teachers, schools and families. Special thanks to Anna Pearson and Yoshi Yonezawa for their valuable comments through reviewing, and the tutors who assisted tutorials during the experiment. We are thankful to Sarah Meng Li for sharing her notes verifying quantum teleportation. We are also grateful to Karen Clayton for her help with logistical organisation. 

MHW would like to thank the Rhodes Trust and Magdalen College, Oxford, for funding his graduate studies.
LY would like to thank the Google PhD Fellowship for funding her graduate studies.

\begin{appendices}

\section{A Crash Course on QPic}
{Our focus is on graphical approaches in physics, which are already widely recognised and align with a longstanding tradition that dates back to the Feynman diagrams  -used to depict interactions among subatomic particles -followed by Penrose’s application of tensor notation for visualizing multilinear functions \citep{penrose1971applications}. 

Later research by \cite{joyal1991geometry, joyal1995geometry} in the 1990s proved that the use of graphical notation for physics could not only be assigned mathematical rigour, but could also be extended to the broader realm of applied and pure mathematics, under the unifying framework of category theory. The resulting field of research of “categorical quantum mechanics (CQM)”, emerging around 2004, fundamentally reformulated  QT in terms of Penrose diagrams \citep{abramsky2004categorical, abramsky2009categorical}.  Under this framework, it was shown that core aspects of quantum mechanics, with finite-dimensional spaces of quantum states, can be derived from the formal properties of dagger-compact categories, as well as from their associated graphical calculus. 

In this direction, the term “Quantum Picturalism”  was coined in 2009  \citep{coecke2010quantum}, although diagrammatic methods employed akin to QPic were initially introduced in 2005 \citep{coecke2006kindergarten}.
QPic transforms the unique features of quantum theory into visual diagrams, providing a tangible tool for conceptualizing core quantum principles, without relying on the traditional prerequisite-intensive HilbS formalism \citep{von2018mathematical}. Its foundation relies on graphical calculus for category theory, a mathematical toolbox aimed at representing categorical operations, along with their real-world applications, via diagrams. Its capacity arises from several key characteristics:

\begin{itemize}
\item All pictures employed in QPic have a direct mathematical analogue in the traditional Hilbert spaces formalism (\textit{soundness}) \citep{coecke2017picturing}.
\item QPic encompasses the entirety of quantum theory, implying that every possible proof in Hilbert space formalism can also be expressed diagrammatically (\textit{completeness}) \citep{hadzihasanovic2018two, poor2024zx}.
\item QPic diagrams can function as independent mathematical entities, without requiring accompanying text or algebraic representations. The only prerequisite to start learning how to perform diagrammatic computations is a basic knowledge of angles (\textit{mathematical rigor}) \citep{coecke2023quantum}.
\item QPic can enhance modularity in reasoning, helping isolate different stages of a process or computation, maintaining clarity and coherence (\textit{compositionality}) \citep{coecke2023compositionality}.  
\item As a consequence of its category-theoretic foundations,  QPic enables us to draw connections between apparently unrelated fields of research, e.g., ranging from quantum computing to linguistics (\textit{interdisciplinarity}) \citep{coecke2017quantum}.
\end{itemize}

We now present a very brief illustration of QPic, by no means attempting to cover its full scope here, but rather only aiming to give a sense of its key features.

The basic pictorial language of QPic consists of wires and boxes. A wire represents the pathway through which particles travel, while a box signifies a location/place where particles enter, undergo modifications and then exit onto other wires. Below is a depiction of a wire and a box.

\begin{center}
    \includegraphics[width=0.4\linewidth]{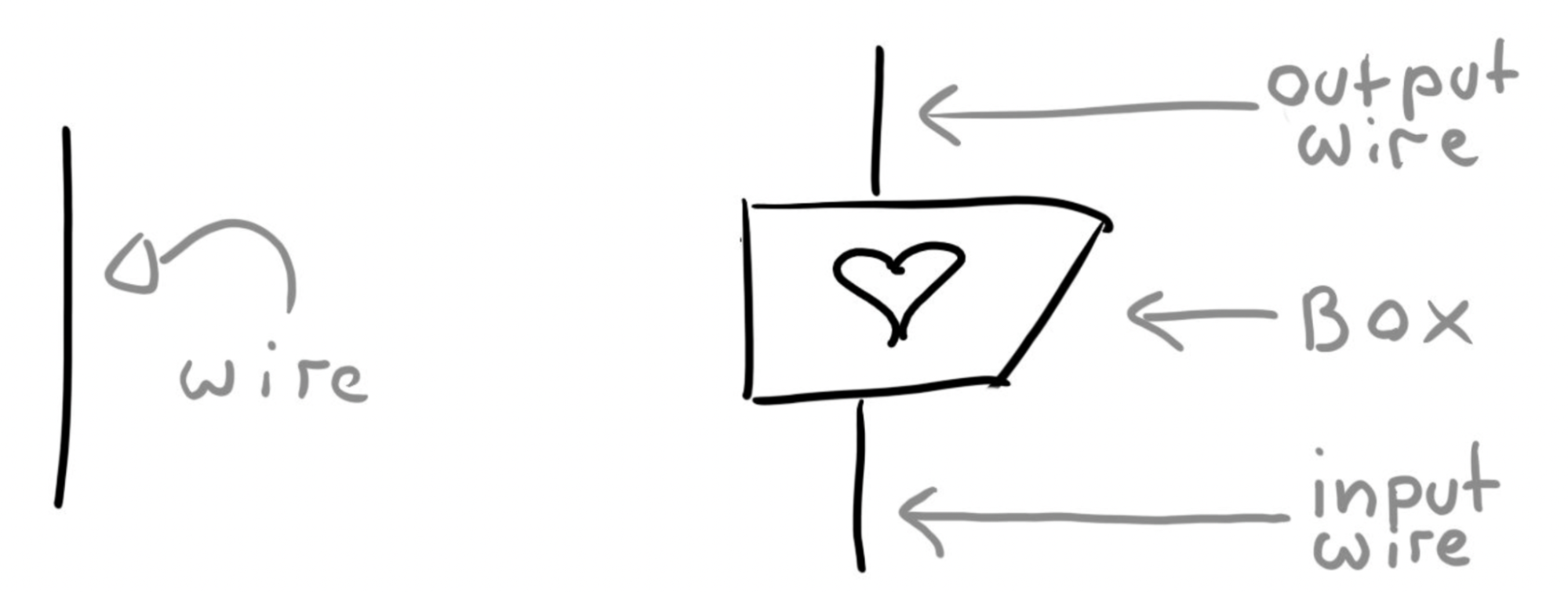}
\end{center}
The pictures above are the direct graphical analogue of their standard formulation in HilbS, where the wire is given by a 2x2 identity matrix, 

$$\begin{pmatrix}
1 & 0 \\
0 & 1 
\end{pmatrix}$$ ,  and the box is given by a 2x2 complex-valued matrix, 

$$\begin{pmatrix}
a & b \\
c & d 
\end{pmatrix}$$ As we anticipated in the previous section, the pictures above entirely replace their algebraic analogues. Therefore, their matrix representations can be ignored in all future computations, which will come down to a straightforward, rule-governed manipulation of pictures.

All the diagrams must be read from bottom to top. So in the box shown above, the particles go into the box from the bottom (via input wire) and leave the box at the top (via output wire). The example box has a single input and a single output. More generally, boxes can have any number of input wires and any number of output wires. Two important examples of boxes are those with no input wires and those with no output wires. These are called \textbf{states} and \textbf{tests} respectively. 

\begin{center}
    \includegraphics[width=0.5\linewidth]{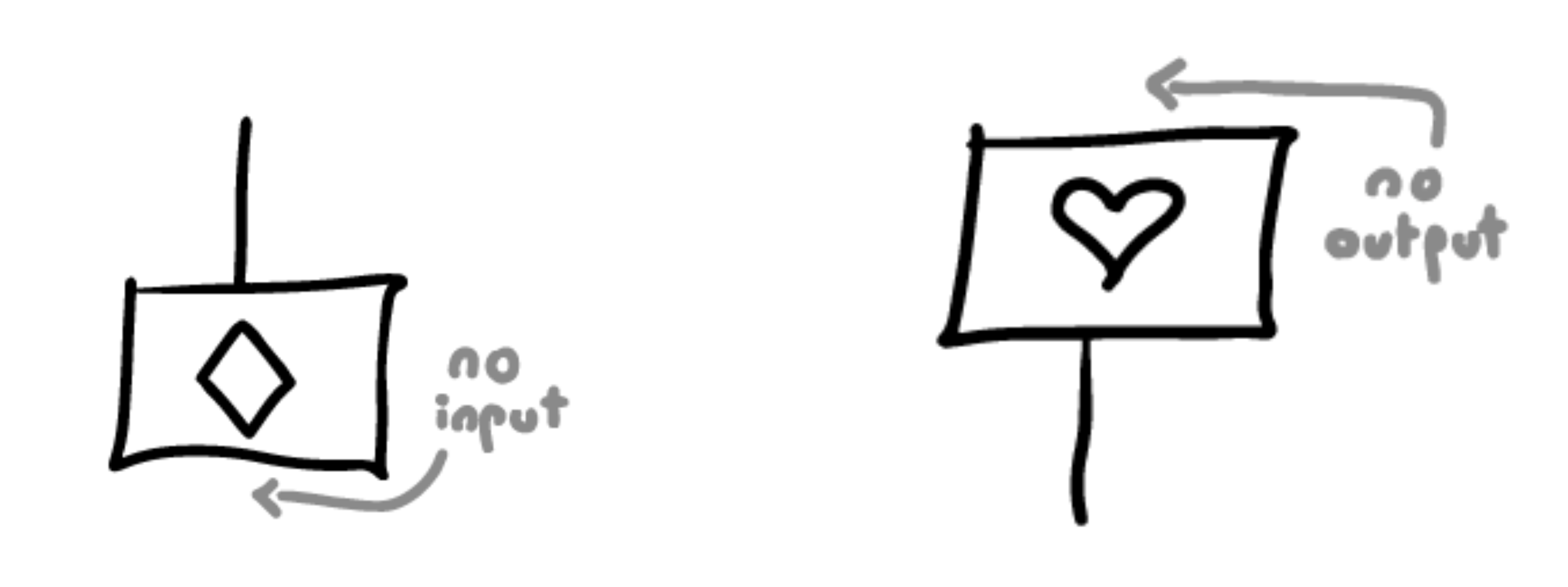}
\end{center}

The above state box prepares particles having the property `diamond’ which leave the box at the top. The test box tests whether the particles entering it from the bottom have the property `heart’.

In HilbS, a state with a single wire corresponds to a 2x1 complex column vector, 

$$\begin{pmatrix}
    a \\
    b
\end{pmatrix}$$  . Its dual test, which tests whether a given state is $$\begin{pmatrix}
    a & b 
\end{pmatrix}$$  is given by the row vector a b. Roughly speaking, states are used to describe quantum systems while tests are used to check whether a given quantum system is in a particular state. 

Wires and boxes alone already showcase the strongly compositional nature of QPic: picture 2 (the one on the right above) can be decomposed into two wires - the input and output wire - and a box. (attach pic of the decomposition if possible). Input and output wires of boxes can be plugged into other wires, giving rise to more complicated pictures, which we call “diagrams”.

\begin{center}
    \includegraphics[width=0.3\linewidth]{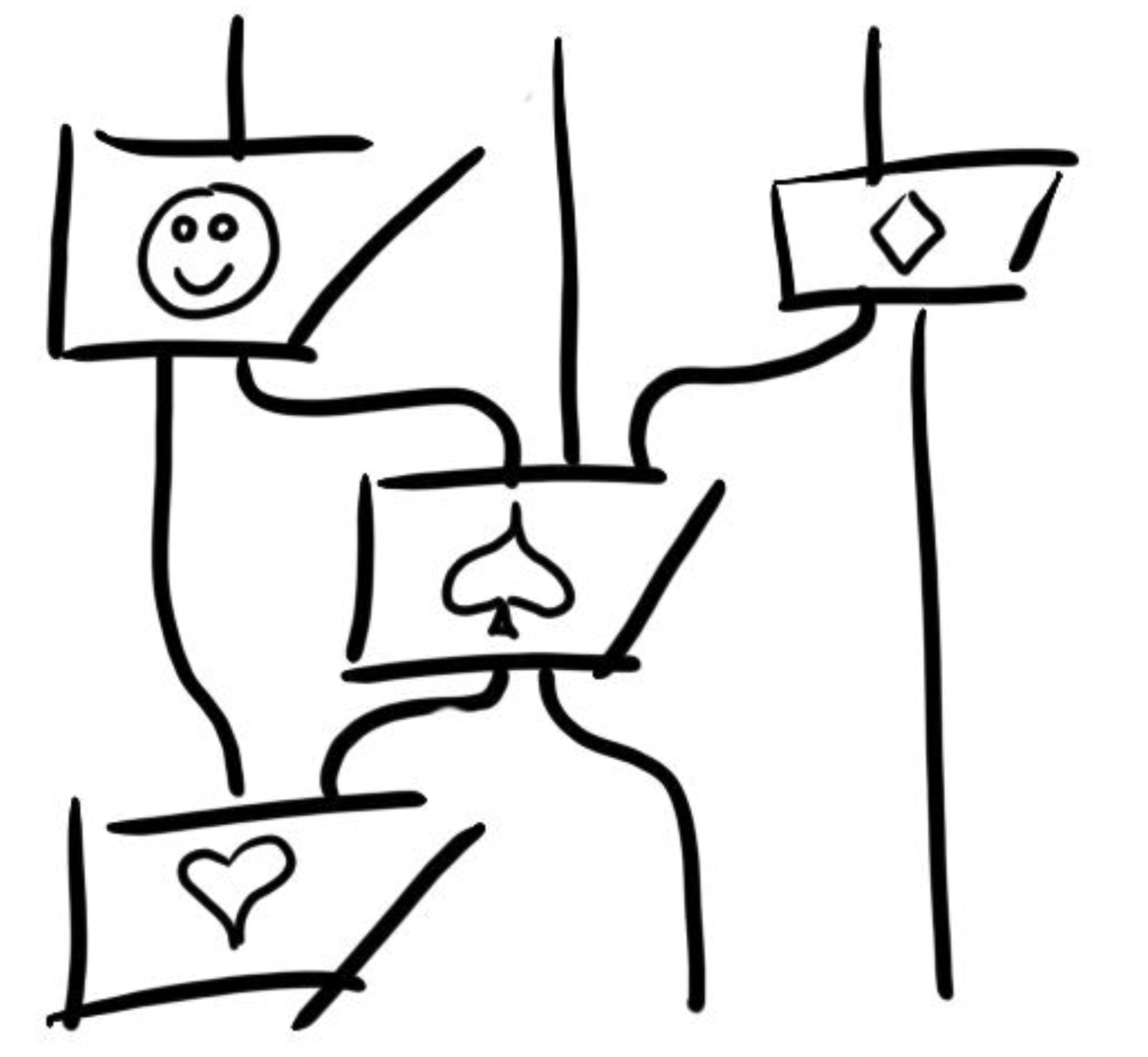}
\end{center}

Once we have diagrams, a key mathematical issue is to clarify when two diagrams are equal, which has the two-fold benefit of avoiding ambiguity as well as simplifying computations. A golden rule is that \textbf{only connectivity matters}: two diagrams made of the same boxes are equal if the boxes are connected in the same way. For example, the following two diagrams are equal even though they look different at first glance.

\begin{center}
    \includegraphics[width=0.45\linewidth]{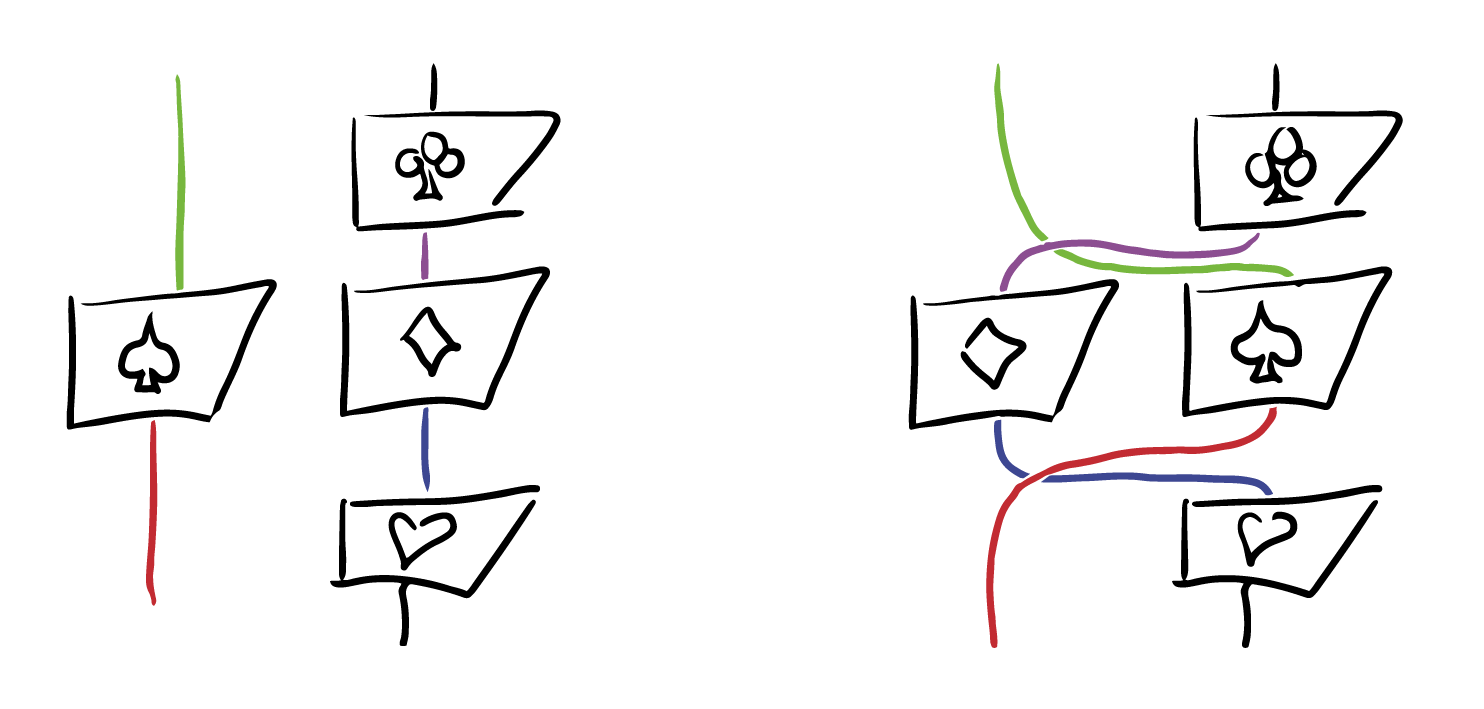}
\end{center}

The wires in the above diagrams are colour-coded to make the same connections apparent. Equality between diagrams is denoted via an `equals’ sign, which gives us an equation. 

\begin{center}
    \includegraphics[width=0.4\linewidth]{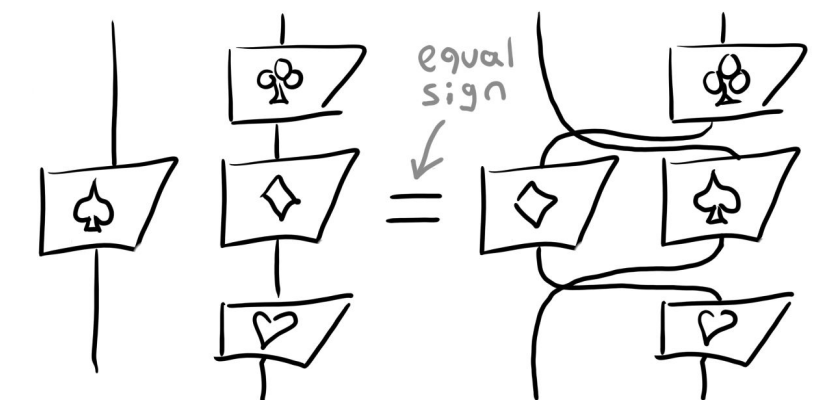}
\end{center}

Diagrams comprising different sets of boxes may also be equal. For example, consider the following equation.

\begin{center}
    \includegraphics[width=0.4\linewidth]{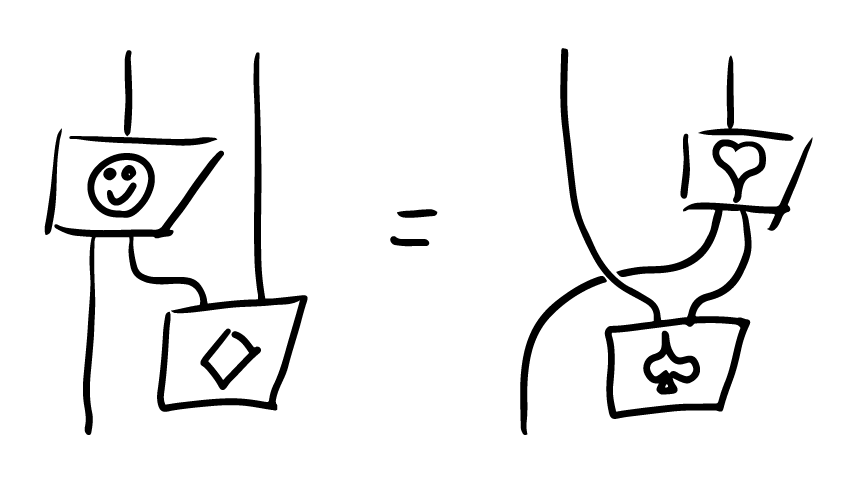}
\end{center}

This equation can be used to obtain more diagram equations by connecting new boxes on both sides. Two examples are given below.

\begin{center}
    \includegraphics[width=0.4\linewidth]{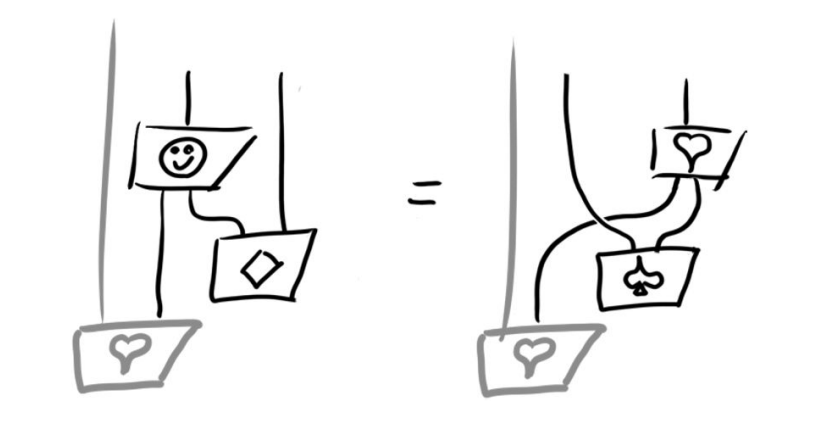}
\end{center}
\begin{center}
    \includegraphics[width=0.4\linewidth]{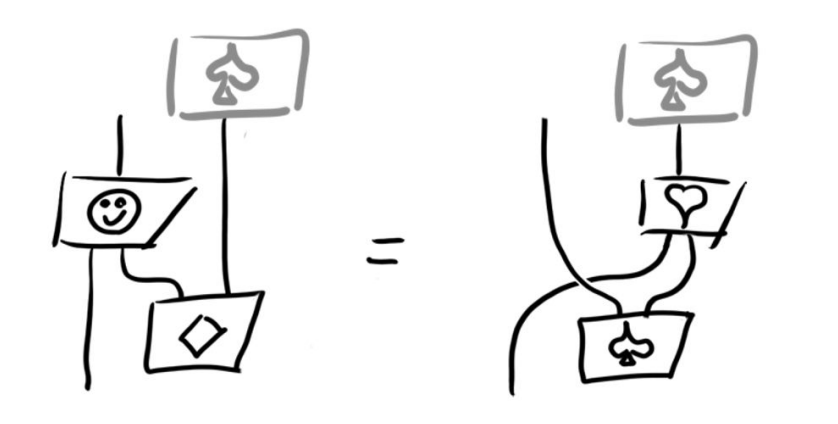}
\end{center}

The additional boxes and wires are depicted in gray colour.
The example above is another instance of QPic compositionality property at play: The only reason we are able to obtain new equations is that we are allowed to perform graphical “cut and paste” operations. 

There are some very special boxes which are themselves made up of wires, and therefore satisfy intuitive equations because of everything just being a wire:

\begin{center}
    \includegraphics[width=0.4\linewidth]{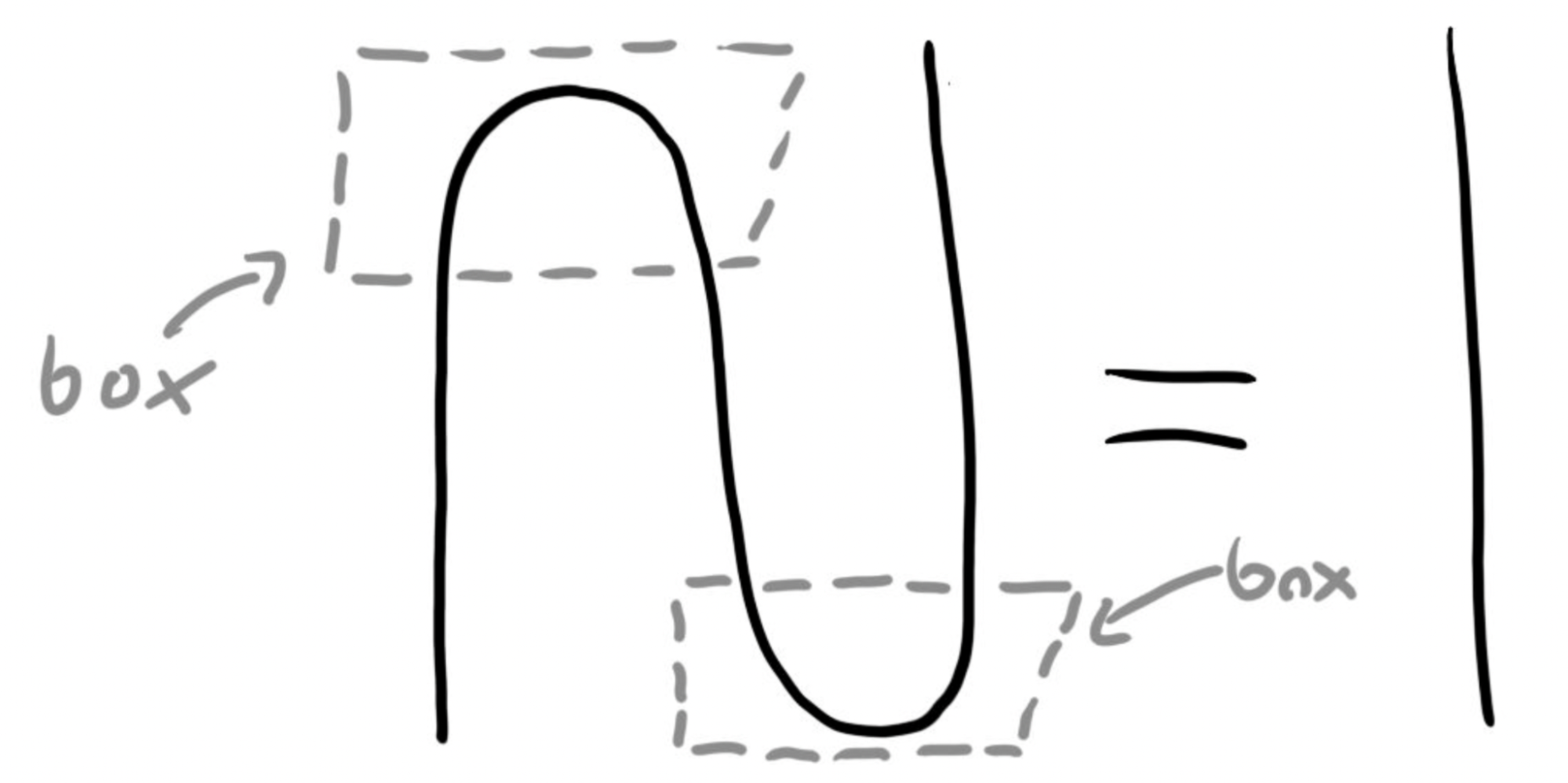}
\end{center}

These boxes are called cups and caps.  

\begin{center}
    \includegraphics[width=0.25\linewidth]{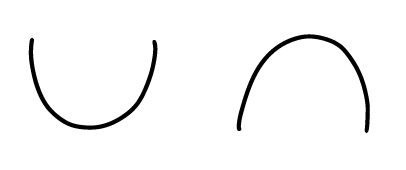}
\end{center}

Notice that the cup has two outputs but no input. Therefore, it is a state. The cap has two inputs but no output. It is, hence, a test.

The cup and the cap in QPic correspond precisely to vectors  $$\begin{pmatrix}
    1 \\
    0 \\
    0 \\
    1
\end{pmatrix}$$  and 
$$\begin{pmatrix}
    1 & 0 & 0 & 1
\end{pmatrix}$$

in HilbS, respectively. As for wires and boxes, their vectorial representations can be ignored in all future computations, which will simply rely on diagrammatic manipulation.
Diagrams obtained by sliding boxes along wires are equal. 

\begin{center}
    \includegraphics[width=0.4\linewidth]{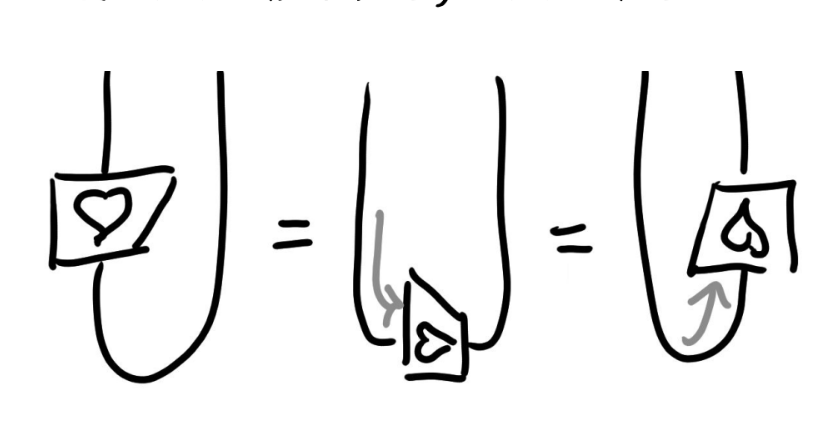}
\end{center}
\begin{center}
    \includegraphics[width=0.4\linewidth]{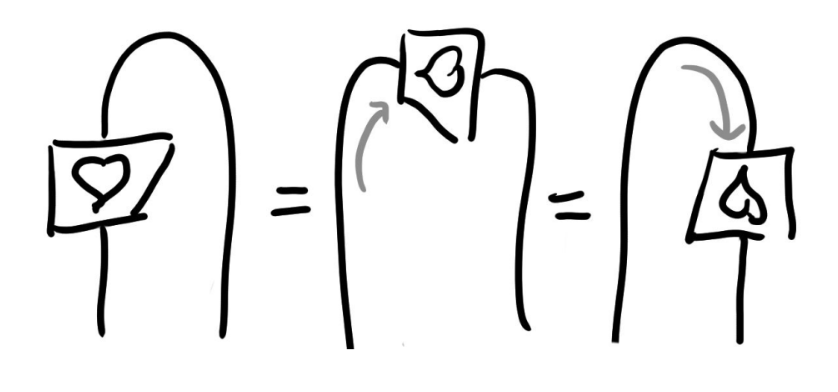}
\end{center}

Another intuitive equation satisfied by diagrams is given below.

\begin{center}
    \includegraphics[width=0.2\linewidth]{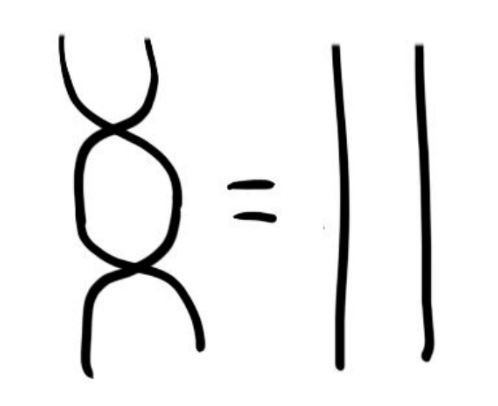}
\end{center}

The left-hand side involves swapping wires. In HilbS, a single swap is represented by the 4x4 matrix:   

$$\begin{pmatrix}
1 & 0 & 0 & 0 \\ 
0 & 0 & 1 & 0 \\  
0 & 1 & 0 & 0 \\
0 & 0 & 0 & 1
\end{pmatrix}$$

The example above showcases QPic’s potential to enhance intuition about quantum processes. While the matrix above (refer to it with a number) obscures conceptual understanding within its structure, QPic diagrams visually convey information by directly picturing the process of swapping wires. 

For these diagrams to represent or describe physical phenomena, there needs to be a notion of space and time incorporated into them. It was mentioned earlier that every diagram is to be read from bottom to top. This means that if we have a diagram with two boxes

\begin{center}
    \includegraphics[width=0.1\linewidth]{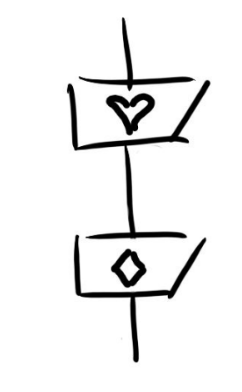}
\end{center}

it is to be understood as depicting that the `diamond’ box is applied first, followed by the `heart’ box. Therefore, time is ordered in the vertical direction of diagrams. A box that is closer to the bottom of the diagram happens earlier in time. 

In HilbS, the above diagram corresponds to multiplying the matrices corresponding to the `diamond’ and `heart’ boxes. Say these matrices are given by

$$\begin{pmatrix}
a & b \\ 
c & d  
\end{pmatrix}$$ 

and 

$$\begin{pmatrix}
e & f \\ 
g & h  
\end{pmatrix}$$

respectively. The above diagram gives the following matrix obtained by multiplication:

\[
\begin{pmatrix}
e & f \\
g & h
\end{pmatrix}
\times
\begin{pmatrix}
a & b \\
c & d
\end{pmatrix}
=
\begin{pmatrix}
ea + fc & eb + fd \\
ga + hc & gb + hd
\end{pmatrix}
\]

One needs to be careful about the order of multiplication. If two boxes are placed in parallel, this means that they are applied to their inputs at the same time but in different places. 

\begin{center}
    \includegraphics[width=0.2\linewidth]{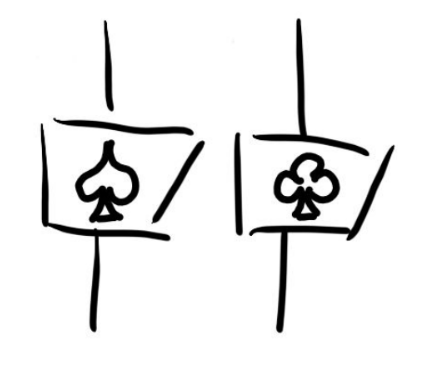}
\end{center}

Hence space is ordered in the horizontal direction of diagrams. In HilbS, the above diagram corresponds to taking the tensor product of matrices corresponding to the `spade’ and `club’ boxes. Say these matrices are given by

$$\begin{pmatrix}
a & b \\ 
c & d  
\end{pmatrix}$$ 

and 

$$\begin{pmatrix}
e & f \\ 
g & h  
\end{pmatrix}$$,

respectively. The above diagram gives the following matrix obtained by tensor:  

\[
\begin{pmatrix}
a & b \\ 
c & d  
\end{pmatrix}
\otimes
\begin{pmatrix}
e & f \\ 
g & h  
\end{pmatrix}
=
\begin{pmatrix}
a \begin{pmatrix} e & f \\ g & h \end{pmatrix} & b \begin{pmatrix} e & f \\ g & h \end{pmatrix} \\
c \begin{pmatrix} e & f \\ g & h \end{pmatrix} & d \begin{pmatrix} e & f \\ g & h \end{pmatrix}
\end{pmatrix}
\]

Notice that the order of the tensor product is different from that of matrix multiplication discussed earlier. 

Without any more ado, one can now derive quantum teleportation within this very minimal language:

\begin{center}
    \includegraphics[width=0.5\linewidth]{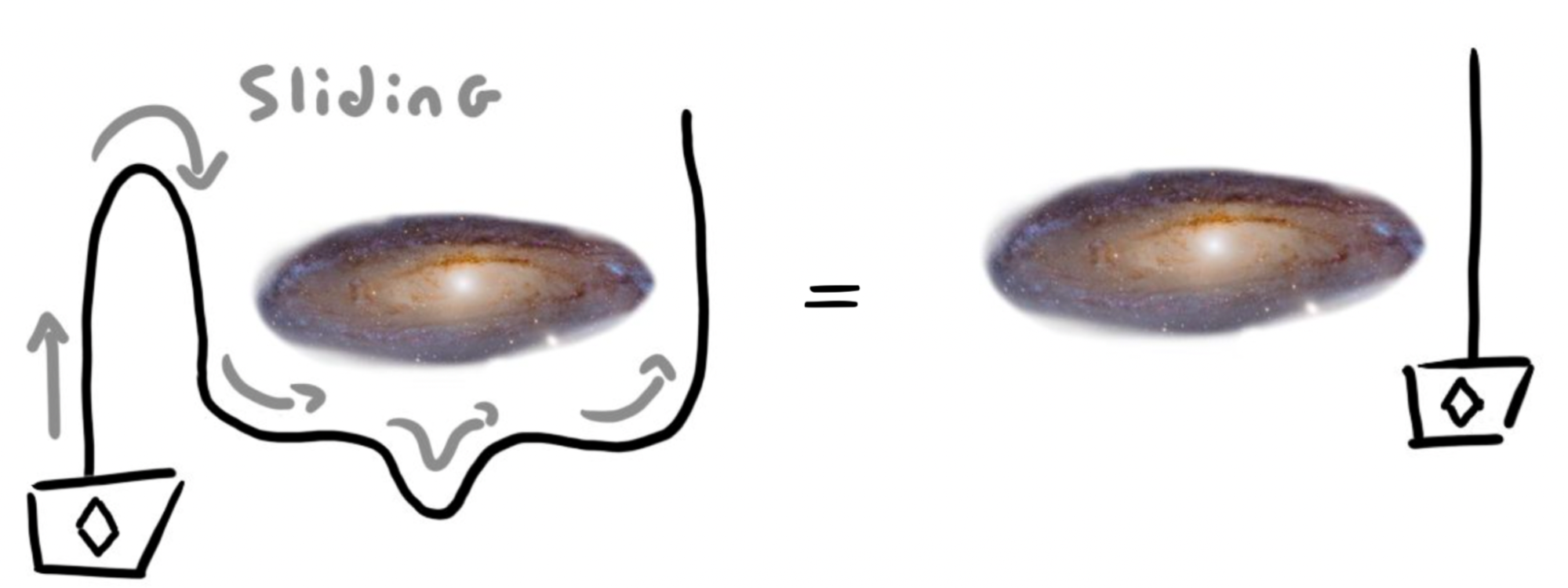}
\end{center}

The above equation can also be verified in HilbS using the corresponding linear algebraic components. For the left-hand side, this would involve taking the tensor product of `diamond’ and cup states, taking the tensor product of a cap test and an identity wire, and finally multiplying the two 8 by 8 matrices corresponding to the two tensor products. The result must be equal to the vector of the `diamond’ state, which is the left-hand side.

As a matter of fact, these cup- and cap-shaped boxes suffice to capture the key notions of linear algebra that are used in the usual quantum formalism. For example, the transpose of box

\begin{center}
    \includegraphics[width=0.1\linewidth]{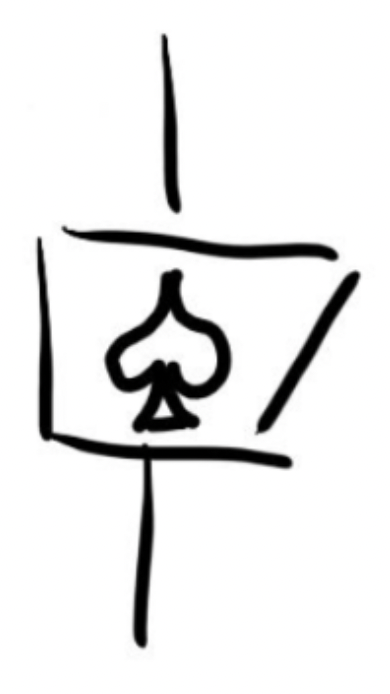}
\end{center}

is obtained by using a cup and a cap, as follows: 

\begin{center}
    \includegraphics[width=0.2\linewidth]{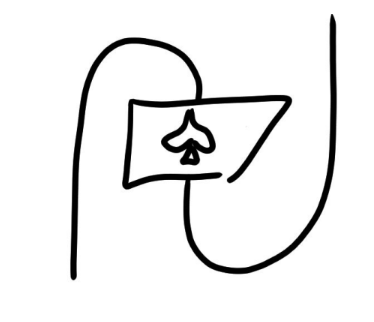}
\end{center}

which intuitively gives 

\begin{center}
    \includegraphics[width=0.4\linewidth]{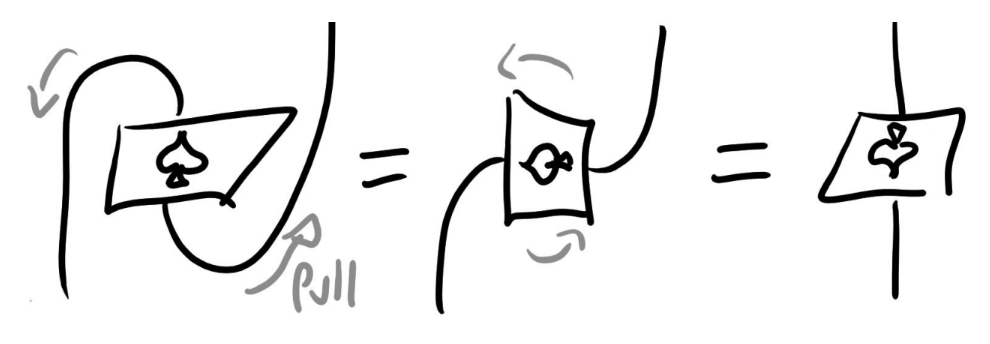}
\end{center}

In HilbS, the transpose of a matrix 

\[
\begin{pmatrix}
a & b \\
c & d
\end{pmatrix}
\]

is given by

\[
\begin{pmatrix}
a & c \\
b & d
\end{pmatrix}
\]

Therefore, whereas transposing a matrix involves rearranging its entries - without providing any intuition on the overall process - QPic diagrams visually convey the underlying operation, now amounting to flipping a box’s inputs with its outputs.

Using the ingredients of HilbS corresponding to diagrams, one can linear algebraically verify all the aforementioned diagrammatic equations.

Other key linear algebraic notions, such as inner product, isometry, unitarity, adjoints, and trace, all have equally simple pictorial interpretations. For readers interested in these pictorial representations and their correspondence to HS, we recommend \cite{coecke2010quantum, coecke2017picturing, coecke2015categorical, coecke2016categorical}.

In order to capture all of quantum theory, we don’t need that much more. The idea is to generalise the notion of wire to what we have called a spider:

\begin{center}
    \includegraphics[width=0.3\linewidth]{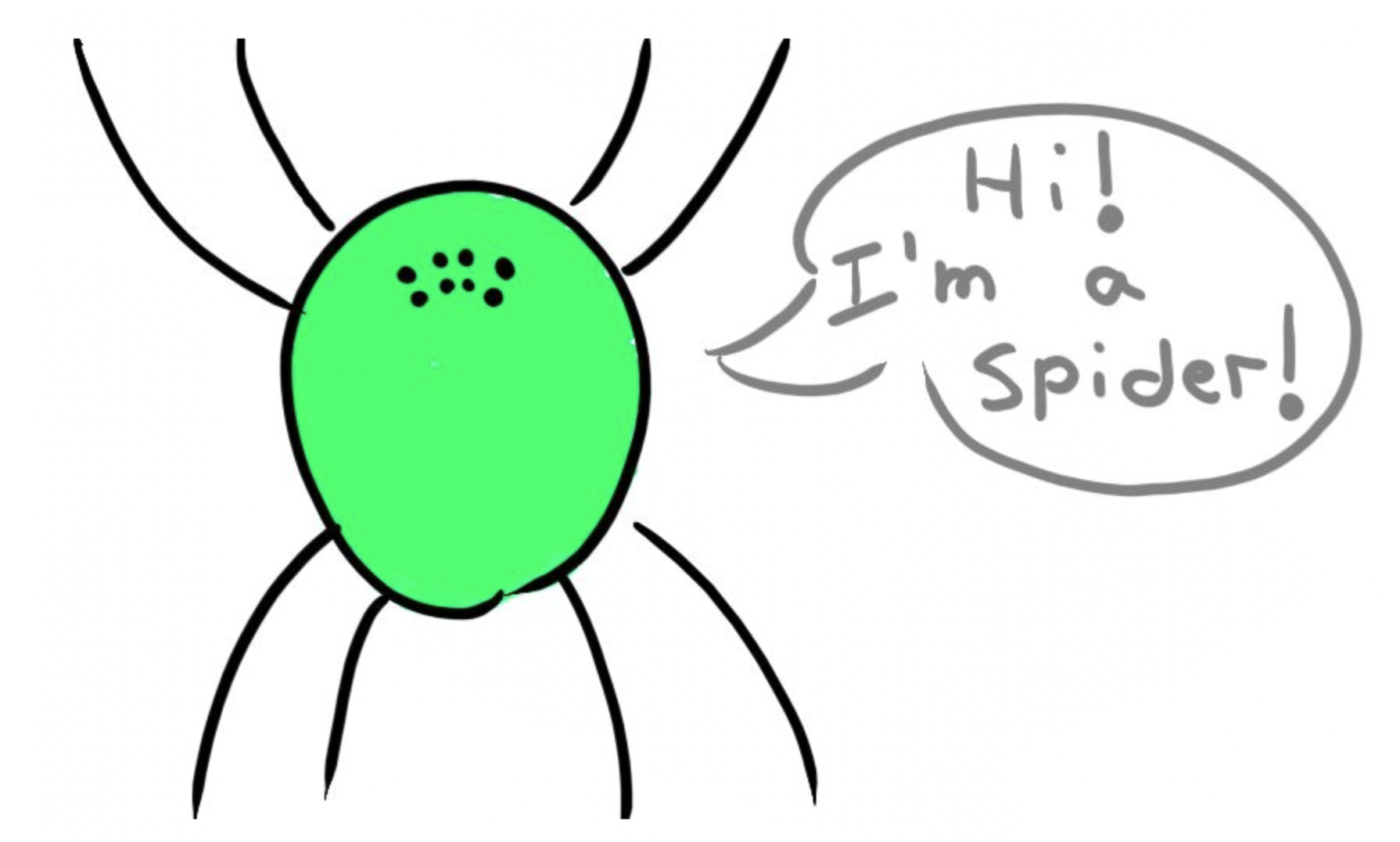}
\end{center}

A spider is a special kind of box that resembles a spider. It generalises a wire in the following sense. One way to think of a wire is something that connects two ends:

\begin{center}
    \includegraphics[width=0.3\linewidth]{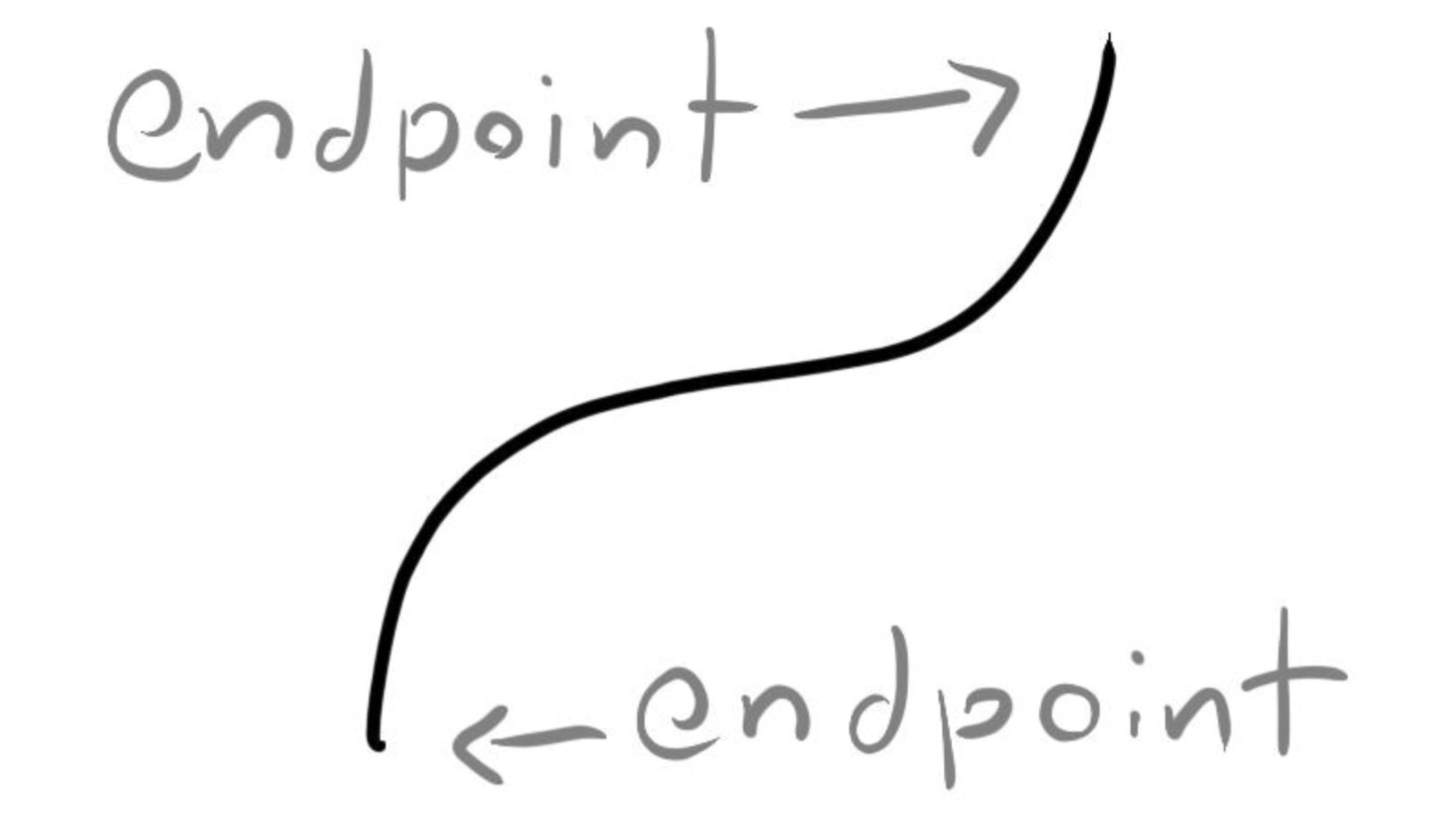}
\end{center}

When we connect two wires, taking one end of each, we again get a wire: 

\begin{center}
    \includegraphics[width=0.3\linewidth]{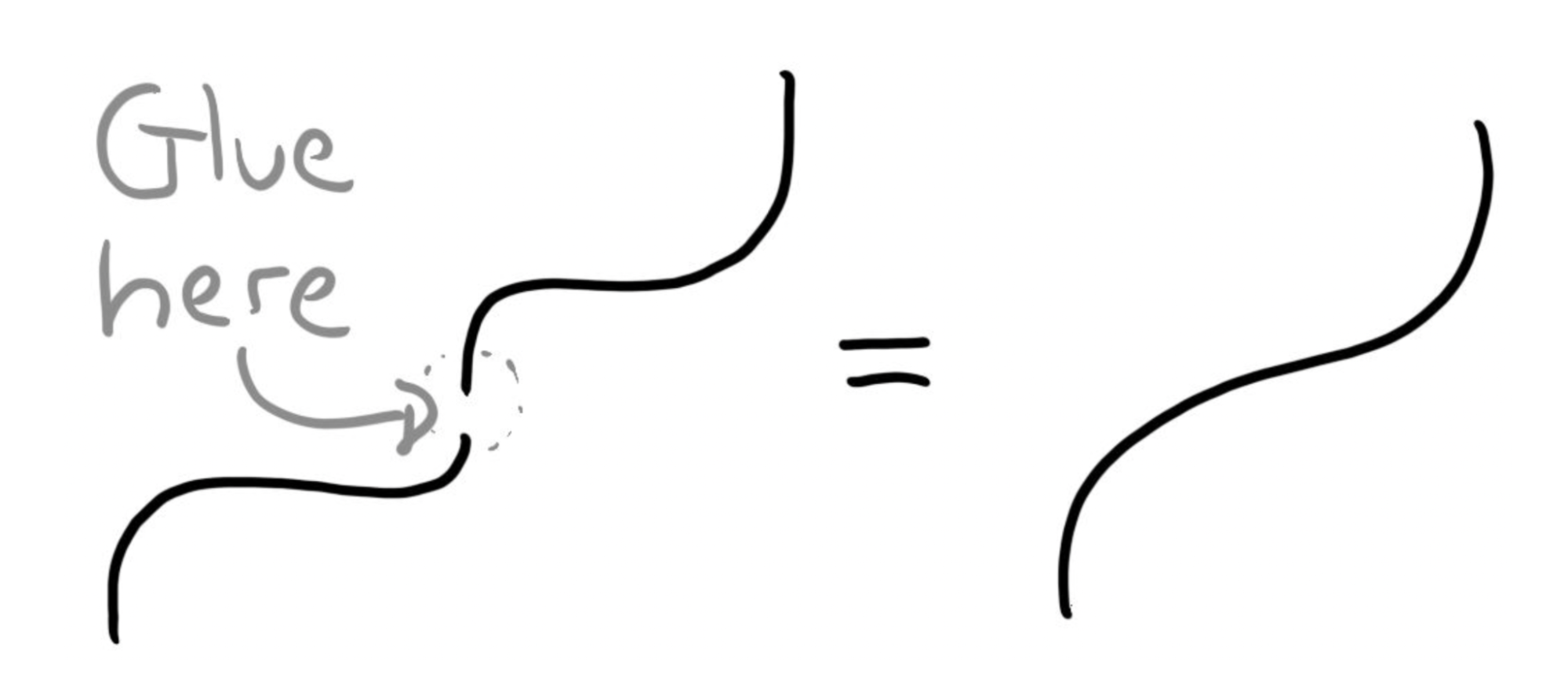}
\end{center}

The same is true for spiders, although we think of them as not having only two ends, but any number. When we connect spiders in the same manner as we described for wires, we again get a spider: 

\begin{center}
    \includegraphics[width=0.4\linewidth]{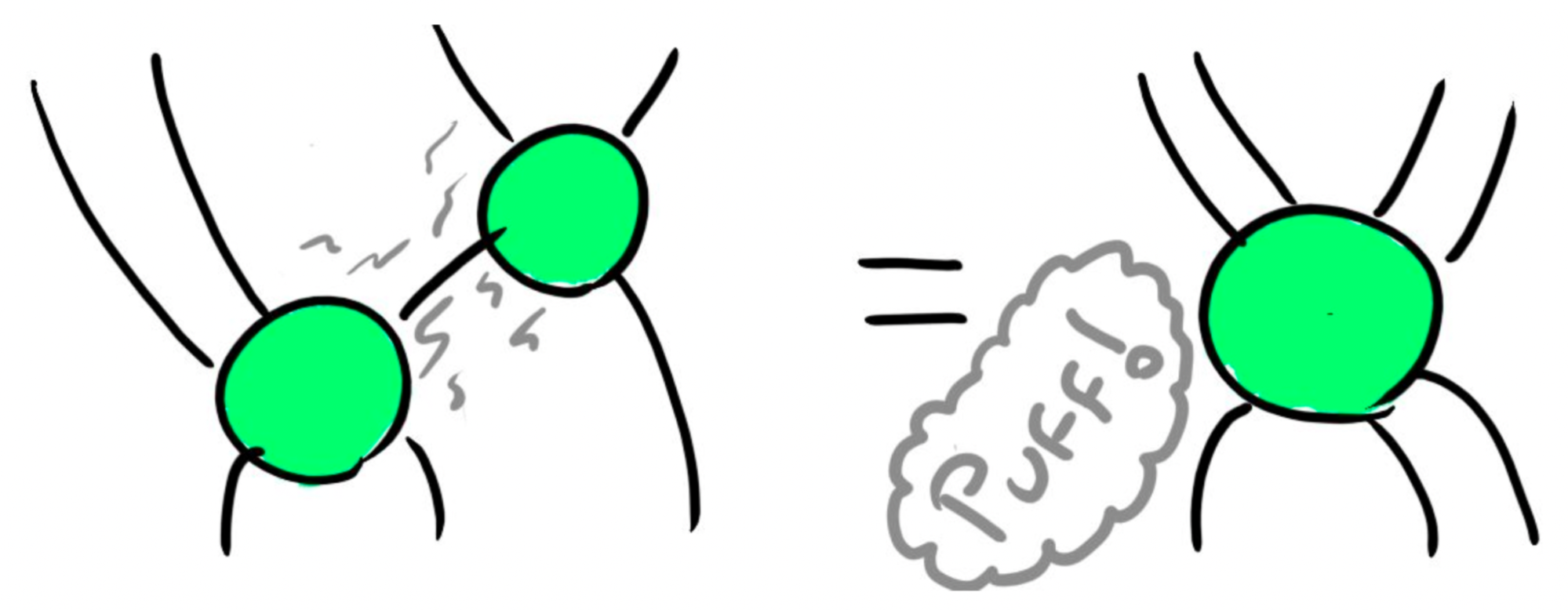}
\end{center}

This rule is called spider fusion. It is one of the ways spiders interact. For this interaction to happen, spiders (of the same colour) need to share at least one leg. The legs they share disappear as they fuse, and all the other legs are collected together. 

What is remarkable about spiders is that this simple rule captures what in the ordinary quantum formalism corresponds to an orthonormal basis, in a manner we won’t go into here \citep{coecke2013new}.

Phases are numerical markings carried by the spiders. When spiders fuse, phases add up as angles. 

\begin{center}
    \includegraphics[width=0.4\linewidth]{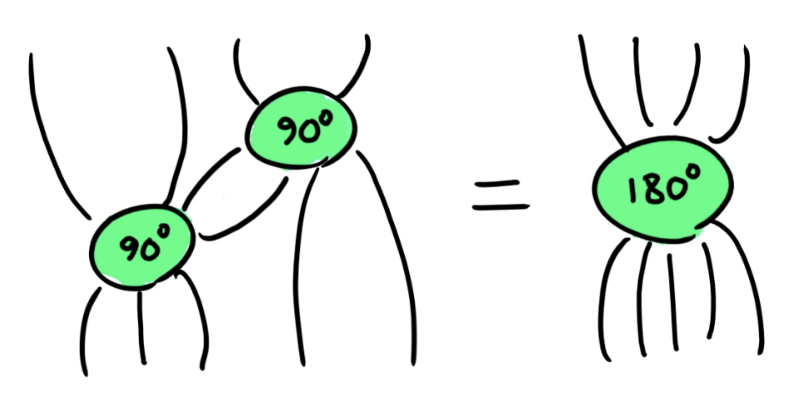}
\end{center}

In order to capture all of quantum theory's concepts, we need to consider two different kinds of spiders, which we indicate by colouring them differently. The rules we discuss below describe the interaction of these two kinds of spiders.

The two kinds of spiders we discuss below correspond to two mutually unbiased orthonormal bases, namely the Z (green) and red (X) bases. 

Spiders of different colours do not fuse together. Instead, the following happens:

\begin{center}
    \includegraphics[width=0.4\linewidth]{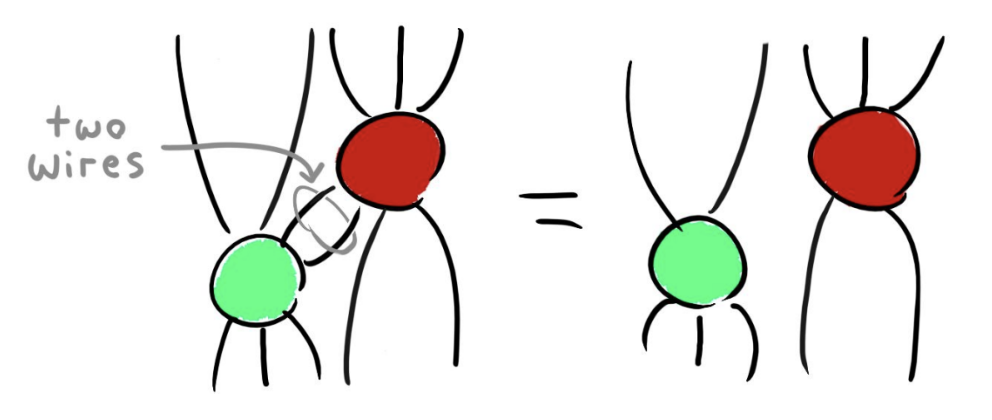}
\end{center}

That is, if they share two legs, they vanish. A funny way to think about that is if each spider hits the other one, that leg falls off.  

Leg-chopping happens when a red spider shares two legs with a green spider: shared legs just “fall off”. 

Spiders with single output legs, such as the following, are special kinds of states called qubits, or quantum bits. Qubits are the basic units of information in quantum computation.

\begin{center}
    \includegraphics[width=0.4\linewidth]{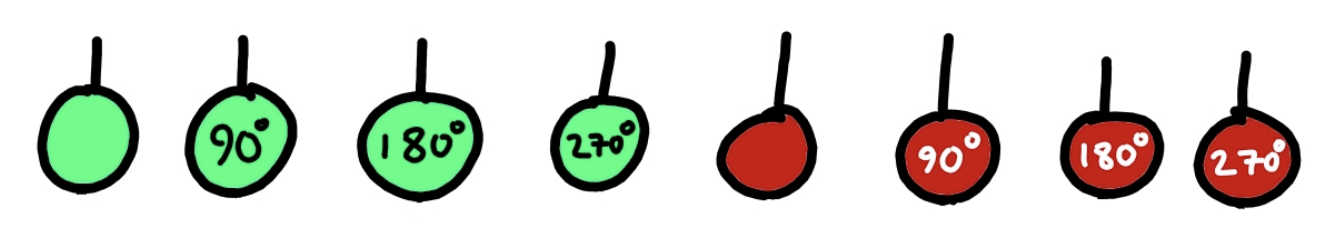}
\end{center}

As can be seen above, qubits can be of two different colours and have phases.

Spiders with phases and having single input and single output wires, such as the following, are called phase gates. They are part of the basic building blocks of quantum circuits and modify the phase of qubits.

\begin{center}
    \includegraphics[width=0.4\linewidth]{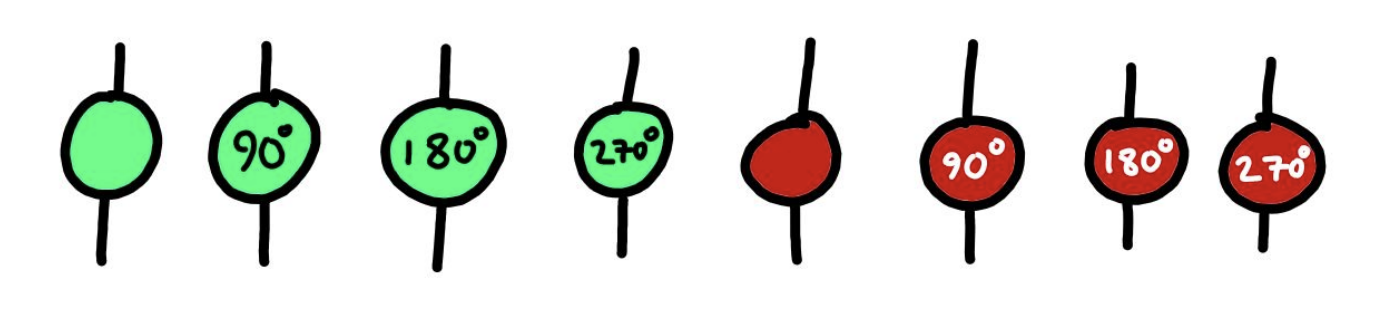}
\end{center}

In HS, a green phase gate with phase (in degrees) corresponds to the matrix

\[
\begin{pmatrix}
1 & 0 \\
0 & e^{i\theta\frac{\pi}{180}}
\end{pmatrix}
\]

Colour-change box is a particular kind of box

\begin{center}
    \includegraphics[width=0.05\linewidth]{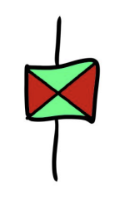}
\end{center}

that changes the colour of a spider when we stick it to all of its legs. 

\begin{center}
    \includegraphics[width=0.5\linewidth]{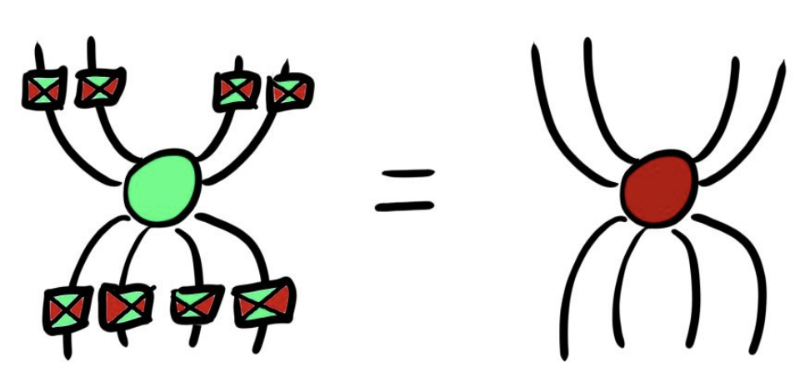}
\end{center}

Applying the colour-change box twice on a wire is equal to just the wire. 

\begin{center}
    \includegraphics[width=0.2\linewidth]{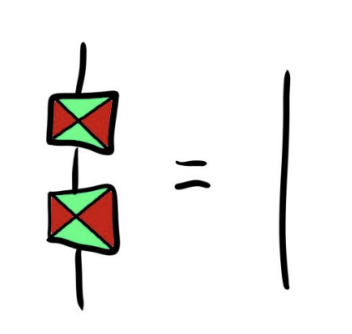}
\end{center}

This is intuitive. Changing the colour twice results in the same colour we started with.

In standard terminology of quantum theory and computation, the colour-change box is called the Hadamard gate. In HilbS, its matrix is given by

\[
H = \frac{1}{\sqrt{2}} \begin{pmatrix}
1 & 1 \\
1 & -1
\end{pmatrix}
\]

The Hadamard gate changes the orthonormal basis for Z into that of X, and vice versa.

Copying of the spiders: single-legged spiders of one colour, whenever their phase is either 0 degrees or 180 degrees, can be copied by spiders of the other colour. 

\begin{center}
    \includegraphics[width=0.5\linewidth]{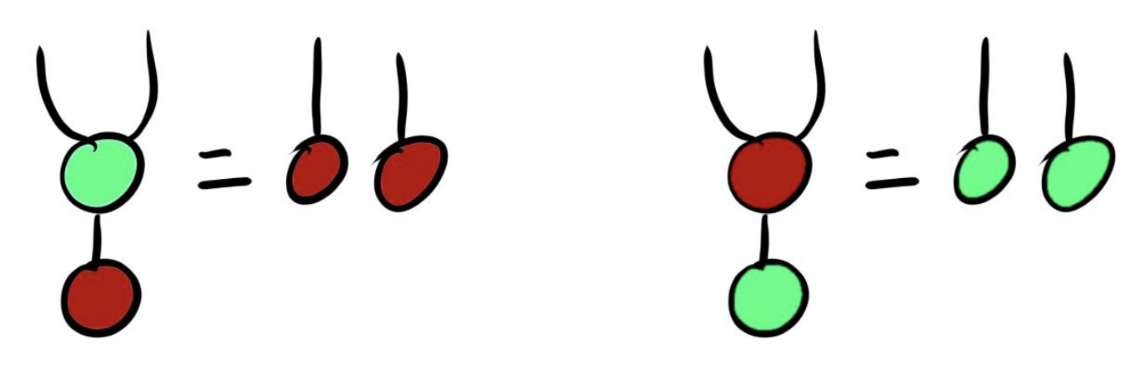}
\end{center}

Square-popping happens when four spiders are connected, forming a square, as in the picture below.

\begin{center}
    \includegraphics[width=0.5\linewidth]{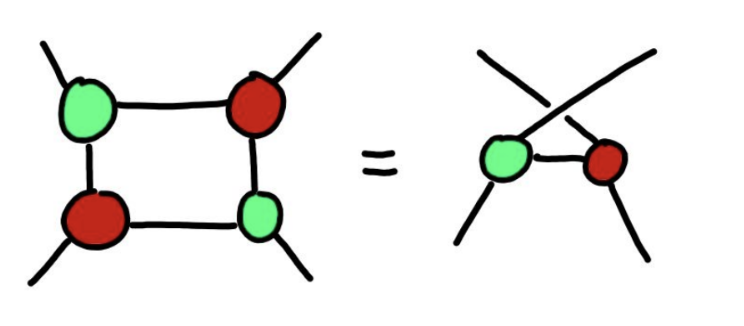}
\end{center}

We can now use these spiders for practical reasons in quantum computing, such as simplifying a quantum program. A quantum program is a circuit made up of certain quantum boxes, usually called gates. We have discussed phase gates and the Hadamard gate. Another essential gate in quantum computing is the controlled-NOT (usually abbreviated as CNOT) gate. The CNOT gate is obtained by composing a red and a green spider as follows. 

\begin{center}
    \includegraphics[width=0.15\linewidth]{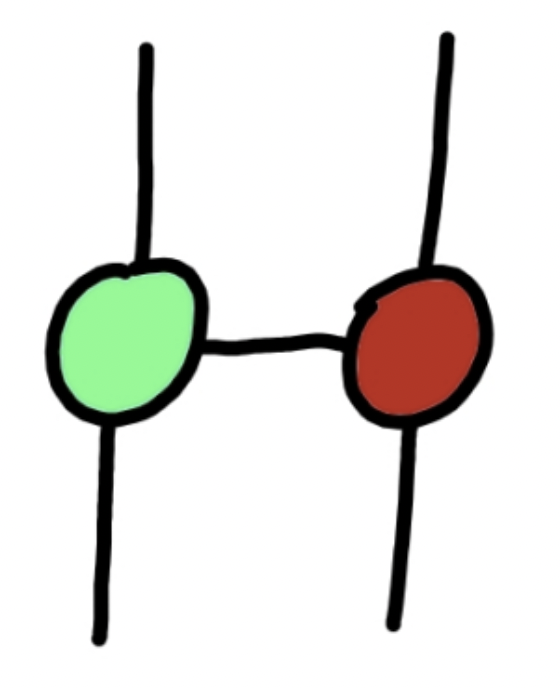}
\end{center}

The CNOT gate has two inputs and two outputs. If the first input has a value of 0, nothing happens to the value of the second input. If the first input has a value of 1, the value of the second input is flipped, i.e., 0 is changed into 1 and vice versa.

In HilbS, the CNOT gate is represented by the 4x4 matrix:

$$\begin{pmatrix}
1 & 0 & 0 & 0 \\ 
0 & 1 & 0 & 0 \\  
0 & 0 & 0 & 1 \\
0 & 0 & 1 & 0
\end{pmatrix}$$

Consider the following circuit that is built of three CNOT gates.

\begin{center}
    \includegraphics[width=0.5\linewidth]{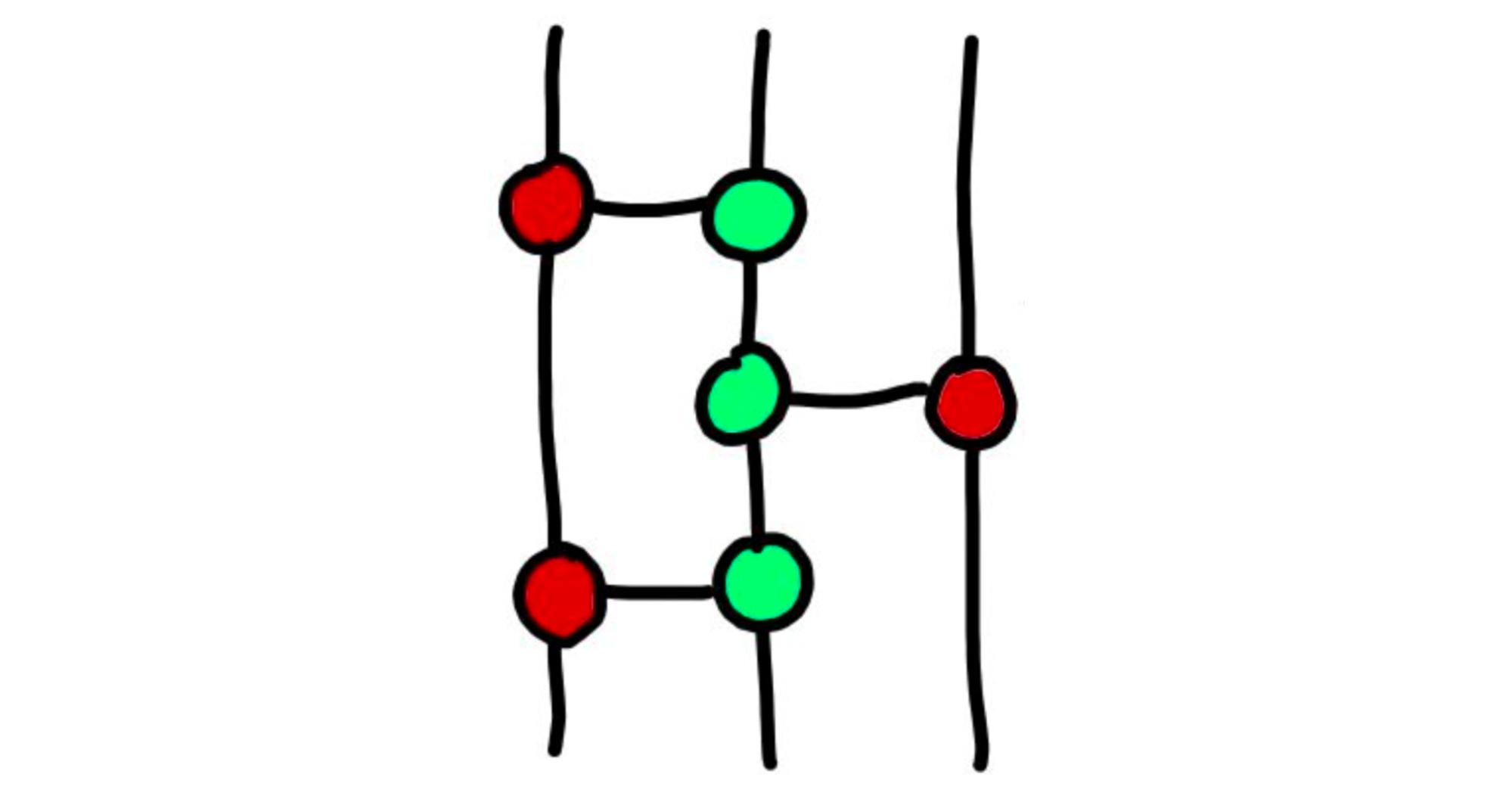}
\end{center}

The vertical wires are quantum particles or qubits. An important task is simplifying this circuit.  We can easily do that now using the graphical rules described above. First, we fuse all adjacent spiders of the same colour: 

\begin{center}
    \includegraphics[width=0.5\linewidth]{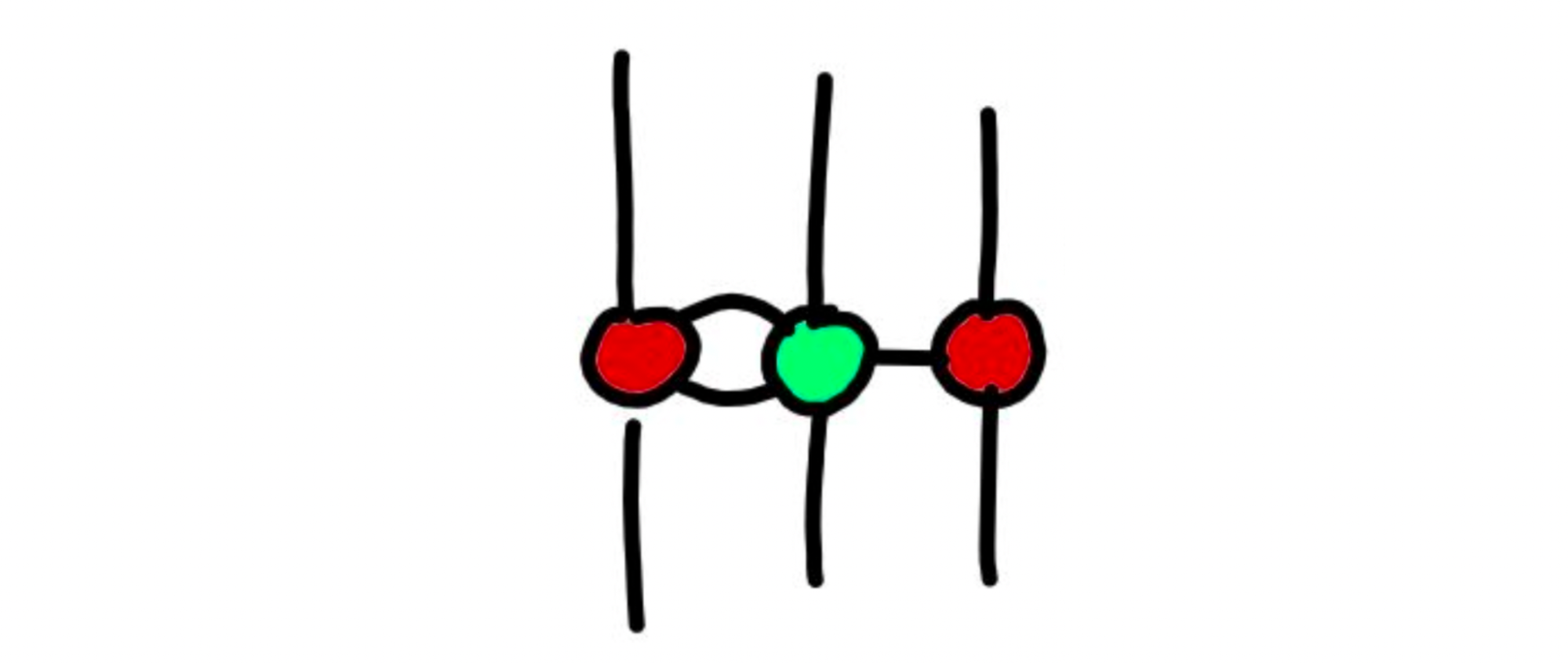}
\end{center}

Then we eliminate the two legs between spiders of different colours: 

\begin{center}
    \includegraphics[width=0.5\linewidth]{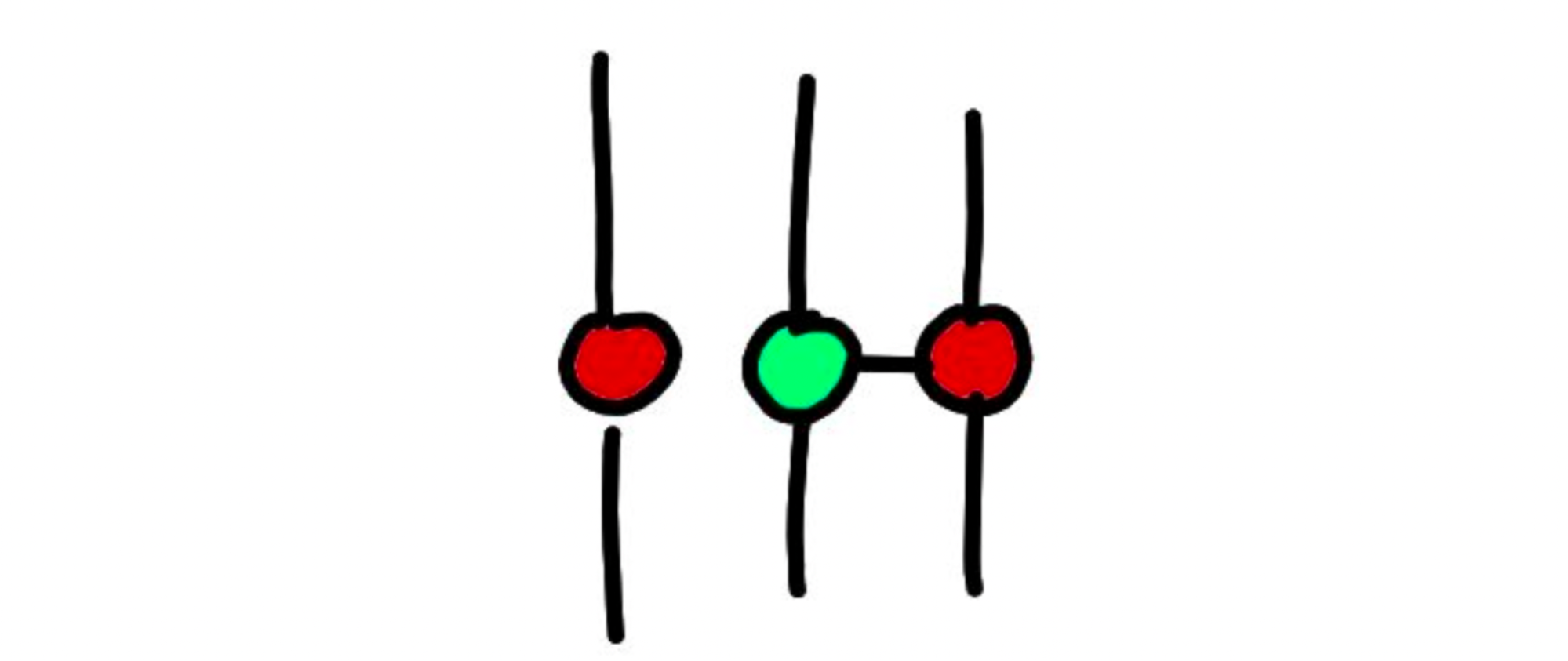}
\end{center}

What remains is a spider with two legs. This is a rotation of 0 degrees i.e. the identity person, which is a single wire: 

\begin{center}
    \includegraphics[width=0.5\linewidth]{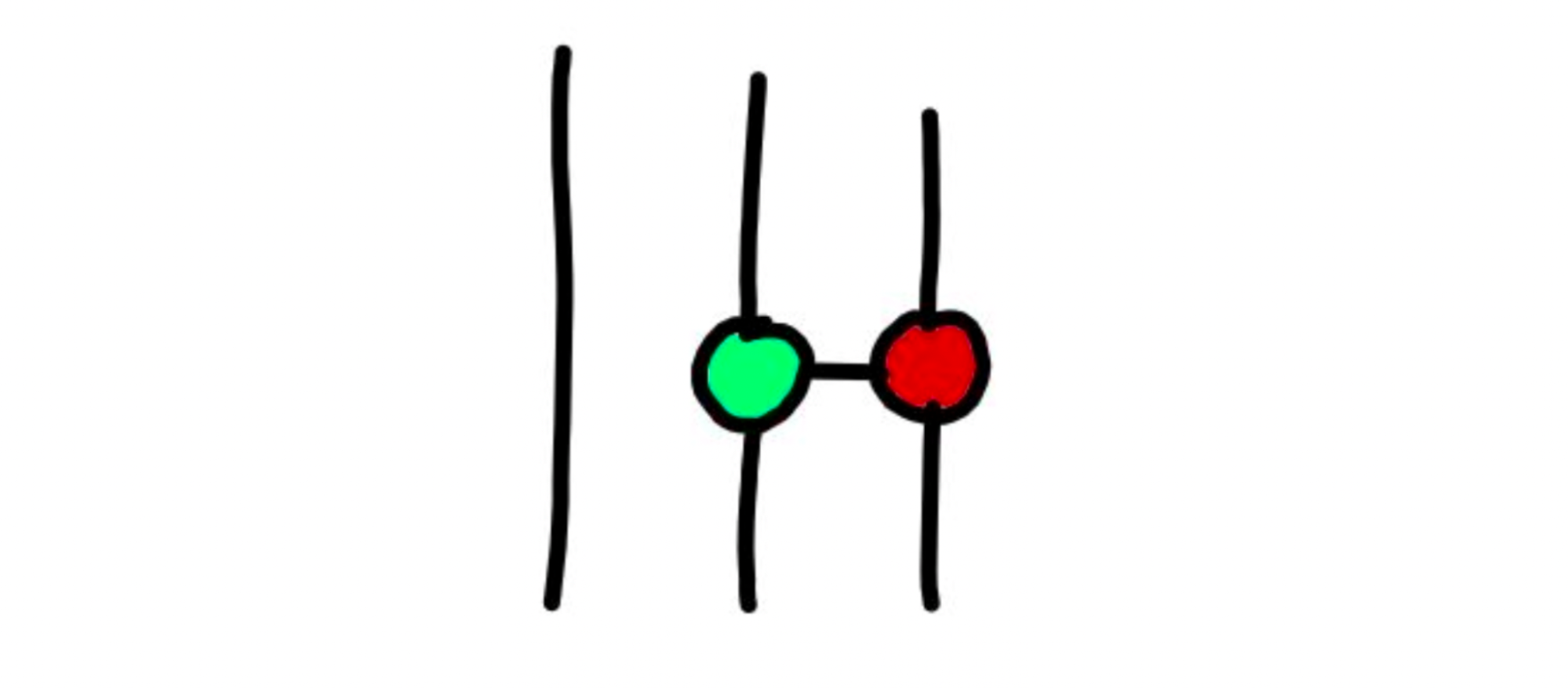}
\end{center}

So we reduced a circuit with three gates, to one with only one gate.  This is only a very simple example, but the same simple technique applies to very complex circuits and is for this purpose widely used for quantum computing \citep{gidney2019flexible, kissinger2019reducing, backens2021there, zhao2021analyzing, coecke2022kindergarden, wille2022basis, litinski2022active, khesin2023graphical}.

The reason for the success of this formalism is that it is, in fact, a valid substitute for the usual formalism, in that any equations that can be derived in the latter can always also be derived using only pictures \citep{hadzihasanovic2018two}.

\section{Methodology}

\subsection{Participants}
We distributed the experiment flyer to approximately 26 high schools across the UK and promoted it via social media platforms. Despite limited outreach due to time and resource constraints, it yielded 734 applications within three weeks. To maintain training quality, we conducted an initial screening to exclude non-high-schoolers, those outside the target age range, and non-UK applicants, as per ethical requirements.

We sent a follow-up email to the remaining applicants, clarifying that attendance at tutorials is compulsory and requesting notification from those unable to meet this requirement. From the pool of positive respondents, we randomly selected 75 candidates based on three criteria: (1) fluency in English, (2) the ability to commit to two hours of weekly study for ten weeks, and (3) gender balance. Anticipating potential dropouts, our target was to have around 50 participants to complete the course.

We then issued a consent form for ethical compliance and excluded any students who indicated they could no longer participate, as well as those who neither attended the first tutorial nor communicated their absence (as outlined in the handbook, which requires students to email if they are aware they will be or were absent). This was done in consideration of the many students on the wait list, ensuring that only committed participants remained in the study.

Typically, high school students are 18 years old when they complete their studies. However, in our study, eligibility was from 16- to 19-year-olds. This broader age range accounted for the possibility that some students may have delayed completing high school, for instance, due to taking a gap year, potentially as a result of the pandemic. Despite this, following the randomized selection process, no 19-year-olds were included in the study, and the oldest participants who completed the experiment were 18 years old. 
\subsection{Materials}
Specific materials were designed for this experiment, encompassing both teaching and evaluation components. 

Teaching materials included: 
\begin{itemize}
    \item The project handbook outlining rules, objectives, requirements, and procedures
\item Pre-recorded video lectures delivered by experienced lecturers
\item Five quizzes, and step-by-step visual demonstrations of corresponding solutions
\item An abridged version of  ‘Quantum in Pictures’ book by \cite{coecke2023quantum}, specifically suited to align with the course contents. 
\item A cheat sheet of quiz questions in preparation for the final exam.
\item Skribbl.io sessions in preparation for the final exam.

\end{itemize}

Evaluation materials consisted of: 
\begin{itemize}
    \item Pre- and post-training surveys to assess changes in knowledge and attitude.
    \item A final exam to evaluate overall learning outcomes.
\end{itemize}

\subsection{Procedure}

All participants needed to attend the weekly online courses specifically designed to teach corresponding content using the QPic method. The content was delivered via pre-recorded lectures, followed by online live tutorials for all groups. This was to help mitigate the teacher effect on training outcomes, as each participant had access to the same lectures during training.

Selected participants received the experiment handbook, which clearly outlines the study’s objectives, requirements, schedule, evaluation methods, data protection and privacy policies, etc. Additionally, they were given a reduced version of the QPic book written by \cite{coecke2023quantum} as their tutorial book, while a more comprehensive textbook by \cite{coecke2017picturing}, used at universities like Oxford, was mentioned as context. 

14 participants withdrew before the actual training, primarily due to exams or job commitments. This required the addition of 15 participants from the waiting list to join additional tutorials at the end of Week 1. First three weeks, we had 62 participants, among these 54 participants had successfully completed the program.

Participants accessed weekly recorded lectures via a secure Microsoft Teams link, and they were also expected to complete half-hour self-study sessions before tutorials to support their learning regardless of tutor variations. Self-study required participants to:
\begin{enumerate}
    \item Watch a pre-recorded lecture ahead of each tutorial. 
\item Answer a quiz on the contents of the preceding tutorials from Week 2 onwards. 
\end{enumerate}
Participants booked tutorials at their convenience, with practice sessions conducted by one of eight tutors, each handling a maximum of 15 participants per session. To minimize tutor variation, a carousel technique was used where tutors rotated weekly, each focusing on a specific week’s material, ensuring they interacted with all groups during the training. We held consultation sessions with our tutors in the 1st, 3rd, and 6th weeks to share experiences and adjust implementations as needed. 

During live tutorials, tutors made every effort to keep participants engaged. Participants were requested to join the sessions with pen and paper, and to draw diagrams by hand. Throughout each session and when possible, participants’ materials and drawings were collected and shared in an online folder, ensuring that everyone stayed actively involved and that their work was documented. Appendix C provides a comprehensive overview of the weekly learning content covered during each tutorial.

After the exam, the course concluded with two final optional events: a fun drawing game, Skribbl.io, and a presentation by Pfaendler about careers in quantum.

One difficulty we encountered is that participants demonstrated a general reluctance to activate their microphones and cameras during live tutorials. To ensure active and continued participation, discussions were encouraged through exercises; students had to solve these exercises during some allotted time. By the end of each tutorial, live solutions were illustrated by the tutor, allowing for potential questions and clarifications.

\subsection{Tutor training}

Prior to training, all tutors underwent a Disclosure and Barring Service (DBS) check in accordance with UK regulations for working with minors. They also completed a one-hour training session covering experiment objectives, student interaction, conflict management, monitoring and encouraging learner progress, handling diverse abilities, and use of Microsoft Teams.

Our tutors were a mix of highly experienced lecturers and doctoral students from the University of Oxford, and researchers from Quantinuum. During their tutorials, tutors were encouraged to create additional examples to reinforce the recorded video lectures and at least one interactive component in their lesson plan. They received continuous support from several experts: Coecke, Kissinger, and Gogioso focused on the types of examples to use, ensuring they were appropriate, up-to-date, and helped develop skills to prepare students for success in the field, should they choose to pursue it. Dundar-Coecke focused on the pedagogical aspects of their presentations and interactions with students. Lead tutors Yeh and Waseem offered technical support for each session, including addressing participants’ emails, Puca managed the booking system.  

\subsection{Assessment and evaluation}
To ensure a comprehensive understanding of the materials, a combination of surveys, quizzes, and a final exam were carefully designed to gauge participants' knowledge, engagement, attitude, and overall learning progress:

\paragraph{Surveys} 
Prior to the experiment, a brief survey collected data on participants’ age, school name, gender, ethnic background, and interest in pursuing a career in the STEM field. This data was anonymised by Kissinger to ensure confidentiality as per ethical requirements and was subsequently used for further analyses.  

The post-experiment ‘exit survey’ was anonymous, consisting of four sections with a total of 49 questions. The first section focused on demographics and socioeconomic status. The second section inquired about participants’ academic background, including their previous knowledge of QT and experiences related to QPic. The final section sought participants’ feedback about the course gathering their opinions, motivations, and suggestions.

\paragraph{Quizzes} 
Five quizzes, designed by Coecke, were created to enhance student engagement in problem-solving after tutorials. The quizzes were not intended to evaluate students' performance, but they aimed to engage students and provide additional examples beyond those covered in the tutorials. This decision was guided by feedback from students who requested additional examples. In response, each quiz was designed to support the tutorials that illustrated each solution step-by-step. 

\paragraph{Final exam} Designing exam questions for such an advanced course targeted at this particular age group, with very limited prior background, required careful consideration to thoroughly evaluate students’ understanding and analytical skills. The draft exam questionnaire mirrored one previously employed by Coecke and Kissinger in the quantum physics course for postgraduate students at the University of Oxford. Once the draft exam was created, it was subsequently reviewed by Dundar-Coecke and Pfaendler to evaluate its applicability, appropriateness, language, difficulty level, presentation style, variability, and representativeness. 

A cheat sheet was provided in preparation for the exam, mirroring a practice commonly adopted during the quantum physics course of reference at the University of Oxford.  Based on the feedback, final exam questions were meticulously structured to include at least three primary constructed-response items. These extended response items were designed to progressively increase in difficulty, starting from basic understanding and moving towards more complex application and synthesis of the material. 
Each extended response item, subdivided into at least three interrelated component prompts,  aimed to comprehensively assess the participants’ grasp of the course content, totaling 11 questions overall. This division was intentional. Participants were required to reflect and respond in various contexts, including writing their response based on prompts, showcasing their stance to solve the problems, and building on their own findings and problem-solving skills at each step. Another reason for structuring the extended response items with sub-categories was the difficulty of isolating concepts and learning content in such a multifaceted field and hence, the sub-questions were designed to build on each other, balancing the difficulty level for both low and high performers, as well as encouraging students to draw connections between different aspects of each week’s topic. 

Questions were crafted to evaluate a broad spectrum of skills, from diagrammatic manipulation calculations to conceptual understanding of quantum information scenarios represented by diagrams. By structuring the exam in this way, we aimed to achieve three primary objectives of the experiment: 
\begin{itemize}
    \item To test not only the students’ knowledge but also their ability to integrate and apply various computational rules and concepts commonly found in QT. This comprehensive approach ensures a thorough assessment of their mastery of the subject matter, emphasizing seamless application and synthesis of the material
\item To evaluate whether the content was learnable for this age group and to determine if the topics were within the Zone of Proximal Development for all levels of learners. This involved evaluating if students can understand and apply the material with appropriate support – limited in our case to online tutorials and video lectures, relying heavily on self-learning. 
\item Despite these disadvantages (c.f. limiting interactions to online tutorials and video lectures), the goal was to ensure that the exam questions were optimally designed to facilitate learning and growth for each student. 
\end{itemize}

The exam was administered as a take-home assignment, allowing students to complete and submit it by the deadline following the University of Oxford policies and guidance around assessment for postgraduate taught courses. The extended time frame enabled students to thoroughly engage with the material, reflect on their understanding, revisit learning resources as necessary, and apply their knowledge and problem solving skills comprehensively. 

All participants who submitted the exam were awarded a certificate of completion, regardless of their performance. Participants were informed that the exam would help us, as researchers, assess the effectiveness of our experimental effort and the comprehensibility of the content. Given that we received submissions from all participants, we are confident in stating that by adopting this learner-friendly method, we encouraged better learning and reduced the pressure from traditional timed exams, thereby fostering a more supportive educational environment. 

In preparing the extended response items, we incorporated questions mainly designed to evaluate procedural knowledge, with some conceptual understanding required. Questions aimed at assessing procedural knowledge were rather computational in nature, focusing on the application of mathematical or logical procedures that required participants to follow specific steps to arrive at a definitive answer. These steps corresponded to the application of specific rules of diagrammatic rewriting, such as spider fusion, leg chopping, and square popping, all of which were introduced during the course. These rules serve as direct analogues to symbolic calculations in traditional mathematical approaches to quantum. Since traditional mathematics was not introduced during the course, diagrammatic rewriting formally and rigorously replaced linear algebra calculations. In this method, diagrams served as the mathematical framework.  

Questions aimed at assessing higher order thinking were conceptual in nature, focusing on understanding and explaining underlying principles. They required participants to demonstrate comprehension of concepts through explanations, definitions or discussions of the implications of certain aspects of quantum theory, enabling them to build procedural knowledge and apply it effectively. 

To provide better insight into the exam structure, we have detailed each question’s difficulty level and the corresponding learning outcomes in Appendix D. The questions were categorized into three difficulty levels: easy, moderate, and challenging, based on our previous experience with postgraduate students. Each difficulty level was explained with one representative example to illustrate the type of questions and expected learning outcomes at that level. 

\subsection{Marking and assessment}
The submitted work underwent a rigorous marking process managed by two experienced raters, Coecke and Kissinger, who have been evaluating similar work for over a decade. Participants’ submission were first anonymised to remove any identifying information and then assessed independently (all submissions were double-marked) to ensure fairness and accuracy. This process involved evaluating each submission without knowledge of the other evaluators' decision, based on the criteria provided in the table in Appendix D. 

The marking criteria outlined in Appendix E focused on specific assessment areas, ensuring alignment with our learning objectives and expected learning outcomes. These criteria are employed alongside discipline-specific standards and mimicked the University of Oxford postgraduate exam marking criteria as a guideline for the overall standards expected at five grade bands as: Distinction (100-70), merit (60-69), pass (50-59), borderline fail (40-49), and fail (0-39).

Some noted advantages of the extended response item design are:
\begin{itemize}
    
\item We developed the scoring rubric following the University of Oxford marking guidelines. This helped ensure that the assessment criteria were clear, consistent, and aligned with the high standards expected of postgraduate coursework. 
\item Time allotment: Given that the majority of the participants submitted their essays in Week 1, we observed that adequate time was allocated for the completion of exam, ensuring that students had sufficient opportunity to demonstrate their understanding and skills. 
\item Collaborative and blinded marking: The marking process was conducted blindly by two primary raters, who have been marking postgraduate students’ exams for more than a decade in the Computer Science Department. In 10 instances where there was a 9\% or more deviation between the marks awarded by the two primary raters, a third rater, Stefano Gogioso, independently reviewed the submission and facilitated further discussion until a consensus was reached. The ratings from the third rater were then combined with the original ratings to create the final score, resolving the discrepancy. This method helped ensure impartiality, accuracy, and fairness of the grading and also aimed to provide participants with clear evaluation criteria and constructive feedback alongside their marks, and maintained consistency in the final grades.
\end{itemize}

Some disadvantages of the extended response item design are:
\begin{itemize}
\item One of the primary disadvantages of the method was that exams required hand-scoring, which was labor-intensive.
\item Additionally, students were not familiar with the specific format of the exams used across the entire experiment, which could have negatively impacted their performance. This potential drawback needs to be considered in future teaching and assessments. 
\item The assessment format might be particularly difficult for students who had weaker analytical or computational skills, and potentially be disadvantageous to these individuals. 
\item Although we did not get any feedback on this point, the necessity for participants to draw diagrams by hand with some due care and neatness could be considered another disadvantage.  
\end{itemize}

\subsection{Measured variables}
The outcome variables for the assessment were derived from the two questionnaires: the exam and pre-study and post-training surveys. The exam focused on the number of problems students successfully solved using QPic based on the criteria outlined in Appendix D.

The pre-training survey obtained participants’ attitudes before training and focused on three aspects: (1) participants’ commitment, (2) interest in pursuing a STEM career, and (3) at least one specific reason for their interest in participating. The post-training survey gathered feedback on the overall setup, demographics, satisfaction with the tutorials, any changes in their career plans, and their post-training/future expectations. Additionally, an optional question regarding their grades in STEM subjects was included to aid in analyzing correlations between exam scores and attitude towards STEM direction. 

After the completion of each week’s tutorial, we gathered weekly observational data in the form of feedback collected by tutors regarding participants’ attendances, active/passive participation, problem solving efforts, requests, and types of questions asked. This was to facilitate the assessment of each student’s progress throughout the training, enable meaningful comparisons, and take appropriate actions when any learning obstacles were identified. 

\subsection{Data exclusion rule}
Participants had been informed at the outset that they could withdraw from the study at any point. Attendance during the training period was monitored, but only data from participants who attended tutorials and completed post-testing were included in the analysis. As per the attendance rules agreed during the recruitment stage, participants had to register for one of five tutorial time options each week. To prevent a spillage effect, participant interactions were discouraged during and after the training until they had submitted their exam answers. To do so, we ensured that participants’ contact details were not shared among them. This measure was intended to uphold the integrity of their training and ensure the validity of the analyses. 

With these precautions in place, no unsupervised interaction or violation of research protocols between participants was observed that would require us to exclude any data from the analyses.

\section{Weekly learning content and procedure}\label{sec:Content}

\emph{Week 1, Quantum in Pictures (wires and boxes)}: In the first session, participants were introduced to diagrams composed of wires and boxes to describe processes and perform mathematical operations. Our tutor, Lia Yeh, briefly highlighted the key aspects of the Week 1 video lecture content and connected her exercises to those each, gradually increasing the difficulty level. Since this week also covered the concepts of space and time in diagrams and the physics they represent, it was crucial to provide a foundational understanding for the rest of the course, our tutor used several examples to clarify the content.

After introducing each key concept and at the end of the tutorial, she conducted a Q\&A session. She first requested participants to draw a simple diagram based on what they had practiced. She then asked participants to share their drawings via the chatbox (or show on screen) and continued the discussion based on each individual's drawings. This approach helped our tutor identify each student's level of understanding and pinpoint the gaps that needed to be addressed for a comprehensive grasp of the course content.

\emph{Week 2, the topic Quantum Teleportation}: In this session, participants were introduced to the concept of the ‘quantum lottery’ to describe quantum measurement. This tutorial built on the foundational knowledge from Week 1, showing diagrams of wires and boxes can be used for advanced mathematical reasoning essential for understanding quantum teleportation, which is a fundamental building block of quantum communication and quantum computing and serves as the first introduction to the concept of quantum entanglement.

Our tutor, Richie Yeung, began by briefly summarizing the key aspects of the corresponding video lecture, ensuring that participants had a clear understanding of the core principles before progressing to more complex applications. He then introduced further examples and illustrated how quantum  measurements can be represented diagrammatically. As quantum teleportation involves complex interactions, the tutorial emphasized the importance of precise diagrammatic representation. Richie demonstrated how wires and boxes can be arranged to depict the complex processes involved in teleportation, making abstract concepts more tangible and accessible. To solidify understanding, he conducted a Q\&A session after each major concept introduction and also at the end of the tutorial. He draw some options for the diagram in multiple choice and the students replied A/B/C/D in the chat box e.g., participants draw a specific quantum teleportation diagram based on the exercises that had practiced, allowing Richie to provide immediate feedback and clarification.

\emph{Week 3, A World of Spiders: Participants} were introduced to 'spiders’ -a special type of box with basic rules for working with them, starting with the ‘fusion’, ‘color change’, and ‘leg-chopping’ rules. These spiders are fundamental building blocks for quantum computing. The two spider colors, green and red, correspond to the Z and X bases, respectively. 

Our tutor Razin Shaikh began by reviewing the key aspects of the corresponding video lecture, as in previous sessions. He provided several examples to illustrate each rule and gradually increased the complexity of the exercises. Razin then conducted interactive Q\&A sessions, asking participants their diagrams, allowing his to provide feedback and address any misconceptions. 

Week 4 focused on using spiders for quantum computing, covering logical operations (such as copying and adding), quantum gates (such as rotations, CNOT gates, and Hadamard gates), and measurements. This tutorial also introduced diagrammatic substitution rules for circuit simplification and other relevant calculations. The tutorial demonstrated that every qubit quantum computation can be drawn using just spiders, simplifying complex operations. 

Our tutor Harny Wang started by summarizing the corresponding video lecture content and then introduced the above key concepts with several examples. During the Q\&A sessions, Harny asked participants to draw specific quantum gate diagrams and provided immediate feedback. 

\emph{Week 5, Quantum Teleportation with Spiders} explores quantum teleportation further and introduces ‘quantum test computing’ [measurement-based quantum computing (MBQC)]. MBQC is an alternative model to the circuit-based approach. MBQC uses ideas from quantum teleportation to shift computation from gates to measurements. 

Our tutor Boldizsár Poór provided an overview of the Week 5 video lecture content and demonstrated how spiders are used in MBQC with several examples. Boldizsár conducted interactive Q\&A sessions, requesting participants to draw MBQC diagrams and combine their previous learnings and share their work for feedback. An optional additional event, “Meet the Professors” was offered, giving participants exposure to diverse pathways to higher education.

\emph{Week 6} introduced notions of relativistic causality within the diagrammatic framework in the episode Keeping Einstein Happy. This episode covers concepts of ‘sure-boxes’ [deterministic processes] and ‘maybe-boxes’ [non-deterministic processes], providing a deeper understanding of causality in quantum systems. 

Our tutor Vincent Wang reviewed the corresponding video lecture content and demonstrated each concept with several examples, gradually increasing the complexity. Similar to the previous sessions, interactive Q\&A sessions enabled participants to draw diagrams and share their works for clarifications.

\textit{Week 7 Quantum vs. Ordinary Particles} introduced the concept of double wires to distinguish the quantum world from the classical world. Measurement and encoding processes were defined in terms of spiders. These processes convert quantum states into ordinary states and vice versa, respectively. Quantum uncertainty and doubled quantum circuits were also introduced.

Mathematically, doubling each wire \citep{selinger2007dagger} generalises from pure state quantum mechanics to mixed state quantum mechanics. This enables the full reasoning theory for mixed state quantum mechanics. Moreover, the complex amplitudes of a quantum process, after doubling, gives the real probabilities of the quantum process occurring. 

Following the corresponding video lecture content, our tutor Muhammad Hamza Waseem reviewed all the diagram rewrite rules the students had seen in the course. He particularly emphasised how double wires and spiders could be drawn as pairs of single wires and spiders. Applications of rewrite rules were demonstrated by solving two exercises in collaboration with the students. The first exercise was taken from the lecture notes and consisted of a circuit made of double and single spiders. The second exercise was novel and more complex. It comprised a circuit containing discarding processes, and double and single spiders with phases. There was a Q\&A session at the end of the tutorial.

\emph{Week 8, Everything Just in Pictures} focused on describing all quantum processes using pictures alone. It introduced the ‘square-popping’ rule [bialgebra between Z and X observables to mathematicians, and the three-CNOTS-make-a-swap relation to quantum programmers] and provided the necessary tools to fully understand quantum processes through diagrams.

Our tutor Cole Comfort summarized the Week 8 video lecture content and demonstrated each concept with several examples followed by the interactive Q\&A sessions. 

In \emph{Week 9}, participants were given a week off before the exam. During this week, students who missed Week 7 or 8 tutorials or wanted additional review could attend makeup tutorials. This session provided an opportunity for participants to revisit challenging concepts and ensure they were well-prepared for the upcoming evaluations. 

The last tutorial, \emph{Week 10}, covered the course content from last week and included a question and answer session to address any remaining questions. Participants received a take home exam to submit after three weeks. They were instructed to work on the exam questions, to apply and demonstrate their understanding of the course material comprehensively.

\section{The characteristics of exam questions and the corresponding exam questions}\label{sec:Exam}

Every question of this high school course exam was from a past exam in a graduate course at University of Oxford---except for Question 2c which was a ``new'' question that modified Question 2a by changing the tripartite entangled state to the Hadamard basis. Table~\ref{table:questions} lists out for each question on this high school course exam, the question on a past University of Oxford postgraduate exam which it reproduces.
The jargon of the high school exam was modified to match the terminology used in the course as to be less intimidating and improve memory retention for the students, but the concepts are unchanged.
Parts about computing probabilities were removed from the high school exam, despite the fact that probabilities can be reasoned about in the QPic formalism. This was in consideration that the students are not explicitly taught probability theory quantitatively at the high school level, to not introduce an additional unknown variable in the study.

\begin{table}[h]
\captionsetup{justification=centering}
\caption{}
\centering
\label{table:questions}
\begin{tabular}{ll}
\toprule
\emph{This high school exam}
& \emph{Postgraduate exam at the University of Oxford}\\
\midrule
Question 1a & Question 2a, MT2018 MSc Quantum Computer Science\\
Question 1b & Question 2b, MT2018 MSc Quantum Computer Science\\
Question 1c & Question 2c, MT2018 MSc Quantum Computer Science\\
\multirow{2}{9em}{Questions 2a \& 2b} & Question 2b, MT2021 MSc Quantum Processes and Computation \&\\
 & Question 4, MT2018 MSc Quantum Computer Science\\
Question 3a & Question 2a, MT2020 MSc Quantum Processes and Computation\\
Question 3b & Question 2, MT2019 MSc Quantum Processes and Computation\\
\bottomrule
\end{tabular}
\end{table}

Direct comparison of high school and graduate student performance was not one of the three focuses of this study; to properly do so would require a larger future study controlling variables across multiple groups.
A notable difference is that the University of Oxford graduate course constitutes significantly more lecture hours---three times more than in this high school course. The usual graduate courseload is three courses per term, whereas this time commitment would be unreasonable to expect from high school students pursuing this extracurricularly. Moreover, the graduate students' grade on the exam constitutes $100\%$ of their course grade, hence a significant proportion of the overall academic performance for Master's degrees. The high school students' priorities included non-extracurricular academics and employment, particularly the final weeks of preparation for exams crucial to their college admission prospects. In consideration of these factors, the fact that 1 in 2 participating high school students achieved a distinction, a mark which Master's students at the University of Oxford would be delighted by, is remarkable.

The exam questions themselves cannot be made public yet, as doing so would preclude future studies from being comparable to this one.
Instead, this section presents learning outcome(s) of the exam questions, which quantum concepts they tested, and some discussion question by question.

\subsection*{Question 1 (required procedural knowledge)}

\emph{Learning outcome}: 
\begin{itemize}
    \item The ability to demonstrate adept computational skills through the accurate application of diagrammatic rewriting techniques.
\end{itemize}

The first question focused on reasoning with quantum circuits. This tested successful processing of quantum computations with a certain degree of complexity, which is a necessary prerequisite to understanding quantum algorithms and quantum information protocols. Specifically: 

\paragraph{Question 1a}
Question 1a required students to apply a four-qubit circuit to a quantum state. In QPict, this task can be solved relatively straightforwardly using fusion and copy rules, covered in Week 3 of the course. In contrast, in HilbS, the circuit is simplified by multiplying three 16x16 matrices one by one with the state which is a 16x1 vector. Furthermore, in HilbS, one additional confounding factor is that it is not just matrix multiplication: at bare minimum, how to calculate tensor product of matrices is also needed to work with even two qubit circuits. Note that, as the number of qubits increases, the complexity for QPict remains manageable, whereas for HilbS using matrices, the complexity grows exponentially. This difference highlights the efficiency of the QPict approach in handling circuit simplification, compared to the more cumbersome approach required in HilbS. 

\paragraph{Question 1b}
Question 1b required students to demonstrate that a three-qubit circuit reduces to the identity, i.e. three plain wires. In QPict, this simplification necessitates additional rules, such as the leg-chopping and colour-change, covered in Week 3 of the course. On the other hand, in HilbS, students just multiply 8x8 matrices to verify the identity, a process that becomes increasingly tedious. Moreover, in HilbS, the 8x8 matrices for all three parts of Question 1 are matrices of complex numbers. This requires the students to additionally need to know how to add and multiply complex number matrices, which is not needed in QPict to do the same reasoning. As in the previous question, the complexity of the question remains manageable for QPict but grows exponentially for HilbS. 

\paragraph{Question 1c}
Question 1c asked what a three-qubit circuit simplifies to. In QPict, solving this requires the application of one more rule, specifically the square-popping/bialgebra rule, covered in Week 8 of the course. In contrast, in HilbS, this question would be virtually possible to solve due to the limitations of matrix representations, which do not facilitate straightforward rewritings. \\

Within QPict, the differential complexity of Question 1a, 1b, and 1c is notable, with 1a presenting a lower difficulty level than 1b because of the latter requiring application of two extra diagram rewrite rules. 1c possesses a greater challenge than 1b due to its requisite application of an additional simplification rule, notably the square popping rule, introduced in the last week of the course. This increases the computational difficulty of 1c. 

Regarding the specific tasks of Question 1b and 1c, students are required to reduce the circuit to three plain wires, but only 1b hints at the result. The absence of guidance towards the final solution in Question 1c contributes to its perceived difficulty relative to 1b.

Overall, Questions 1a, 1b, and 1c are interconnected as students could apply similar simplification rules, fostering a cumulative learning process. The successful resolution of 1c serves as an indicative milestone, signifying students' higher level grasps/ understanding of the course material. 

\subsection*{Question 2 (required both higher order thinking and procedural knowledge)}

\emph{Learning outcome}
\begin{itemize}
    \item Understanding and solving questions about  
teleportation protocols, 
multi-partite entanglement, 
quantum communication protocols 
\end{itemize}

The second question was the most conceptual on the exam. The setup of the problem is: One person performs a quantum measurement and classically communicates the measurement outcome to another person, who then decides which quantum operation to perform based on that outcome. This is a quantum communication protocol that builds upon the students' understanding of quantum teleportation, entanglement, measurement, and classical feed-forward of measurement outcomes. It challenges the student to come up with protocols to send quantum information deterministically, in spite of the non-deterministic quantum measurement outcomes.

Question 2 primarily concerns quantum communication and corresponding content covered in Weeks 2 and 5. While the applied rule is mainly spider fusion, the real focus is on the conceptual understanding, as the order in which rules are applied is crucial. Fundamentally, this question addresses the nature of quantum observables. Participants need to comprehend entanglement thoroughly before they can start thinking about devising a solution.

Note that Week 6 was not directly tested in the exam, but it introduced the key concepts of quantum processes, which are closely related to the material covered on Question 2. While we did not test specific definitions or ask participants to rewrite rules, we did assess the kind of abstract process-theoric reasoning that was the focus of Week 6, which means that although the content was not directly examined, the skills and understanding developed during this lecture were essential for answering the second question.

Question 2 requires participants to understand and implement a conceptual, multistep process involving communication protocols. The complexity here lies in the higher number of possible combinations and the necessity of performing more than one round of measurement and correction. It also evaluates whether participants have engaged with the course material, as a similar but simpler question was introduced in one of the quizzes provided.

\paragraph{Question 2a}
Question 2a presents a three-qubit entangled GHZ state shared between three parties and asks the student to design a communication protocol that results in a two-qubit entangled state shared between two of the parties. In HilbS, the state in Q2a would be represented by an 8x8 matrix. The ‘maybe-tests’ [non-deterministic quantum effects] would correspond to 2x2 observables and the corrections would be 2x2 Pauli operators. In HilbS, both a schematic diagram of the protocol, and a computation of what is being done, are separately needed. In QPict, these separate pieces are unified, by using diagrams to keep track of both the protocol and what it computes to. Therefore, in HilbS any attempt at coming up with a protocol needs a separate calculation keeping track of the steps in order to determine whether it is correct, whereas in QPict the correctness of the protocol can be deduced from the diagram for the protocol itself. 
\paragraph{Question 2b}
Question 2b adds another level of difficulty to Q2a by introducing a four-qubit entangled state and four parties. The task now is to design a protocol in which the final state is a two-qubit entangled state shared between two of the parties. The increased complexity of this question is due to two rounds of measurement and error correction.  This questions tests participants’ understanding of multipartite entanglement and the multiple steps involved in quantum communication protocols. 

In HilbS, the state in Q2b would be a 16x16 matrix. The maybe-tests would correspond to products of 2x2 observables and the corrections would be compositions of 2x2 Pauli operators. While in Hilb, such problems are solved using a combination of both algebra and schematic diagrams of the protocol. In QPict, everything is done diagrammatically.

\paragraph{Question 2c}
Question 2c is a variant of Q2a but the state is on a different basis. It examines a different concept: complementary observables. In this scenario, the shared entangled state is diagonal in a different basis than the measurements and correction gates, resulting in the measurements destroying all entanglement. This question highlights the importance of understanding the interplay between measurement bases and entanglement. In HilbS, the state in Q2c would be represented by an 8x8 matrix. The maybe-tests would correspond to 2x2 observables and the corrections would be 2x2 Pauli operators. In QPict, application of diagram rewrite rules show that the task of creating a two-qubit entangled state between the two parties is not possible. \\

Overall, Questions 2a, 2b, and 2c are interconnected, as they all require a deep understanding of teleportation protocols and the application of related quantum principles. Successfully navigating these questions indicated a robust conceptual grasp of the course material. Regarding the difficulty level, both 2a and 2b require students to formulate conceptual multi-step processes. The number of possible combinations is higher in 2b than in 2a, requiring more than one round of measurement and error correction. 

\subsection*{Question 3 (required procedural knowledge)}

This question comprises two subquestions, 3a and 3b. However, subquestion 3a is multifaceted, requiring participants to navigate and process through four distinct layers to arrive at a solution, 3a1, 3a2, 3a3, 3a4. 

\emph{Learning outcomes}
\begin{itemize}
\item Conceptualize and reason with a mathematical representation of entangled states in mixed state quantum mechanics.
\item Measure qubits in mixed-state quantum mechanics, recognizing that same basis measurements on a Bell state results in a pure maximally entangled state, whereas complementary basis measurements on a Bell state collapses to a product (i.e. no entanglement) state.
\item Discard (aka tracing out aka apply the maximally mixed effect) a subsystem of a mixed quantum state, recovering the state after discarding.
\item Reason with more complex entangled mixed states versus product states.
\end{itemize}

Overall, Question 3 corresponds to Chapter 5 in the course book (Week 8) and references Section 4.1 for the use of ordinary cup states (Week 7). 
\paragraph{Questions 3a1 and 3a2}
Question 3a1 and 3a2 are canary questions, designed to be the simplest on the exam and evaluated together. They are very similar questions, requiring the application of one simplification rule each, namely spider fusion and leg chopping. The purpose of canary questions is to test basic understanding of two frequently used simplification rules, therefore serving as a litmus test for general understanding of the course material.      

They also require participants to recall and repeat two equations introduced in the second half of the course (Week 7: Quantum vs ordinary). Since attendance for the pre-recorded lectures was not monitored, these questions also served as a benchmark for assessing whether participants watched all the tutorials. Additionally, these questions were crafted to prevent students from simply searching for answers online, thus evaluating their retention of pre-recorded lecture material. 3a and 3b also covertly assess understanding of measuring the Bell state in identical or different bases, which is a fundamental concept in QT. 

In HilbS, Q3a would require checking whether there would be classical correlations between two qubits. In Q3a1, there is a Bell state measured in the same basis. In Q3a2, it is a Bell state measured in complementary bases. Q3b3 involves a GHZ state with one qubit discarded and the rest measured in the same basis as the GHZ state. Q3b4 involves a GHZ with one qubit discarded and the rest measured in a basis complementary to the previous one.
\paragraph{Questions 3a3 and 3a4}
Questions  3a3 and 3a4 - just like canary questions -  require performing calculations in order to verify if measurement of two qubits leads to an ordinary cup-state. As opposed to canary though, the presence of discard boxes and the necessity to undouble/double notation - cfr. Week 6 and 7 of the course  - add a level of complexity in the calculation procedure.

\paragraph{Question 3b}
Question 3b is a sentinel question, representing the most challenging question on the exam. A correct answer to this indicates an excellent understanding, modular reasoning abilities, and creative problem solving. One of the exam markers, Bob Coecke, contends that only participants who are exceptionally well-prepared during the experiment period can clove the sentinel question, which primarily requires familiarity with (un)doubling notation and application of the square-popping rule taught in Week 8, the final week of the course. 

Question 3b focuses on four entangled qubits in mixed-state quantum mechanics. In HilbS, representing these qubits as a matrix would result in a 16x16 matrix, which equates i.e. 256 complex numbers. As in Question 3a, the task is to calculate measurement outcomes,  testing the correct use of quantum vs ordinary notation. 

Question 3b's higher level of complexity can be attributed to a variety of concurring factors, including: (1)  higher number of qubits involved (2) application of the square-popping rule (3) harder drawing skills and clarity required to picture the solution.

\section{Marking criteria}\label{sec:Marking}

The marking criteria outlined in Table \ref{tab:marking_criteria} below are meticulously designed to target key assessment areas, aligning closely with the course's learning objectives and expected outcomes. These criteria provide a clear and structured framework for evaluating student performance, ensuring that assessments accurately reflect the core competencies and skills intended for development in conjunction with discipline-specific criteria, and should be viewed as guidance on the overall standards
expected at different grade bands, aligning with the taught postgraduate generic marking criteria used at the University of Oxford, a 0-100\% grading structure in line with the current university regulations.

\renewcommand{\arraystretch}{2.5}
\makeatletter
\renewcommand{\fnum@figure}{Table \thefigure}
\makeatother

\begin{figure*}[!t]
    \centering
\tabulinesep=1mm
\resizebox{8cm}{!} {
\begin{tabu}{|X[7.5, c]|X[7.5c]|X[7.5c]|X[10.5c]|}
\hline
\multicolumn4{|c|}{\textbf{Distinction $>$70}} \\
\hline
\textit{Understanding} & \textit{Use of knowledge} & \textit{Structure} & \textit{Grade bands} \\
\hline
Advanced, in-depth, authoritative, full understanding of key ideas. Originality of the solutions, legitimacy of chain of reasoning in the answers provided.
&
Complex work and key problems solved. Correct application of concepts and techniques (e.g. the appropriate use of diagrammatic rewriting rules), the ability to use proper terminology (e.g. ``square-popping'', ``leg-chopping'')
&
Coherent and compelling work.
Logical and concise presentation. The solution drawn/written in a clear and unambiguous way (e.g. the proper use of notations, the difference between ``quantum'' and ``ordinary'' diagrams).
&
\textbf{(90-100)} insightful work displaying in-depth knowledge. Outstanding work, independent thought, highest standards of problem solving

\medskip

\textbf{(80-89)} insightful work displaying in-depth knowledge. Good quality of work, independent thought 

\medskip

\textbf{(70-79)} thoughtful work displaying in-depth knowledge, good standards of problem solving 
\\
\hline
\multicolumn4{|c|}{\textbf{Merit 60-69}} \\
\hline
\textit{Understanding} & \textit{Use of knowledge} & \textit{Structure} & \textit{Grade bands} \\
\hline
In-depth understanding of key ideas with evidence of some originality
&
Key problems solved. Correct application of most concepts, techniques, and correct use of terminology
&
Coherent work, logically presented. Clear solutions
&
\textbf{(65-69)} thoughtful work displaying good knowledge and accuracy. Evidence for the ability to solve problems

\medskip

\textbf{(60-64)} work displays good knowledge, some evidence for problem solving
\\
\hline
\multicolumn4{|c|}{\textbf{Pass 50-59}} \\
\hline
\textit{Understanding} & \textit{Use of knowledge} & \textit{Structure} & \textit{Grade bands} \\
\hline
Understanding of some key ideas with evidence of ability to reflect critically 
&
Some key problems solved. Correct application of some concepts, techniques, and terminology
&
Competent work in places but lacks coherence 
&
\textbf{(55-59)} work displays some understanding in most areas, but standard of work is variable

\medskip

\textbf{(50-54)} work displays knowledge and understanding of some areas, but some key problems are not solved
\\
\hline
\multicolumn4{|c|}{\textbf{Fail 40-49}} \\
\hline
\textit{Understanding} & \textit{Use of knowledge} & \textit{Structure} & \textit{Grade bands} \\
\hline
Superficial understanding of some key ideas, lack of focus 
&
Key problems are not solved/understood, gaps in application of concepts, techniques, and terminology
&
Weaknesses in structure and/or coherence 
&
\textbf{(40-49)} work displays patchy knowledge and understanding, most key problems are not solved
\\
\hline
\multicolumn4{|c|}{\textbf{Fail 0-39}} \\
\hline
\textit{Understanding} & \textit{Use of knowledge} & \textit{Structure} & \textit{Grade bands} \\
\hline
Lack of understanding
&
Key problems misunderstood/unanswered, limited/incorrect application of concepts, techniques and terminology 
&
Work is confused and incoherent 
&
\textbf{(33-39)} incomplete answers with some superficial knowledge

\medskip

\textbf{(20-32)} some attempt to write something relevant but many flaws

\medskip

\textbf{(0-19)} serious errors, irrelevant answers
\\
\hline
\end{tabu}}
    \caption{Marking Criteria}
    \label{tab:marking_criteria}
\end{figure*}

\section{Strategies for student retention}\label{sec:Strategies}
A number of steps were taken to enhance the student experience and create an environment conducive to students to be engaged in learning. In this section, we describe measures implemented to this end in the Week 1 tutorial, across the course to enable students to keep up the material, and enriching activities for professional development and confidence building.

\subsection*{Drawing connections to everyday examples in Week 1}
The first week of tutorials was carefully thought out, as it constituted the students’ first interaction with the course in real time. It was a requirement for students to attend a Week 1 tutorial (or one of its makeup tutorials) to participate in the course, in order to count the total number of admitted students who were unresponsive, in consideration of the wait list from which that number of students (15) could then be swiftly admitted. The focus being on processes and their visual representation as diagrams, the Week 1 tutorial involved little calculation and did not directly teach quantum concepts, but rather introduced conceptual grounding for the mathematics in the ensuing weeks to follow. Following a review of the lecture material on boxes and wires and a variety of examples, the tutor asked each student to draw ``any diagram made of boxes and wires''. The question was intentionally open-ended to enable students to draw from their creativity, and the diagrams shared by each student demonstrated that they understood how to combine boxes and wires to express more abstract concepts. It was of highest importance to ensure each student in attendance had the opportunity to present their diagram and explain its meaning, as this set the tone for students to engage actively in future tutorials. Additionally, this enabled the students to try several available options to share their diagram during the tutorial in real time, such as sharing video holding up a paper drawing, sharing screen drawing onscreen, or uploading a photo along with typing their explanation to the Microsoft Teams chat. In spite of some technical difficulties, each student managed to get at least one of these options working for them during their tutorial.

We posit that a conceptual, rather than a calculation-intensive, initial exposure can make the material less intimidating. Below, we provide selected examples of the diagrams drawn by each student during the Week 1 tutorial, in response to the tutor's question. Seeing a range of examples reinforced understanding, yet each example was relatable as it came from their peers. The students presented a diversity of examples, drawn from their everyday experiences, mathematics, or a topic they found interesting.

\begin{figure}[H]
    \centering
    \includegraphics[width=0.5\linewidth]{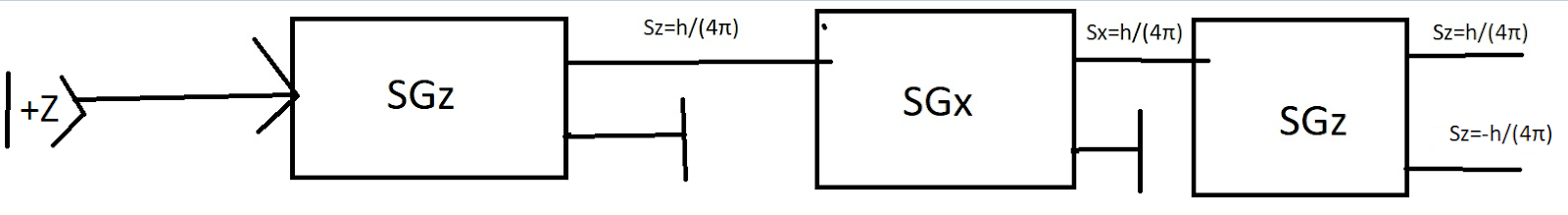}
    \caption{A student drew this and wrote, ``Experiment 3 of the Stern-Gerlach experiments with all initial particles being in a state of spin-up''. This shows that this student had prior exposure to quantum mechanics, likely from the Picturing Quantum Processes textbook by ~\cite{coecke2017picturing}, which shows a likewise diagram of the Stern-Gerlach experiment as an example of measurement in quantum mechanics.}
    \label{fig:student-diagram-stern-gerlach}
\end{figure}

\begin{figure}[H]
    \centering
    \includegraphics[width=0.5\linewidth]{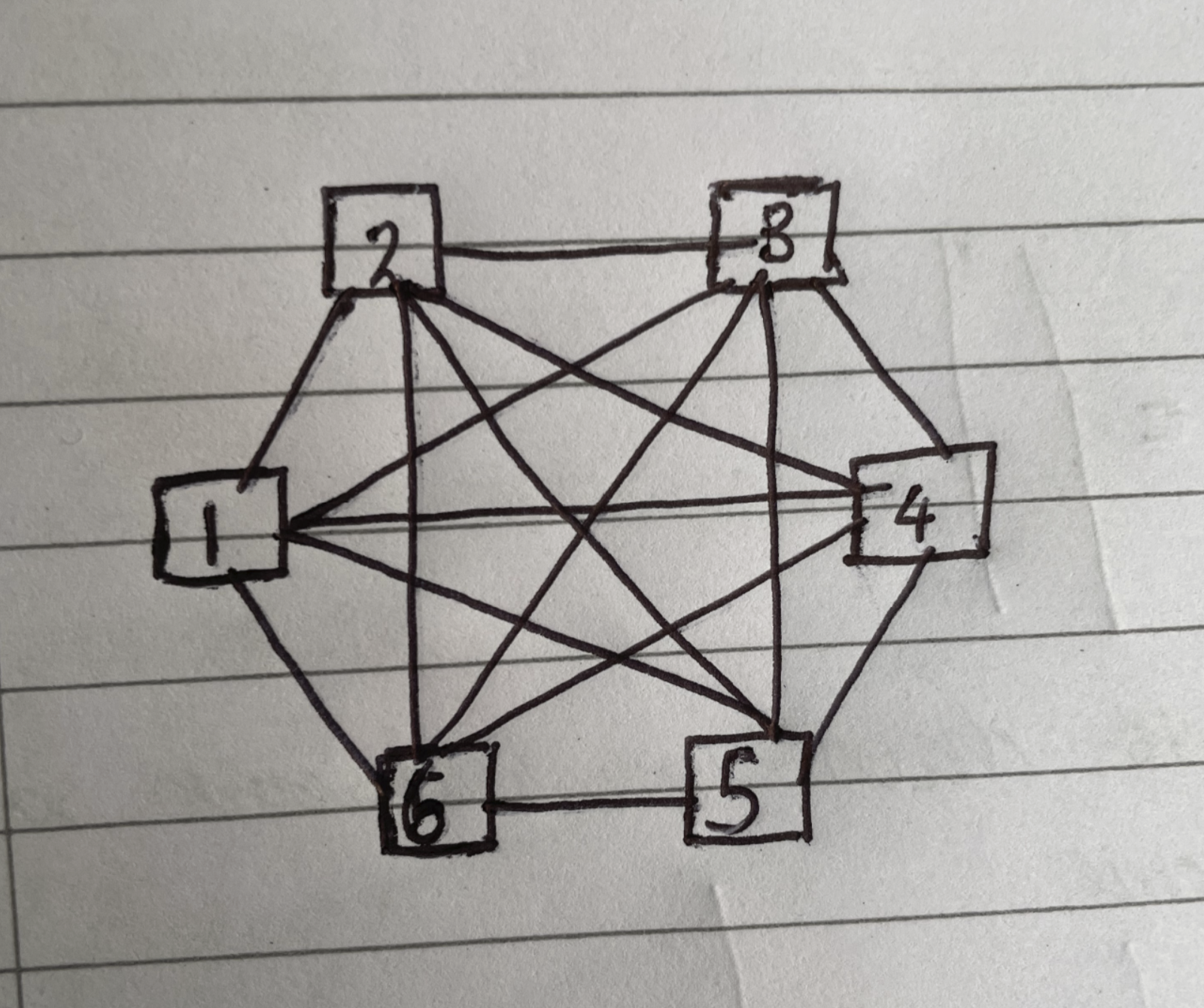}
    \caption{Although graphs were not mentioned in the lecture or the tutorial, a student made the connection with graphs as an example of boxes and wires, sharing a complete graph on six vertices.}
    \label{fig:student-diagram-complete-graph}
\end{figure}

\begin{figure}[H]
\centering
\begin{subfigure}{.5\textwidth}
  \centering
  \includegraphics[width=.7\linewidth]{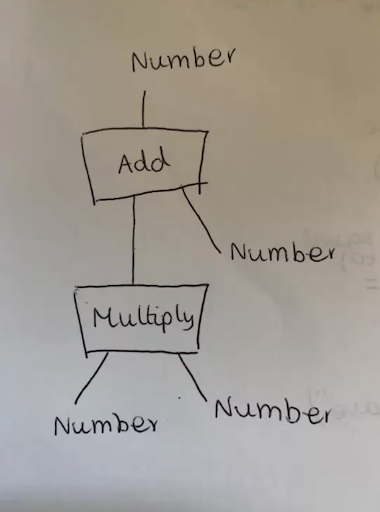}
  \caption{A student drew two numbers being multiplied, the product of which is added to a third number.}
\end{subfigure}%
\begin{subfigure}{.5\textwidth}
  \centering
  \includegraphics[width=.9\linewidth]{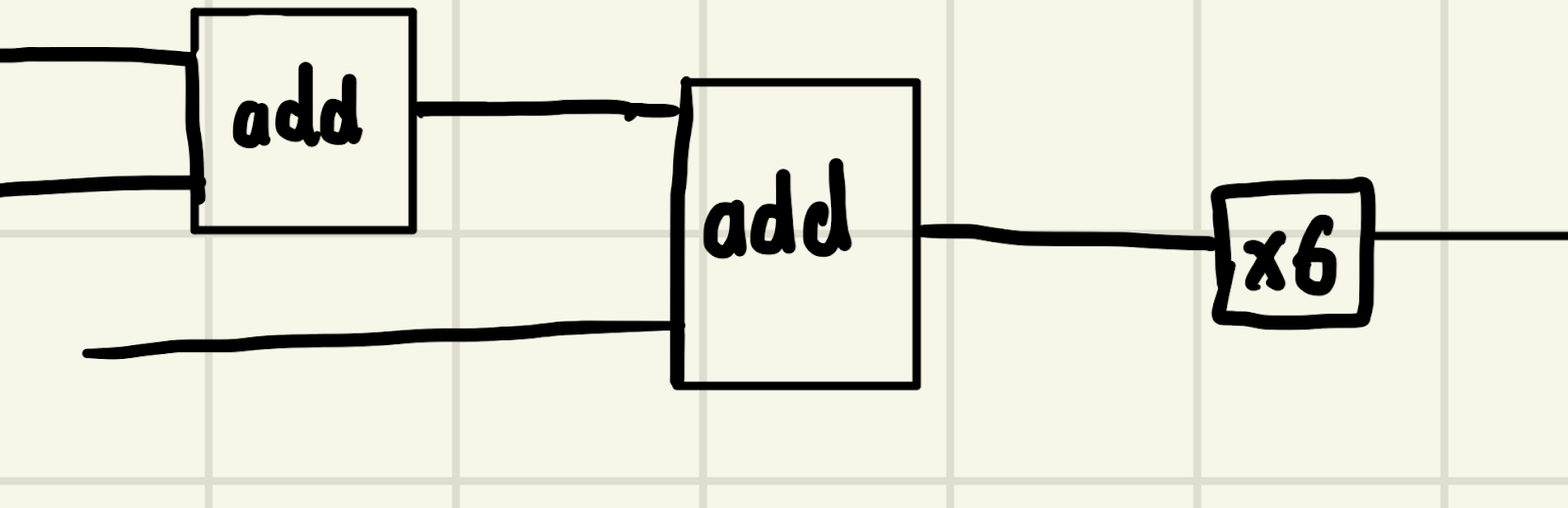}
  \caption{A student described this process as ``a simple diagram on adding 3 numbers then multiplying it by 6''.}
\end{subfigure}
\caption{These two students, during two different Week 1 tutorial sessions, independently realized that wire types could be numbers, which can undergo arithmetic operations.}
\label{fig:student-diagrams-arithmetic}
\end{figure}

\begin{figure}[H]
    \centering
    \includegraphics[width=0.5\linewidth]{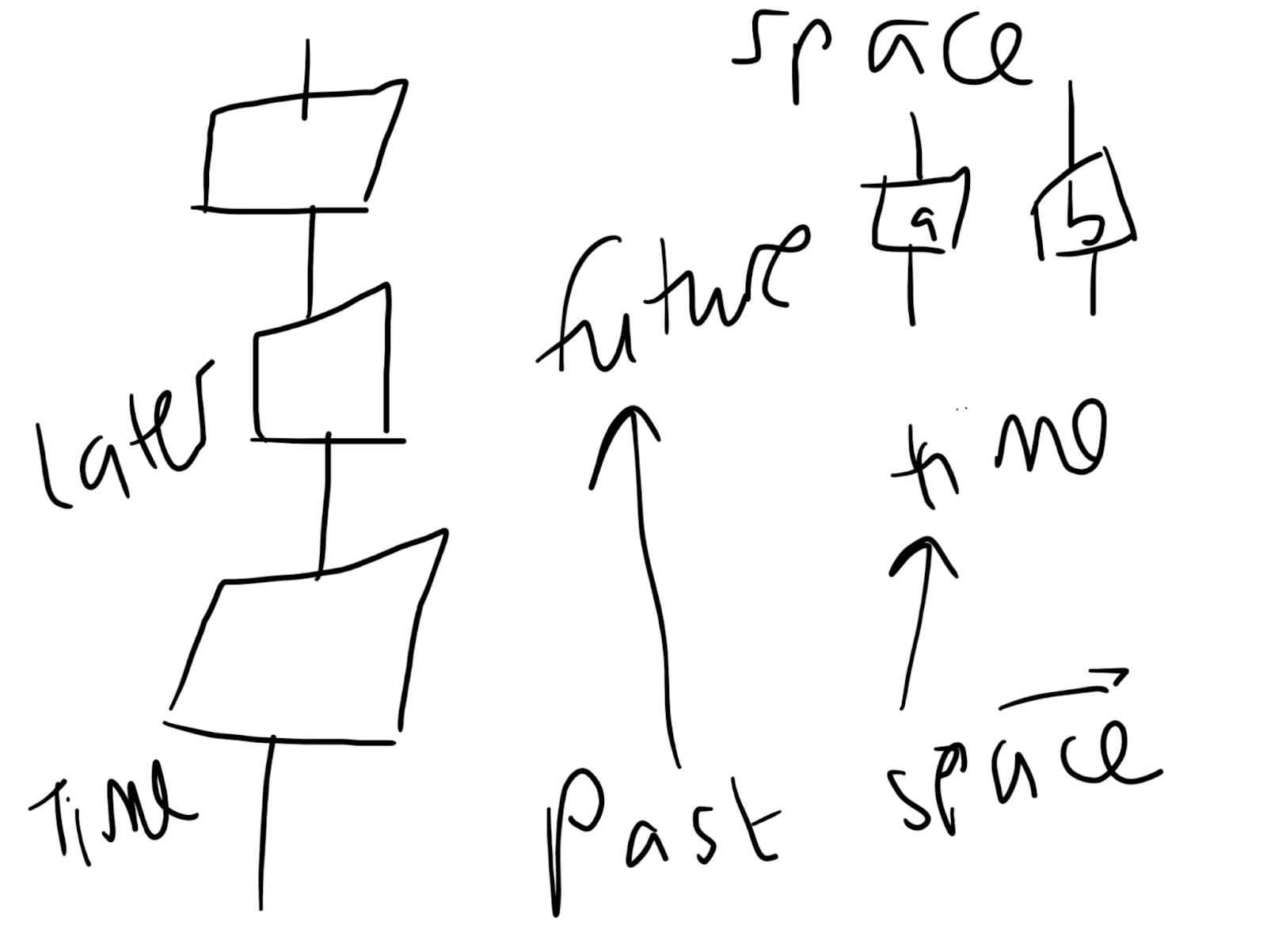}
    \caption{A student wished to better understand the Week 1 material, by challenging themself to explain in their own words a diagram from their notes about abstracting processes occurring across space (horizontal) and time (vertical). Processes happening at the same time but spatially apart are drawn in parallel, while processes happening at different times at the same location are drawn in sequence.}
    \label{fig:student-diagram-space-time}
\end{figure}

\begin{figure}[H]
    \centering
    \includegraphics[width=0.5\linewidth]{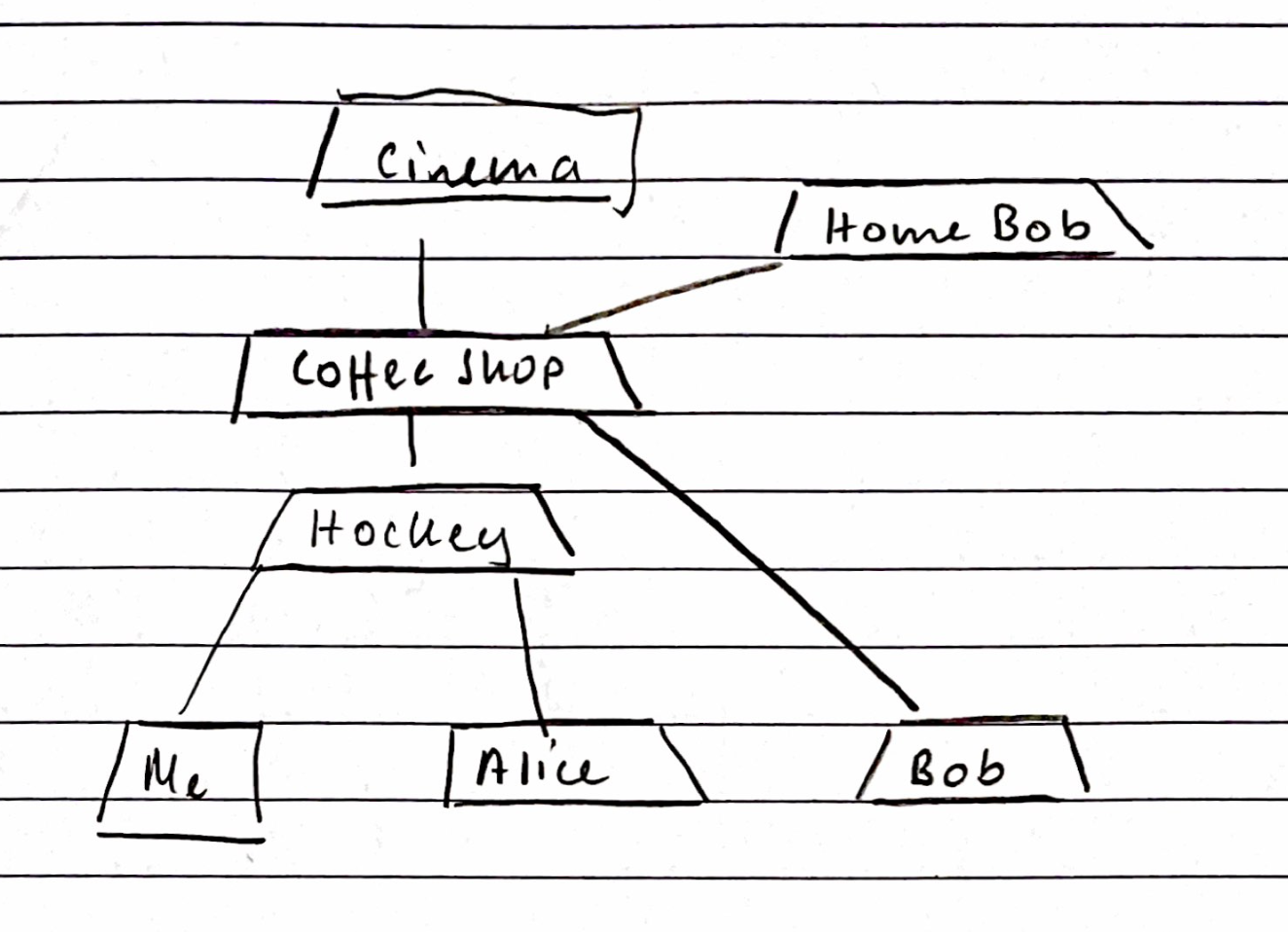}
    \caption{The student sharing this diagram described it as ``Having fun with friends in the afternoon''. Through diagrams such as this, students connected what they learned about process diagrams with their daily lives.}
    \label{fig:student-diagram-fun-with-friends}
\end{figure}

\begin{figure}[H]
    \centering
    \includegraphics[width=0.5\linewidth]{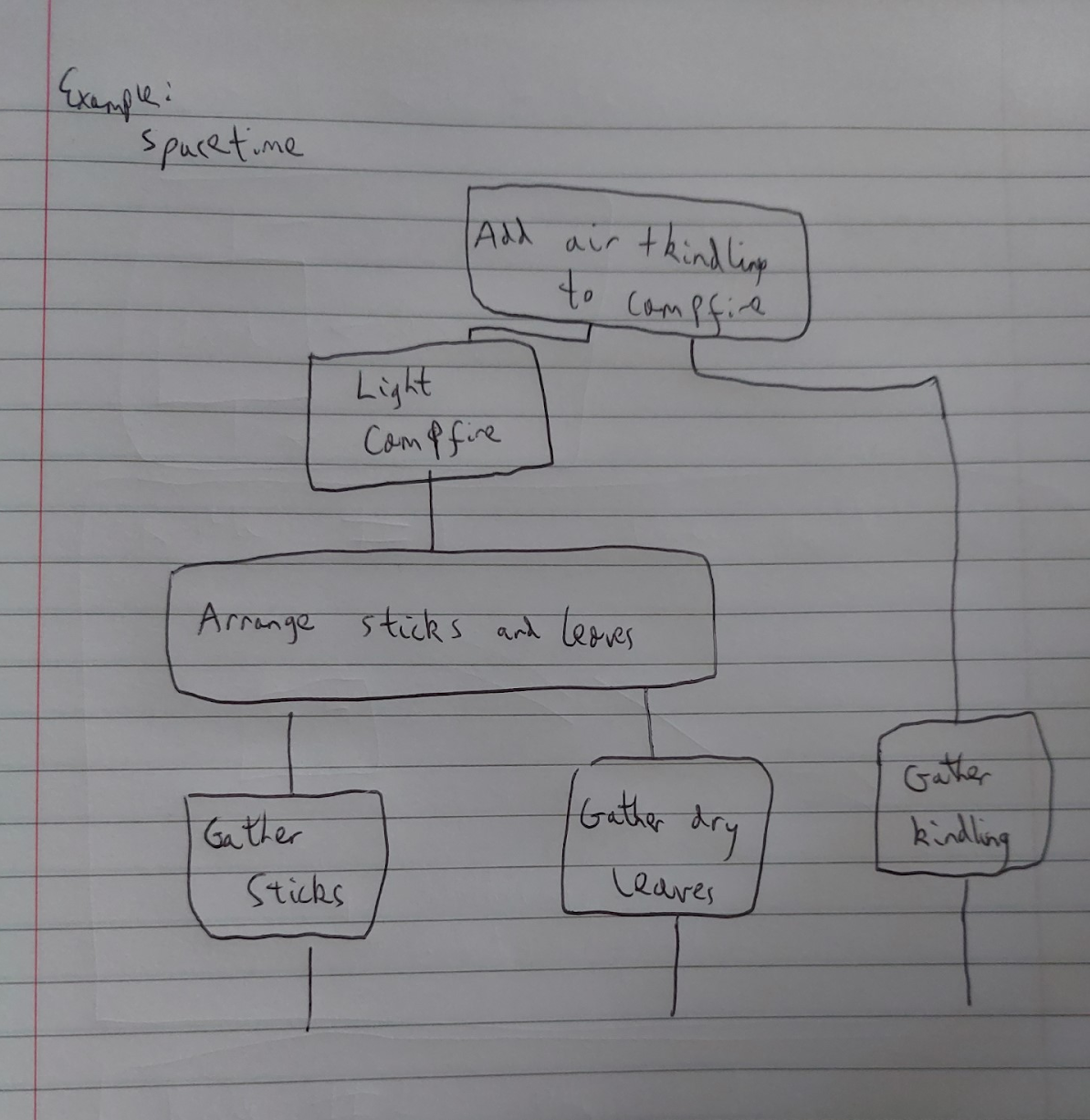}
    \caption{A student shared a diagram representing the process of how to build a fire. This demonstrates the understanding that different wires can represent different types: Here, the sticks and leaves are arranged to become the lit campfire, before the kindling is added to it.}
    \label{fig:student-diagram-campfire}
\end{figure}

\begin{figure}[H]
    \centering
    \includegraphics[width=0.5\linewidth]{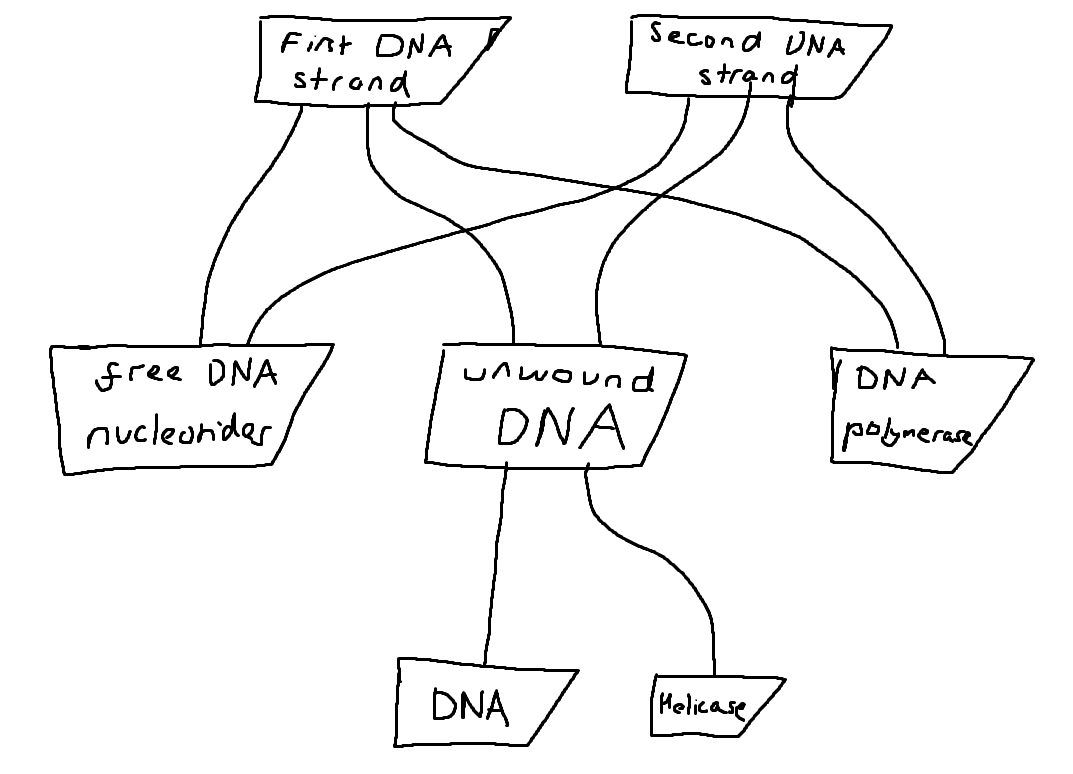}
    \caption{A student shared this diagram and wrote, ``I've been learning about the process of DNA replication in cells, and it has quite a nice process that can be represented across multiple steps using multiple boxes.''}
    \label{fig:student-diagram-dna}
\end{figure}

\begin{figure}[H]
    \centering
    \includegraphics[width=0.5\linewidth]{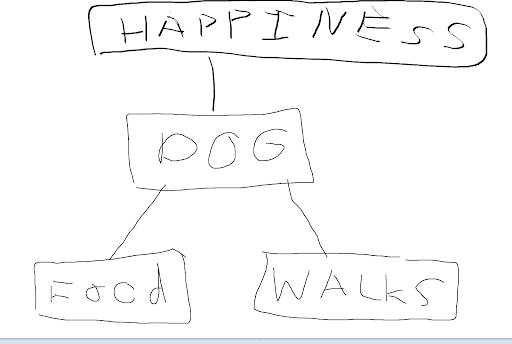}
    \caption{This ``very simple diagram about a dogs life'' brought a smile to the class. The student thought about how they provide food and walks, and the dog brings them happiness.}
    \label{fig:student-diagram-dog-happiness}
\end{figure}

\FloatBarrier
\subsection*{Interventions for students to keep up with the material}
The tutorial sign up was capped at 15 students per tutorial, with exceptions on a case-by-case basis to accommodate students’ availability. These small tutorial class sizes enabled an environment for students to ask questions by raising their hand in Microsoft Teams or by typing in the chat box. More importantly, it enabled the tutor to ask questions of the students, which for some material (such as Week 1, as discussed in the previous section) allowed for sufficient time to hear from each student, while for other material was better suited to request students to volunteer their answer.

Due to the sequential nature of the course and that it took place in summer time and exams for some students, it was accounted for that students who were away for one course week could have difficulty catching up upon their return. To address this, we scheduled 1-2 makeup tutorials for each week’s content during the following week. Students who had attended the week's tutorial were also welcome to attend the makeup tutorial, which could be for reasons including desiring additional review or having previously experienced internet connectivity issues. Although the makeup tutorials were for only a few students per week, rather than losing those few students per week (with regards to either their participation or understanding), we believe this additional safeguard increased the weekly student retention.

Office hours (i.e. optional drop-in sessions) were provided on a daily basis in the hour prior to the tutorials, to create more opportunities for students to ask questions about the material. However, very few students attended these, with a majority of the office hours having no students in attendance. This could be due to a number of reasons, such as it being intimidating, unclear as to its intended purpose, or overlapping in functionality with the tutorials in which students did ask questions. This suggests that office hours are an ineffective way to engage students in an online course format.

Tutors were encouraged to take notes, and to communicate with the adjacent weeks’ tutors on what material was covered time permitting, as well as what could benefit from review on. Each tutor was tasked to incorporate at least one interactive component in their tutorials, defined as being not purely lecture but rather involving student participation.

In addition to serving as material for student practice, the quizzes provided feedback to the course organizers as to the status of students’ conceptual understanding of the material, and which concepts to reinforce in the upcoming tutorials.

\subsection*{Enrichment of the learning material and beyond}
At the halfway point of the course, in Week 5, we held a ``Meet the Professors'' event where the students had the opportunity to meet the professors organizing the course whom they have been seeing in the recorded lectures and whom authored the course textbook. The outcome was for the students to realize that professors are people too, gaining exposure to non-curricular aspects of higher education. The students’ questions were in themselves interesting, so we have provided them verbatim to shed insight on what the students wanted to know most:
\begin{itemize}
    \item Why did you guys go into quantum computing?
    \item What ways are there to get more involved in this field? Are there any articles or Youtubers that you would recommend?
    \item Do you need much group theory for the linear algebra in quantum computing?
    \item Is there much opportunity to get into quantum computing through the Oxford physics course? I know there is a quantum mechanics module and a 4th-year option for mathematics and physics, but there's not much information on computing
    \item why is quantum computing important, how can we benefit from it in the future
    \item Would you consider quantum computers and classical computers incompatible, or do you see them working in conjunction?
    \item I came across a Hadamard gate in some of your work. From what I've seen they can turn collapsed states into superposed ones. Is there some interesting way that these work actual computing?
    \item whats stopping us from reaching functional and efficient quantum computers is it the lack of hardware or the lack of the theory that goes behind them?
    \item Do you think quantum computers will ever follow a path similar to classical computers where they can become common household items? or due to the sensitivity and complexity behind them they will be reserved for research uses
    \item With Quantum teleportation how do you ensure it goes to the right location? [Follow up question:] If you send it through light (I'm thinking of like an optic cable) does that mean there is a slight logistical challenge in space as you may require a cable? Otherwise surely the light may get blocked/absorbed
\end{itemize}

In Week 6, the tutor experimented with the website skribbl.io as a medium to gamify the material. Skribbl.io’s description of itself is that it is ``is a free online multiplayer drawing and guessing pictionary game. A normal game consists of a few rounds, where every round a player has to draw their chosen word and others have to guess it to gain points!''. The incorporation of this game to draw diagrams prompted by a custom word bank of the learned concepts, was well-received by the students. Some students even requested additional skribbl.io sessions to be held for fun, a few of which (accounting for the limit on the number of participants a skribbl.io game can accommodate) were organized following the end of the course as a fun cumulative review to celebrate the new knowledge gained.

Finally, also after the end of the course, a Careers in Quantum informational presentation and Q\&A session was organized. This gave the students context for what real quantum computers look like, applications of quantum computation and quantum technologies, and careers in quantum science and technology and how to pursue them.
}
\end{appendices}


\end{document}